\documentclass[useAMS,usenatbib,usegraphicx]{mn2e}
\usepackage{psfig,amssymb}
\usepackage{times,color}
\usepackage{amsmath}
\begin{document}
\title[Dark matter annihilation in dwarf spheroidal galaxies and $\gamma$-ray
observatories: I. Classical dSphs]{Dark matter profiles and
  annihilation
  in dwarf spheroidal galaxies: prospectives for present and future $\gamma$-ray observatories\\
  I. The classical dSphs} \author[Charbonnier, Combet, Daniel et
al.]{A. Charbonnier$^{1}$, C. Combet$^{2}$, M. Daniel$^{4}$,
  S. Funk$^{5}$,
  J.A. Hinton$^{2}$\thanks{E-mails:jah85@leicester.ac.uk (JH),
    dmaurin@lpsc.in2p3.fr (DM), walker@ast.cam.ac.uk (MGW)},
  D. Maurin$^{6,1,2,7}$\footnotemark[1], \newauthor C. Power$^{2,3}$,
  J. I. Read$^{2,11}$, S. Sarkar$^{8}$,
  M. G. Walker$^{9,10}$\footnotemark[1], M. I. Wilkinson$^{2}$
  \\
  $^1$Laboratoire de Physique Nucl\'eaire et Hautes Energies,
  CNRS-IN2P3/Universit\'es Paris VI et Paris VII,
  4 place Jussieu, Tour 33, 75252 Paris Cedex 05, France\\
  $^2$Dept. of Physics and Astronomy, University of Leicester, Leicester, LE1 7RH, UK\\
  $^3$International Centre for Radio Astronomy Research, University of Western
   Australia, 35 Stirling Highway, Crawley, Western Australia 6009,
   Australia\\
  $^4$Dept. of Physics, Durham University, South Road, Durham, DH1 3LE, UK\\
  $^5$W. W. Hansen Experimental Physics Laboratory, Kavli Institute for Particle Astrophysics and Cosmology, Department of Physics and SLAC National Accelerator Laboratory, Stanford University, Stanford, CA 94305, USA\\
  $^6$Laboratoire de Physique Subatomique et de Cosmologie, CNRS/IN2P3/INPG/Universit\'e Joseph Fourier Grenoble 1,53 avenue des Martyrs, 38026 Grenoble, France\\
  $^7$Institut d'Astrophysique de Paris, UMR7095 CNRS, Universit\'e Pierre et Marie Curie, 98 bis bd Arago, 75014 Paris, France\\
  $^8$Rudolf Peierls Centre for Theoretical Physics, University of Oxford, 1 Keble Road, Oxford, OX1 3NP, UK\\
  $^9$Institute of Astronomy, University of Cambridge, Madingley Road, Cambridge, CB3 0HA, UK\\
  $^{10}$Harvard-Smithsonian Center for Astrophysics, 60 Garden St., Cambridge, MA 02138, USA\\
  $^{11}$Institute for Astronomy, Department of Physics, ETH Z\"urich, Wolfgang-Pauli-Strasse 16, CH-8093 Z\"urich, Switzerland
}
\pagerange{\pageref{firstpage}--\pageref{lastpage}} \pubyear{Xxxx}
\date{Accepted Xxxx. Received Xxxx; in original form Xxxx}
\label{firstpage}

\maketitle

\begin{abstract}
  Due to their large dynamical mass-to-light ratios, dwarf spheroidal
  galaxies (dSphs) are promising targets for the indirect detection of
  dark matter (DM) in $\gamma$-rays. We examine their
  detectability by present and future $\gamma$-ray observatories. The
  key innovative features of our analysis are: (i) We take into account the {\it angular size} of the dSphs; while nearby objects have higher
  $\gamma$ ray flux, their larger angular extent can make them
less attractive targets for background-dominated instruments. (ii) We
  derive DM profiles and the astrophysical $J$-factor (which
  parameterises the expected $\gamma$-ray flux, independently of the
  choice of DM particle model) for the classical dSphs
  {\em directly} from photometric and kinematic data. We assume very little
  about the DM profile, modelling this as a smooth
  split-power law distribution, with and without  sub-clumps. (iii) We use
  a Markov Chain Monte Carlo (MCMC) technique to marginalise over
  unknown parameters and determine the sensitivity of our derived
  $J$-factors to both model and measurement uncertainties. (iv) We use
  simulated DM profiles to demonstrate that our $J$-factor
  determinations recover the correct solution within our quoted
  uncertainties.
   
  Our key findings are: (i) Sub-clumps in the dSphs do {\em not} usefully
  boost the signal; (ii) The sensitivity of atmospheric Cherenkov
  telescopes to dSphs within $\sim\!20$\,kpc with cored halos can
  be up to $\sim\!50$ times worse than when estimated assuming them to be 
  point-like. Even for the satellite-borne Fermi-LAT the sensitivity is significantly degraded
  on the relevant angular scales for long exposures, hence it is
  vital to consider the angular extent of the dSphs when selecting
  targets; (iii) {\em No} DM profile has been ruled out by current data, but
  using a prior on the inner dark matter cusp slope $0\leq\gamma_{\rm
    prior}\leq1$ provides $J$-factor estimates accurate to a factor of
  a few if an appropriate angular scale is chosen; (iv) The $J$-factor is best constrained
  at a critical
  integration angle $\alpha_c=2r_\mathrm{half}/d$ (where $r_\mathrm{half}$ is the half
  light radius and $d$ is the distance to the dwarf) and we estimate
  the corresponding sensitivity of $\gamma$-ray observatories; (v) The
  `classical' dSphs can be grouped into three categories:
  well-constrained and promising (Ursa Minor, Sculptor, and Draco),
  well-constrained but less promising (Carina, Fornax, and Leo I), and
  poorly constrained (Sextans and Leo II);  (vi) Observations of
  classical dSphs with Fermi-LAT integrated over the mission
  lifetime are more promising than observations with the planned
  Cherenkov Telescope Array for DM particle mass $\lesssim\! 700$~GeV. However, even Fermi-LAT will {\em not} have sufficient integrated signal from the classical dwarfs to detect DM in the `vanilla' Minimal  Supersymmetric Standard Model. Both the Galactic centre and the `ultra-faint' dwarfs are likely to 
 be better targets and will be considered in future work.
\end{abstract}

\begin{keywords}
astroparticle physics ---
(cosmology:) dark matter --- 
Galaxy: kinematics and dynamics ---
$\gamma$-rays: general ---
methods: miscellaneous
\end{keywords}

\section{Introduction\label{sec:intro}}

The detection of $\gamma$-rays from dark matter (DM) annihilation is one of
the most promising channels for indirect detection
\citep{1978ApJ...223.1015G,1978ApJ...223.1032S}. Since the signal goes
as the DM density squared, the Galactic centre seems to be
the obvious location to search for such a signal
\citep{1987ApJ...313L..47S}. However, it is plagued by a confusing
background of astrophysical sources
\citep[e.g.][]{2004A&A...425L..13A}. For this reason, the dwarf
spheroidal galaxies (dSphs) orbiting the Milky Way have been flagged
as favoured targets given their potentially high DM densities and
small astrophysical backgrounds
\citep{1990Natur.346...39L,Evans:2003sc}.

Despite the growing amount of kinematic data from the classical dSphs,
the inner parts of their DM profiles remain poorly
constrained and can generally accommodate both cored or cuspy
solutions
\citep[e.g.][]{koch07b,2007ApJ...669..676S,2009ApJ...704.1274W}.
There are two dSphs|Fornax and Ursa Minor|that show indirect hints of
a cored distribution \citep{2003ApJ...588L..21K,2006MNRAS.368.1073G};
however, in both cases the presence of a core is inferred based on a
timing argument that assumes we are not catching the dSph at a
special moment. Theoretical expectations remain similarly
uncertain. Cusps are favoured by cosmological models that model the
DM alone, assuming it is cold and collisionless
\citep[e.g.][]{navarro96}. However, the complex dynamical interplay
between stars, gas and DM during galaxy formation could erase
such cusps leading to cored distributions
\citep[e.g.][]{1996MNRAS.283L..72N,2005MNRAS.356..107R,2008Sci...319..174M,2010ApJ...725.1707G,2010Natur.463..203G,2011arXiv1105.4050C}.
Cores could also be an indication of other possibilities such as self-interacting dark matter
\citep[e.g.][]{2000PhRvD..62f3511H,Moore:2000fp}.

Knowledge of the inner slope of the DM profile is of critical
importance as most of the annihilation flux comes from that
region. Lacking this information, several  studies have
focused on the detectability of these dSphs by current
$\gamma$-ray observatories such as the satellite-borne Fermi-LAT and atmospheric Cherenkov telescopes (ACTs) such as H.E.S.S., MAGIC and VERITAS,
using a small sample of cusped and cored profiles (generally one of
each). Most studies rely on standard core and cusp profiles fitted to
the kinematic data of the dSph of interest
\citep{2006PhRvD..73f3510B,2007PhRvD..76l3509S,2009JCAP...01..016B,2009MNRAS.399.2033P,2009A&A...496..351P}. Other
authors use a `cosmological prior' from large scale cosmological
simulations \citep[e.g.][]{2010AdAst2010E..45K}. Both approaches may
be combined, such as in \citet{2007PhRvD..75h3526S} and
\citet{2009JCAP...06..014M} who rely partially on the results of
structure formation simulations to constrain the inner slope and then
perform a fit to the data to derive the other parameters. However such
cosmological priors remain sufficiently uncertain that their use is
inappropriate for guiding observational strategies. 
There have been only a few studies \citep[e.g.,][]{2009PhRvD..80b3506E} 
which have \emph{not} assumed strong priors for the DM profiles.

In this work, we revisit the question of the detectability of dark
matter annihilation in the classical Milky Way dSphs, motivated by ambitious plans for
next-generation ACTs such as the Cherenkov
Telescope Array (CTA). We rely {\em solely} on published kinematic
data to derive the properties of the dSphs, making minimal assumptions
about the underlying DM distribution.  Most importantly, we
do not restrict our survey of DM profiles to those suggested by
cosmological simulations. We also consider the effect of the spatial
extent of the dSphs, which becomes important for nearby systems
observed by background-limited instruments such as ACTs.

This paper extends the earlier study of \citet{letter_dsph} which showed that there is a critical integration angle (twice the
half-light radius divided by the dSph distance) where we can obtain a
robust estimate of the $J$-factor (that parameterises the expected
$\gamma$-ray flux from a dSph independently of the choice of dark
matter particle model; see Section \ref{sec:method}), regardless of
the value of the central DM cusp slope $\gamma$. Here, we
focus on the full radial dependence of the $J$-factor. We consider the
effect of DM sub-lumps within the dSphs, discuss which
dSphs are the best candidates for an observing programme, and examine
the competitiveness of next-generation ACTs as dark
matter probes.

This paper is organised as follows. In Section~\ref{sec:method}, we
present a study of the annihilation $\gamma$-ray flux, focusing on
which parameters critically affect the expected signal. In
Section~\ref{sec:detectability}, we discuss the sensitivity of
present/future $\gamma$-ray observatories. In Section~\ref{sec:mcmc},
we present our method for the dynamical modelling of the observed
kinematics of stars in dSphs. In Section~\ref{sec:results}, we
derive DM density profiles for the classical dSphs using an
MCMC analysis, from which the detection potential of future
$\gamma$-ray observatories can be assessed. We present our
conclusions in Section~\ref{sec:conclusions}.\footnote{Technical details are deferred to Appendices. In
  Appendix~\ref{app:defs}, we comment on the various notations used in
  similar studies and provide conversion factors to help compare 
  results. In Appendix~\ref{app:toyJ}, we provide a toy model for quick
  estimates of the $J$-factor. In Appendix~\ref{app:boost}, we calculate
  in a more systematic fashion the range of the possible `boost factor'
  (due to DM clumps within the dSphs) for
  generic dSphs.  In Appendix~\ref{app:PSF}, we show that convolving the
  signal by the PSF of the instrument is equivalent to a cruder
  quadrature sum approximation. In Appendix~\ref{app:CLs}, we discuss some
  technical issues related to confidence level determination from the
  MCMC analysis.  In Appendix~\ref{app:fake}, the reconstruction method is
  validated on simulated dSphs. In Appendix~\ref{app:biases}, we discuss
  the impact of the choice of the binning of the stars and of the shape
  of the light profile on the $J$-factor determination.}

This paper includes detailed analyses from both high-energy
astrophysics and stellar dynamical modelling. To assist readers from
these different fields in navigating the key sections, we
suggest that those who are primarily interested in the high-energy
calculations may wish to focus their attention on
Sections~\ref{sec:method},~\ref{sec:detectability} and
\ref{sec:results} before moving to the conclusions. Readers from the
dynamics community may instead prefer to read
Sections~\ref{sec:method},~\ref{sec:mcmc}
and~\ref{sec:results}. Finally, those who are willing to trust
the underlying modelling should proceed to Section~\ref{sec:results}
where our main results regarding the detectability of dSphs
are presented in Figs.~\ref{fig:JdSph_Jdm_bkgd},
~\ref{fig:Jd_gamma_fixed}, ~\ref{fig:min_detect_sigmav} and
\ref{fig:mssm}.

\section{The dark matter annihilation signal: key parameters}
\label{sec:method}

\subsection{The $\gamma$-ray flux}

The $\gamma$-ray flux $\Phi_{\gamma}$ (photons cm$^{-2}$~s$^{-1}$~GeV$^{-1})$ from DM annihilations
in a dSph, as seen within a solid angle $\Delta\Omega$, is given by (see
Appendix~\ref{app:defs} for definitions and conventions used in the
literature):

\begin{equation}
     \frac{\mathrm{d}\Phi_{\gamma}}{\mathrm{d}E_{\gamma}}(E_{\gamma},\Delta\Omega)
        = \Phi^{\rm pp}(E_{\gamma}) \times J(\Delta\Omega)\,,
\label{eqn:dmanihil}
 \end{equation}
 The first factor encodes the (unknown) particle physics of DM annihilations which we wish to measure. The second factor encodes
 the astrophysics {\em viz.}
 the l.o.s. integral of the DM
 density-squared over solid angle $\Delta\Omega$ in the dSph --- this is
 called the `$J$-factor'. We now discuss each factor in turn.

\subsubsection{The particle physics factor} 
\label{sec:dNdE_xsec}
The particle physics factor ($\Phi^{\rm pp}$) is given by: 

\begin{equation}
     \Phi^{\rm pp}(E_{\gamma})\equiv  \frac{\mathrm{d}\Phi_{\gamma}}{\mathrm{d}E_{\gamma}}
        = \frac{1}{4\pi}\frac{\langle\sigma_{\rm
ann}v\rangle}{2m_{\chi}^{2}}
          \times \frac{dN_{\gamma}}{dE_{\gamma}}\;,
 \end{equation}
 where $m_{\chi}$ is the mass of the DM particle,
 $\sigma_{\rm ann}$ is its self-annihilation cross-section and $\langle
 \sigma_{\rm ann} v\rangle$ the average over its
 velocity distribution, and $\mathrm{d}N_{\gamma}/\mathrm{d} E_{\gamma}$ is the
 differential photon yield per annihilation. A benchmark value is $\langle \sigma_\mathrm{ann} v\rangle\sim
 3\times 10^{-26}\, {\rm cm}^3\, {\rm s}^{-1}$
 \citep{1996PhR...267..195J}, which would result in a present-day DM
 abundance satisfying cosmological constraints. 

 Unlike the annihilation cross section and particle mass, the
 differential annihilation spectrum ($\mathrm{d}N_{\gamma}/\mathrm{d}
 E_{\gamma}(E_\gamma)$) requires us to adopt a specific DM
 particle model. We focus on a well-motivated class of models that are within
 reach of up-coming direct and indirect experiments: the Minimal
 Supersymmetric Standard Model (MSSM). In this framework, the
 neutralino is typically the lightest stable particle and therefore one
 of the most favoured DM candidates (see
 e.g. \citealt{2005PhR...405..279B}). A $\gamma$-ray continuum is
 produced from the decay of hadrons  (e.g. $\pi^0 \rightarrow \gamma\gamma$) resulting from the
 DM annihilation. Neutralino
 annihilations can also directly produce mono-energetic $\gamma$-ray
 lines through loop processes, with the formation of either a pair of
 $\gamma$-rays ($\chi\chi\rightarrow \gamma\gamma$;
 \citealp{1997NuPhB.504...27B}), or a $Z^0$ boson and a $\gamma$-ray
 ($\chi\chi\rightarrow \gamma Z^0$; \citealp{1998PhRvD..57.1962U}). We
 do not take into account such line production processes since they
 are usually sub-dominant and very model dependent
 \citep{2008JHEP...01..049B}. The differential photon spectrum we use
 is restricted to the continuum contribution and is written as:

\begin{equation}
\frac{\mathrm{d} N_{\gamma}}{\mathrm{d} E_{\gamma}} (E_\gamma)
= \sum_i b_i \, \frac{\mathrm{d} N_\gamma^i}{\mathrm{d} E_\gamma}
(E_\gamma, m_\chi) \; ,
\label{eq:3contribs}
\end{equation}
%
where the different annihilation final states $i$ are characterised by
a branching ratio $b_i$.

\begin{figure}
\includegraphics[width=\linewidth]{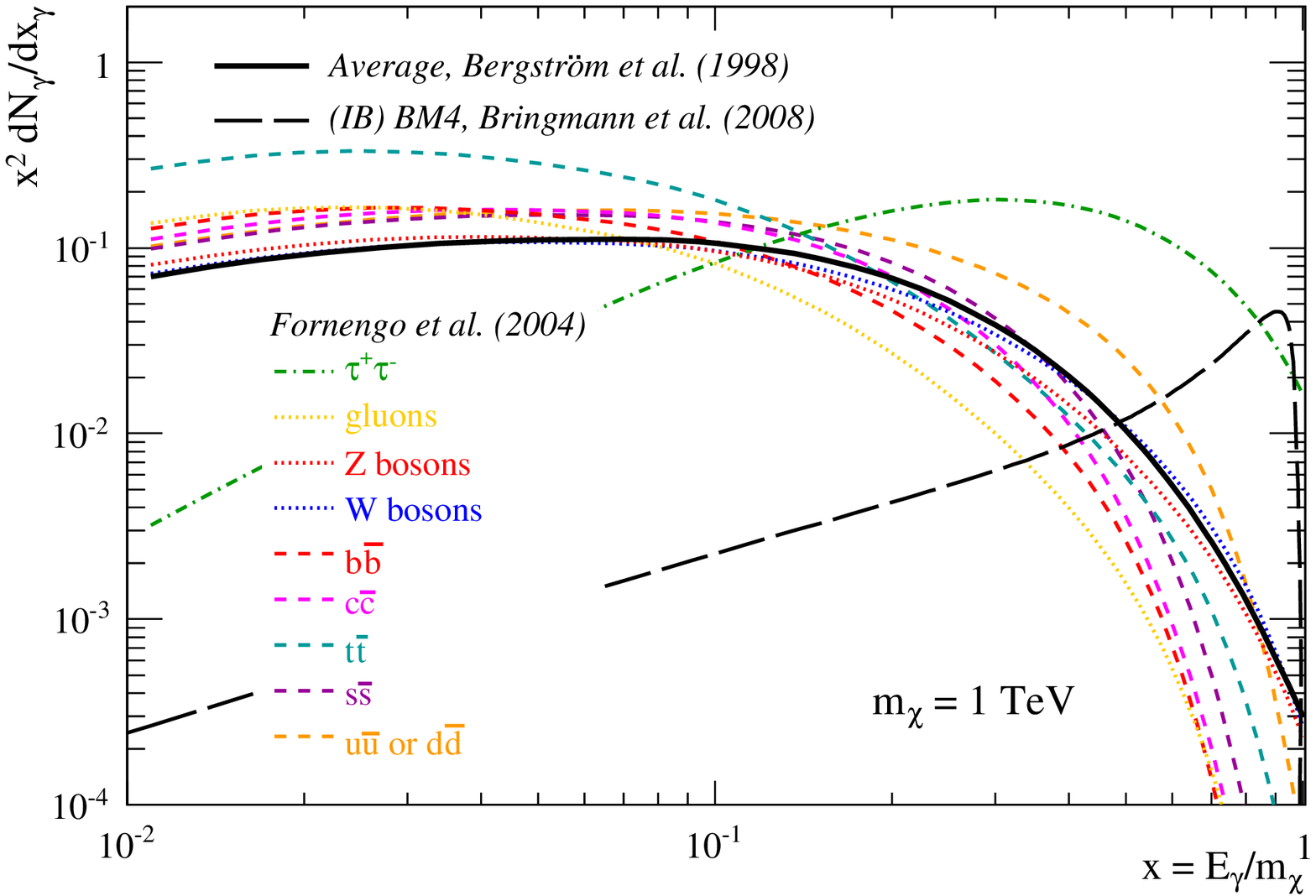}
\caption{Differential spectra (multiplied by $x^2$) of $\gamma$-rays
  from the fragmentation of neutrino annihilation products (here for a
  DM particle mass of $m_\chi = 1$~TeV). Several different
  channels are shown, taken from \citet{2004PhRvD..70j3529F} and an
  average parametrisation \citet{1998APh.....9..137B} is marked by the
  black solid line; this is what we adopt throughout this paper. The
  black dashed line is the benchmark model BM4
  \citep{2008JHEP...01..049B} which includes internal bremsstrahlung
  and serves to illustrate that very different spectra are
  possible. However, the example shown here is dominated by line
  emission and therefore highly model dependent; for this reason, we
  do not consider such effects in this paper.}
\label{fig:phi-susy}
\end{figure}

Using the parameters in \citet{2004PhRvD..70j3529F}, we plot the
continuum spectra calculated for a 1 TeV mass neutralino in
Fig. ~\ref{fig:phi-susy}.
Apart from the $\tau^+\tau^-$ channel (dash-dotted line), all the
annihilation channels in the continuum result in very similar spectra
of $\gamma$-rays (dashed lines). For charged annihilation products,
internal bremsstrahlung (IB) has recently been investigated and 
found to enhance the spectrum close to the kinematic cut-off (e.g.,
\citealp{2008JHEP...01..049B}). As an illustration, the long-dashed
line in Fig.~\ref{fig:phi-susy} corresponds to the benchmark
configuration for a wino-like neutralino taken from
\citet{2008JHEP...01..049B}. However, the shape and amplitude of this
spectrum are strongly model dependent \citep{2009JCAP...01..016B} and,
as argued in \citet{2010PhRvD..81j7303C}, this contribution is
relevant only for models (and at energies) where the line contribution
is dominant over the secondary photons.

We wish to be as model-independent as
possible, and so do not consider internal bremsstrahlung. In the
remainder of this paper, all our results will be based on an {\em
  average} spectrum taken from the parametrisation
(\citealp{1998APh.....9..137B}, solid line in
Fig.~\ref{fig:phi-susy}):
\begin{equation}
\frac{\mathrm{d} N_{\gamma}}{\mathrm{d} E_{\gamma}} =
\frac{1}{m_\chi}\frac{\mathrm{d} N_{\gamma}}{\mathrm{d} x} =
\frac{1}{m_\chi} \frac{0.73 \, e^{-7.8x}}{x^{1.5}}\, ,
\label{eq:param-berg}
\end{equation}
with $x \equiv E_{\gamma}/m_{\chi}$. Finally, in order to be conservative in deriving
detection limits, we also do not consider the possible `Sommerfeld
enhancement' of the DM annihilation cross-section
\citep{2004PhRvL..92c1303H,2005PhRvD..71f3528H}.\footnote{This effect
  depends on the mass and the velocity of the particle; the resulting
  boost of the signal and the impact on detectability of the dSphs has
  been discussed, e.g., in \citet{2009MNRAS.399.2033P}.} This
 depends inversely on the DM particle velocity, and thus
requires precise modelling of the velocity distribution of the DM
within the dSph; we will investigate this in a separate study.

\subsubsection{The J-factor}

The second term in Eq.~(\ref{eqn:dmanihil}) is the astrophysical
{\it J-factor} which depends on the spatial distribution of
DM as well as on the beam size. It corresponds to
the l.o.s. integration of the DM density
squared over solid angle $\Delta\Omega$ in the dSph:
 
 \begin{equation}
      J = \int_{\Delta\Omega}\int \rho_{\rm DM}^2 (l,\Omega) \,dld\Omega.
      \label{eq:J}
 \end{equation}
 The solid angle is simply related to the integration angle
 $\alpha_{\rm int}$ by
\[
  \Delta\Omega = 2\pi\cdot(1-\cos(\alpha_{\rm int})) \,.
\]
The $J$-factor is useful because it allows us to rank the dSphs by
their expected $\gamma$-ray flux, independently of any assumed DM
particle physics model. Moreover, the knowledge of the relative
$J$-factors would also help us to evaluate the validity of any
potential detection of a given dSph, because for a given particle
physics model we could then scale the signal to what we should expect
to see in the other dSphs.

All calculations of $J$ presented in this paper were performed
using the publicly available {\tt CLUMPY} package
(Charbonnier, Combet, Maurin, in preparation) which includes models
for a smooth DM density profile for the dSph, clumpy dark
matter sub-structures inside the dSph, and a smooth and clumpy Galactic
DM distribution. \footnote{In Appendix~\ref{app:toyJ}, we provide
approximate formulae for quick estimates of the $J$-factor and
cross-checks with the numerical results.}

\subsubsection{DM profiles}
For the DM halo we use a generalised ($\alpha,\beta,\gamma$)
Hernquist profile given by
\citep{hernquist90,1993MNRAS.265..250D,zhao96}:
\begin{equation}
  \rho(r)=\rho_\mathrm{s}\biggl (\frac{r}{r_\mathrm{s}}\biggr )^{-\gamma}\biggl [1+\biggl (\frac{r}{r_\mathrm{s}}\biggr )^{\alpha}\biggr ]^{\frac{\gamma-\beta}{\alpha}},
  \label{eq:hernquist1}
\end{equation}
where the parameter $\alpha$ controls the sharpness of the transition
from inner slope, $\lim_{r\rightarrow 0}\mathrm{d}\ln(\rho)/\mathrm{d}\ln(r)= -\gamma$, to
outer slope $\lim_{r\rightarrow \infty}\mathrm{d}\ln(\rho)/\mathrm{d}\ln(r)= -\beta$, and $r_\mathrm{s}$ is a characteristic scale. 
In principle we could add an additional parameter in order to introduce an exponential cut-off in the
profile of Eq~(\ref{eq:hernquist1}) to mimic the effects of tidal truncation, as proposed in e.g., 
the Aquarius  \citep{2008MNRAS.391.1685S} or Via Lactea II \citep{2008Natur.454..735D} simulations.
However, the freedom to vary parameters $r_s$, $\alpha$ and $\beta$ in Eq~(\ref{eq:hernquist1}) 
already allows for density profiles that fall arbitrarily steeply at large radius.
Moreover, given that our MCMC analysis later shows that the outer slope $\beta$ is unconstrained by the
available data and that the J-factor does not correlate with $\beta$, we choose not to add further shape
parameters.

For profiles such as $\gamma\geq1.5$, the quantity $J$ from the inner
regions diverges. This can be avoided by introducing a saturation
scale $r_{\rm sat}$, that  corresponds physically to the typical scale
where the annihilation rate $[\langle \sigma v\rangle \rho(r_{\rm
  sat})/m_\chi]^{-1}$ balances the gravitational infall rate of DM
particles $(G\bar{\rho})^{-1/2}$ \citep{1992PhLB..294..221B}. Taking
$\bar{\rho}$ to be about 200 times the critical density gives
\begin{equation}
  \rho_{\rm sat}\approx 3 \times 10^{18}
     \left(\frac{m_\chi}{100~\rm GeV}\right) \times
     \left( \frac{10^{-26} {\rm cm}^3~{\rm s}^{-1}}
	  {\langle \sigma v\rangle}\right)
   M_{\odot}~{\rm kpc}^{-3}.
  \label{eq:rho_sat} 
\end{equation}
The associated saturation radius is given by
\begin{equation}
  r_{\rm sat} = r_\mathrm{s} \left(\frac{\rho_\mathrm{s}}{\rho_{\rm sat}}\right)^{1/\gamma}\ll r_\mathrm{s}\;.
 \label{eq:r_sat}
\end{equation}
This limit is used for all of our calculations. 

\subsection{Motivation for a generic approach and reference models}

In many studies, the $\gamma$-ray flux (from DM annihilations) is
calculated using the point-source approximation
\citep[e.g.,][]{2006PhRvD..73f3510B,2010AdAst2010E..45K}.  This is
valid so long as the inner profile is steep, in which case the total
luminosity of the dSph is dominated by a very small central
region. However, if the profile is shallow and/or the dSph is nearby,
the effective size of the dSph on the sky is larger than the point
spread function (PSF) of the detector, and the point-source
approximation breaks down. For upcoming instruments and particularly
shallow DM profiles, the effective size of the dSph may even be
comparable to the field of view of the instrument. This difference in
the radial extent of the signal does matter in terms of detection (see
Section~\ref{sec:detectability}). Hence we do not assume
that the dSph is a point-source but rather derive
sky-maps for the expected $\gamma$-ray flux.

\subsubsection{Illustration: a cored vs cusped profile}

\begin{figure}
\includegraphics[width=\linewidth]{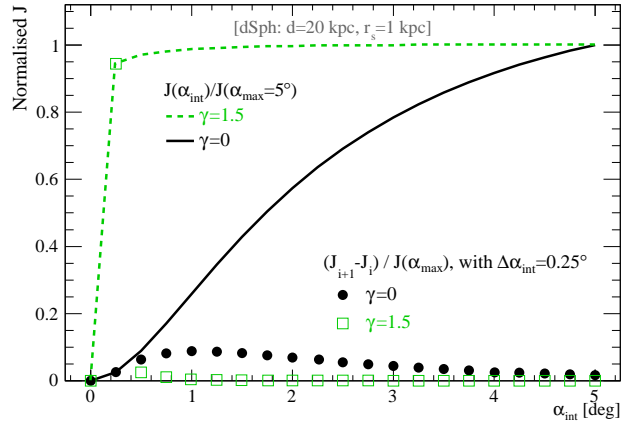}
\caption{Finite size effects: $J$ as a function of the integration
  angle $\alpha_{\rm int}$ for a dSph at 20\,kpc (pointing towards
  the centre of the dSph). The black solid line is for a cored
  profile ($\gamma=0)$ and the green dashed line is for a cuspy
  profile ($\gamma=1.5$); both are normalised to unity at $\alpha_{\rm
    int}=5^{\circ}$.}
\label{fig:cumul_generic_norm}
\end{figure}

Fig.~\ref{fig:cumul_generic_norm} shows $J$ as a function of the
integration angle $\alpha_{\rm int}$ for a dSph at 20\,kpc (looking
towards its centre). The black solid line is for a cored
profile ($\gamma=0)$ and the green dashed line is for a cuspy profile
($\gamma=1.5$); both are normalised to unity at $\alpha_{\rm
  int}=5^{\circ}$. For the cuspy profile, $\sim\!100\%$ of the signal
is in the first bin while for the cored profile,
$J$  builds up slowly with $\alpha_{\rm int}$, and $80\%$ of the signal
(w.r.t. the value for $\alpha_{\rm int}=5^{\circ}$) is obtained for
$\alpha^{80\%}\approx 3^\circ$.
This is also indicated by the symbols which show the contribution
of DM {\em shells} in two angular bins --- whereas the (green) hollow squares
have a spiky distribution in the first bin ($\gamma=1.5$), the (black) filled
circles ($\gamma=0)$ show a very broad distribution for $J$.

The integration angle required to have a sizeable fraction of the
signal depends on several parameters: the distance $d$ of the dSph,
the inner profile slope $\gamma$, and the scale radius
$r_\mathrm{s}$. Small integration angles are desirable since this
minimises contaminating background $\gamma$-ray photons and maximises
the signal to noise. Thus the true detectability of a dSph will
depend on its spatial extent on the sky, and thus also on $d$,
$\gamma$ and $r_\mathrm{s}$.

\subsubsection{Generic dSph profiles}
\label{sec:generic}

As will be seen in Section~\ref{sec:results}, the errors on the density profiles of the Milky
Way dSphs are large, making it difficult to
disentangle the interplay between the key parameters for
detectability. Hence we select some `generic profiles'
to illustrate the key dependencies.

The most constrained quantity is the mass within the
half-light radius $r_\mathrm{\rm half}$ (typically a few tenths of a
kpc), as this is where most of the kinematic data
come from~\citep[e.g.,][]{2009ApJ...704.1274W,2010MNRAS.406.1220W}. For the
classical Milky Way dSphs, the typical mass within $r_\mathrm{\rm
  half} \sim 300$\,pc is found to be $M_{300} \sim 10^7 M_\odot$
\citep[][--- see also the bottom panel of
Fig.~\ref{fig:cor_par5}]{2008Natur.454.1096S}. If the DM
scale radius  is significantly larger than this
($r_\mathrm{s} \gg r_\mathrm{\rm half}$) and the inner slope
$\gamma\gtrsim 0.5$, we can approximate the enclosed mass by:

\begin{equation}
  M_\mathrm{300} \simeq \frac{4 \pi \rho_\mathrm{s} r_\mathrm{s}^3}{3-\gamma}
  \left(\frac{300\,\mathrm{pc}}{r_\mathrm{s}}\right)^{3-\gamma}
  \approx 10^7 M_\odot \,.
\label{eq:M300}
\end{equation}
The parameter $\rho_\mathrm{s}$ is thus  determined completely by the
above condition, if we choose the scale radius $r_\mathrm{s}$ and cusp
slope $\gamma$.

Table~\ref{tab:generic_model_definition} shows, for several values of
$r_\mathrm{s}$ and $\gamma$, the value required for $\rho_\mathrm{s}$
to obtain the assumed $M_{300}$ mass. We fix $\alpha=1, \beta=3$ but our
results are not sensitive to these choices. \footnote{For a different mass for
the dSph, the results  for $J$ below  have to be rescaled by a
factor $(M_{300}^{\rm new}/10^7 M_\odot)^2$ since the density is
proportional to $M_{300}$, while $J$ goes as the density squared.} The
values of $r_\mathrm{s}$ are chosen to encompass the range of
$r_\mathrm{s}$ found in the MCMC analysis (see
Section~\ref{sec:results}).  
To further convince ourselves that the generic profiles we present here
are a possible description of real dSphs, we checked (not shown) using typical
stellar profiles and properties of these objects (i.e., halflight radius of a
few 100 pc), that a flat $\sim 10$ km~s$^{-1}$ velocity dispersion profile
within the error bars is recovered. We also study below the effect of moving
these dSphs from a distance of 10\,kpc to 300\,kpc, corresponding to the
typical range covered by these objects.

\begin{table}
\begin{center}
  \caption{The required normalisation $\rho_\mathrm{s}$ to have
    $M_{300}=10^7 M_{\odot}$ for a sample of $(1,3,\gamma)$ profiles
    with varying scale radius $r_\mathrm{s}$.}
\label{tab:generic_model_definition}
\begin{tabular}{lcccc} \hline\hline
         &   \multicolumn{3}{c}{$\rho_\mathrm{s}$ ($10^{7} M_\odot$ kpc$^{-3}$)}\\ 
$\gamma~~\backslash~~r_\mathrm{s}$ [kpc]&  0.10  &   0.50  &   1.0    \\ 
  \hline
  0.00   &       224   &  25.8  &  16.02  \\
  0.25   &       196   &  18.6  &  10.22  \\
  0.50   &       170   &  13.4  &   6.47  \\
  0.75   &       146   &   9.5  &   4.06  \\
  1.00   &       125   &   6.7  &   2.52  \\
  1.25   &       106   &   4.7  &   1.54  \\
  1.50   &        88   &   3.2  &   0.92  \\
\hline
\end{tabular}
\end{center}
\end{table}

\subsubsection{Sub-structures within the dSph}
\label{sec:sub_reference}
Structure formation simulations in the currently favoured
$\Lambda$CDM (cold DM plus a cosmological constant)
cosmology find that DM halos are self-similar, containing a
wealth of smaller `sub-structure' halos down to Earth-mass halos
\citep[e.g.][]{2005Natur.433..389D}. However, as emphasised in the
introduction, such simulations typically neglect the influence of the
baryonic matter during galaxy formation. It is not clear what effect
these have on the DM sub-structure distribution.  For this
reason, we adopt a more generic approach. We assess the importance of
clumps using the following recipe:\footnote{More details about the
  clump distributions can be found in Appendix~\ref{sec:sub-clumps}. See
  also, e.g., Section 2 in \citet{2008A&A...479..427L} and references
  therein, as we use the same definitions as those given in that
  paper.}
\begin{enumerate}
   \item we take a fraction $f=20\%$ of DM mass in the form of clumps;
   \item the spatial distribution of clumps follows the smooth one;
   \item the clump profiles are calculated {\em \`a la}
     \citet{2001MNRAS.321..559B} (hereafter B01), i.e.  an `NFW'
     profile \citep{navarro96} with concentration related to the mass
     of the clumps.
 \item the clump mass distribution is $\propto M^{-a}$ ($a=-1.9$),
   within a mass range $M_{\rm min}-M_{\rm max} =
   [10^{-6}-10^{6}]~M_\odot$.
\end{enumerate}

 Although these parameters are very uncertain, they allow us to
investigate the impact of substructures on the J-factor. They are varied within
reasonable bounds in Section~\ref{sec:boost} (and Appendix~\ref{app:boost})
to determine whether the sub-clump contribution can boost the signal. Note
that a 20\% clump mass fraction is about twice as large as the fraction
obtained from numerical simulations (see, e.g., \citealt{2008MNRAS.391.1685S}).
This generous fraction does not affect our conclusions, as discussed below.

\subsection{$J_{\rm sm}$ and $J_{\rm subcl}$ for the generic models}
\label{sec:gen_dep}

\begin{figure}
\begin{center}
\includegraphics[width=\linewidth]{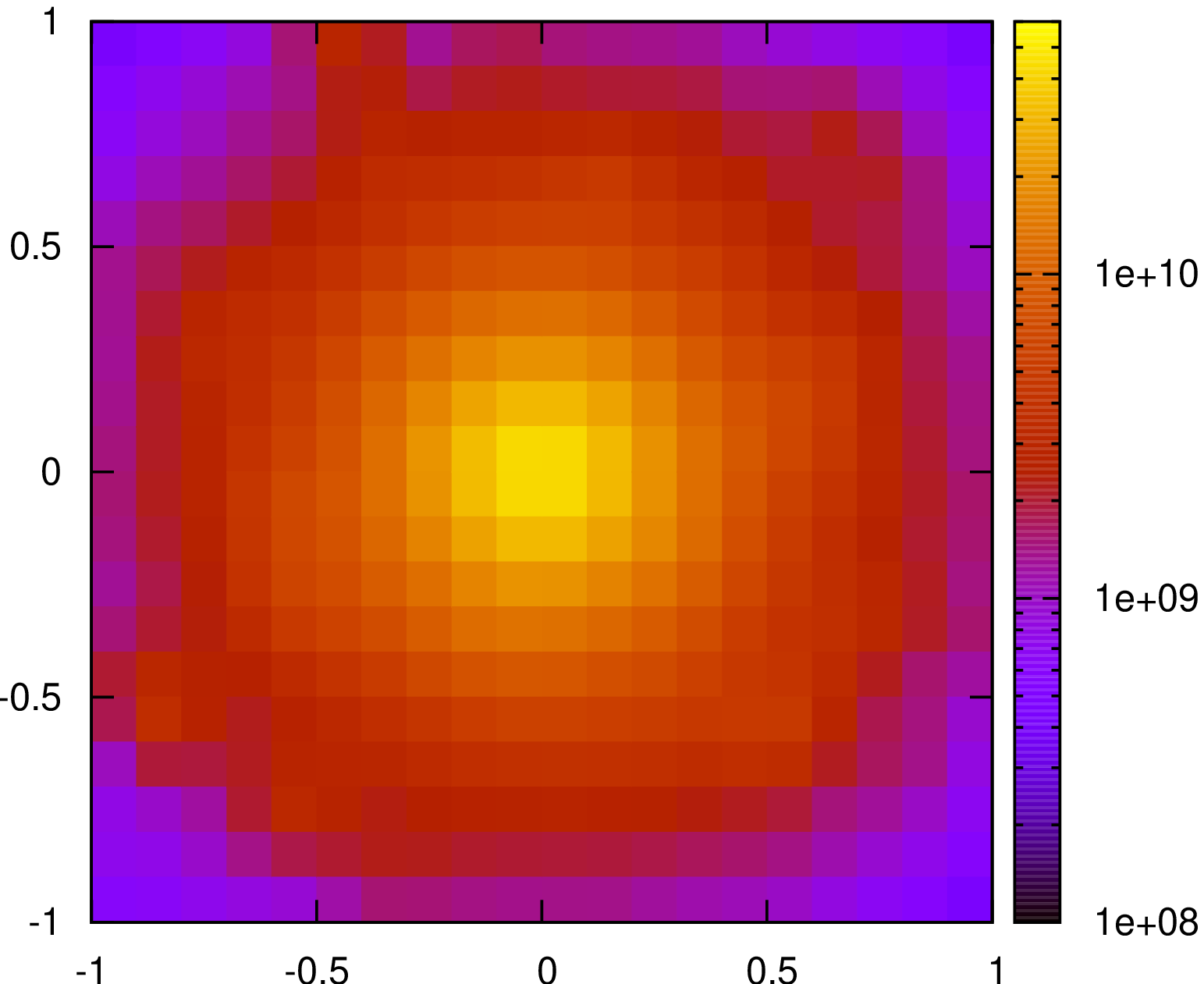}
\includegraphics[width=\linewidth]{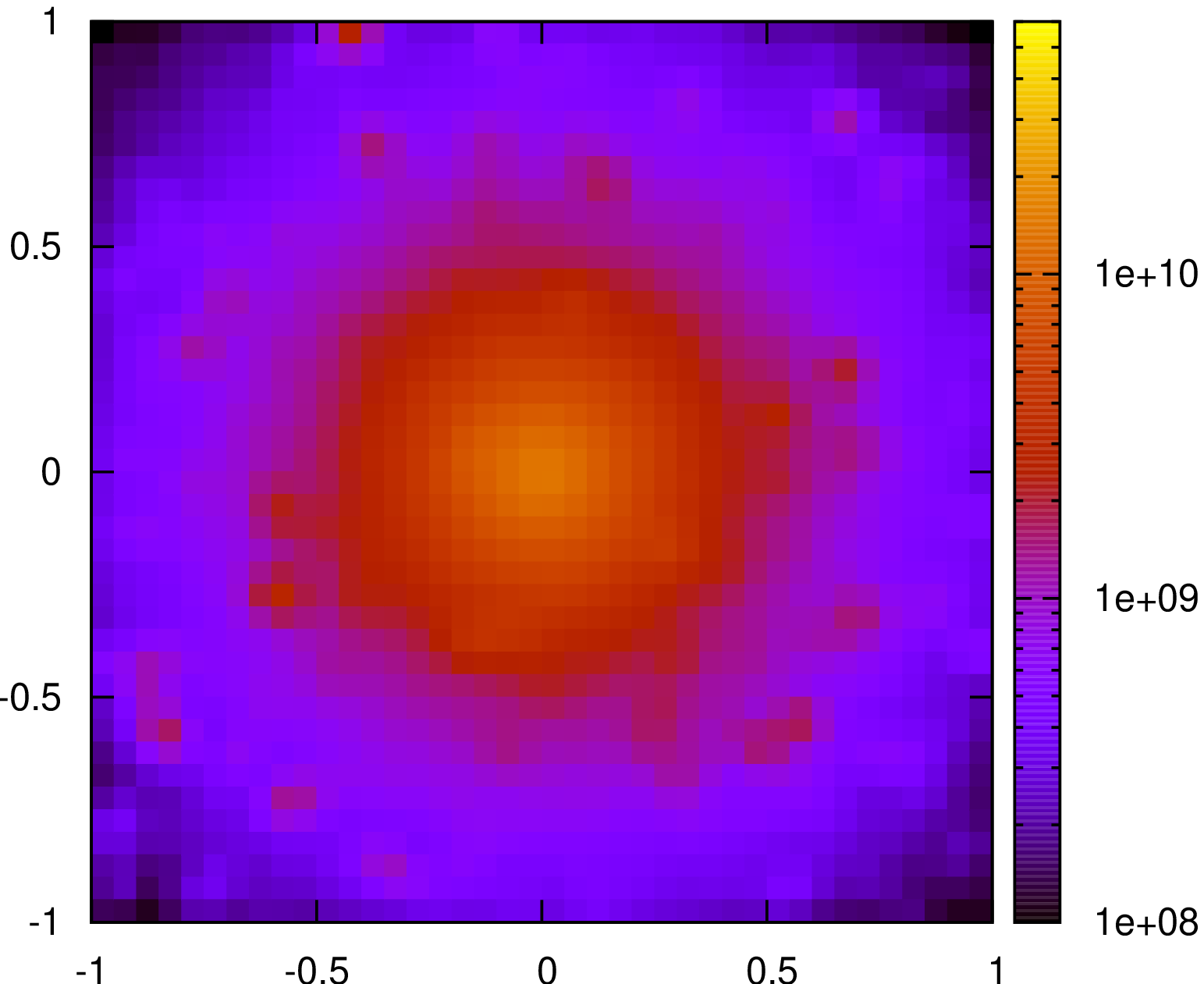}
\includegraphics[width=\linewidth]{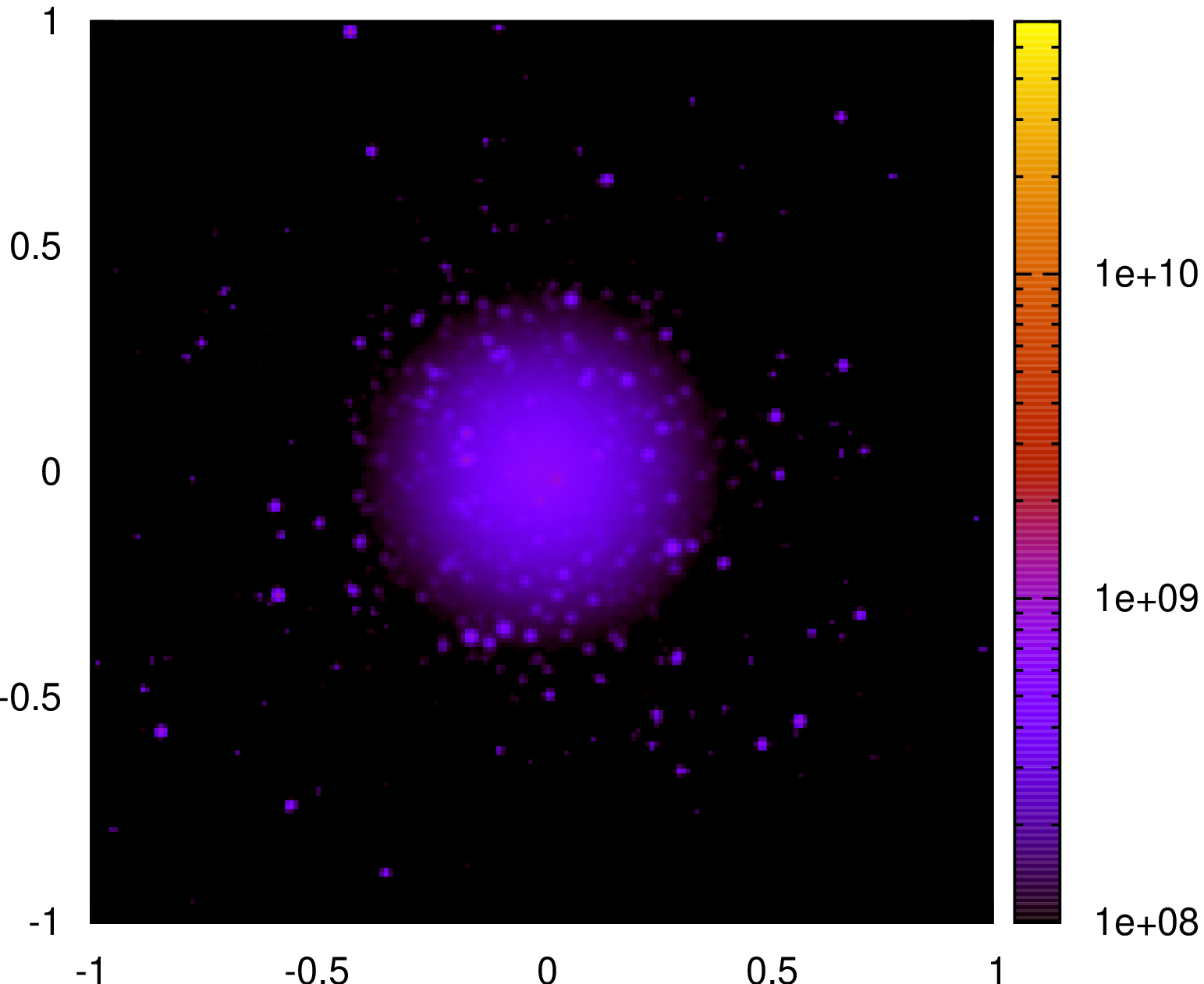}
\caption{2D view ($x$ and $y$ axis are in degrees) of $J$ for the
  generic dSph with $\gamma=0$ and $r_\mathrm{s}=1$~kpc at $d=100$~kpc
  ($M_{300} = 10^7 M_\odot$). The sub-clumps are drawn from the
  reference model described in Section \ref{sec:sub_reference},
  i.e. f=20\%, sub-clump distribution follows smooth, and sub-clump
  inner profiles have NFW with B01 concentration.  From top to bottom
  panel: $\alpha_{\rm int}=0.1^\circ$, $0.05^\circ$, and
  $0.01^\circ$. For the sake of comparison, the same colour scale is
  taken for the three integration angles ($J$ is in units   of
  $M_\odot^2$~kpc$^{-5}$).}
\label{fig:generic_2D_Jsm_and_subcl}
\end{center}
\end{figure}

As an illustration, we show in Fig.~\ref{fig:generic_2D_Jsm_and_subcl}
one realisation of the 2D distribution of $J$ from a generic core
profile ($\gamma=0$) with $r_{s}=1$ kpc (sub-clump parameters are as
described in Section~\ref{sec:sub_reference}). The dSph is at
$d=100$~kpc.  We note that our consideration of a $\gamma=0$ smooth
component with NFW sub-clumps is plausible if, e.g., baryon-dynamical
processes erase cusps in the smooth halo but cannot do so in the
sub-subhalos.  The total $J$ is the sum of the smooth and sub-clump
distributions. The centre is dominated by the smooth component,
whereas some graininess appears in the outskirts of the dSph.
In this particular configuration, the `extended' signal from the core
profile, when integrated over a very small solid angle, could be
sub-dominant compared with the signal of NFW sub-clumps that it
hosts. The discussion of cross-constraints between detectability of
sub-halos of the Galaxy vs. sub-clumps in the dSph is left for a
future study.

In the remainder of the paper, we will replace for simplicity the
calculation of $J_{\rm subcl}(\alpha_{\rm int})$ by its mean value, as
we are primarily interested in `unresolved' observations. Hence clumps
are not drawn from their distribution function, but rather $\langle J_{\rm
  subcl}\rangle$ is calculated from the integration of the spatial and
luminosity (as a function of the mass) distributions (see
Appendix~\ref{sec:sub-clumps}).

\subsubsection{Radial dependence $J(\theta)$}
\label{sec:radial_dep}

\begin{figure}
\includegraphics[width=\linewidth]{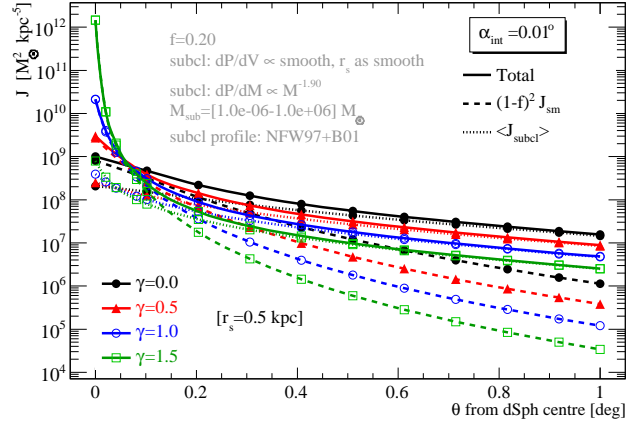}
\caption{$J$ as a function of the angle $\theta$ away from the dSph
  centre for a dSph at $100$ kpc with $r_\mathrm{s}=0.5$~kpc
  ($\rho_\mathrm{s}$ is given in
  Table~\ref{tab:generic_model_definition}). The integration angle is
  $\alpha_{\rm int}=0.01^\circ$. For the four inner slope values
  $\gamma$, the various contributions to $J$ are shown as
  solid (total), dashed (smooth), and dotted lines (sub-clumps).}
\label{fig:generic_Jtheta_sm_and_subcl}
\end{figure}

The radial dependence of $J$ is shown in
Fig.~\ref{fig:generic_Jtheta_sm_and_subcl} for four values of $\gamma$
(for an integration angle $\alpha_{\rm int}=0.01^\circ$). 
The dashed lines show the result for the
smooth distribution; the dotted lines show the sub-clump contribution;
and the solid lines are the sum of the two. The peak of the signal is
towards the dSph centre. As long as the distribution of clumps is
assumed to follow the smooth one, regardless of the value of $\gamma$,
the quantity $(1-f)^{2}J_{\rm sm}(0)$ always dominates (at least by a
factor of a few) over $\langle J_{\rm subcl}(0)\rangle$. (Recall that
in our generic models, all dSphs have the same $M_{300}$.) The scatter
in $J_{\rm tot}(0)$ is about 4 orders of magnitude for
$\gamma\in[0.0-1.5]$, but only a factor of 20 for
$\gamma\in[0.0-1.0]$. Beyond a few tenths of degrees, $\langle J_{\rm
  subcl}\rangle$ dominates. The crossing point depends on a combination
of the clump
mass fraction $f$, $\gamma$, $r_\mathrm{s}$, $d$, $\alpha_{\rm int}$. The dependence of $J$ on the 
two latter parameters are discussed in Appendix~\ref{app:dep_generic}. 
The radial dependence is as expected: the smooth contribution decreases faster than that of
the sub-clump one, because the signal is proportional to the squared
spatial distribution in the first case, but directly proportional to
the spatial distribution in the second case. 
Halving $f$ to match the fraction from N-body simulations would have a 
~25\% effect on  $(1-f)^2J_{\rm sm}$, but decrease $J_{\rm subcl}$ by a
factor $4$, so that the cross-over between the two components would occur
at a larger angle in Fig.~\ref{fig:generic_Jtheta_sm_and_subcl}.

\subsubsection{Boost factor}
\label{sec:boost}

Whether or not the signal is boosted by the sub-clump population is
still  debated  in the literature
\citep{2007PhRvD..75h3526S,2008ApJ...686..262K,2008MNRAS.384.1627P,2009A&A...496..351P}.
As underlined in the previous sections, the sub-clump contribution
towards the dSph centre never dominates over the smooth one if the
spatial profile of the sub-clumps follows that of the smooth distribution, and if
the integration angle remains below some critical angle discussed
below.

Let us first define properly the parameters with respect to which this
boost is calculated, as there is sometimes some confusion about
this. Here, we define it with respect to the integration angle
$\alpha_{\rm int}$ (the pointing direction is still towards the dSph
centre):
\begin{equation}
   B(\alpha_{\rm int}) \equiv 
   \frac{ (1-f)^2 J_{\rm sm}(\alpha_{\rm int}) 
    + J_{\rm subcl}(\alpha_{\rm int})}{J_{\rm sm}(\alpha_{\rm int})}\,.
    \label{eq:fig:generic_boost}
\end{equation}
In most studies, the boost has been calculated by integrating out to the
clump boundary (i.e., $\alpha_{\rm int}^{\rm all} = R_{\rm
  vir}/d$). But the boost depends crucially on $\alpha_{\rm int}$ (the
radial dependence of the smooth and sub-clump contributions differ, see
Section~\ref{sec:radial_dep}).

\begin{figure}
\includegraphics[width=\linewidth]{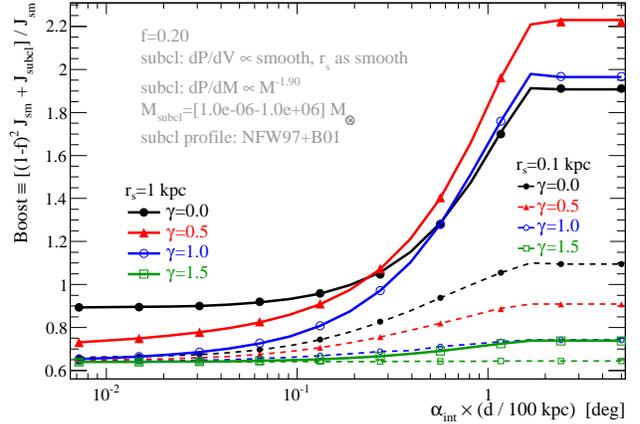}
\caption{Boost factor as a function of $\alpha_{\rm int} \times (d/100$~kpc)
for profiles sub-clumps {\em follow} smooth (see Section~\ref{sec:sub_reference}):
the dSph is at $d=100$~kpc (lines) or $d=10$~kpc (symbols).}
\label{fig:generic_boost_default}
\end{figure}
We plot in Fig.~\ref{fig:generic_boost_default} the boost  for
different inner slopes $\gamma$, where a direct consequence of
Eq.~(\ref{eq:rescale_Jalpha}) is the $\alpha_{\rm int}\times d$
rescaling. For $r_\mathrm{s}\lesssim 0.1$ kpc (regardless of $\gamma$),
or for $\gamma\gtrsim1.5$ (regardless of $r_{\rm s}$), the signal is
never boosted.  \footnote{The difference between the level of boost observed
for $r_\mathrm{s}=0.1$~kpc or $r_\mathrm{s}=1$~kpc can be understood
if we recall that the total mass of the clump is fixed at 300 pc,
regardless of the value of $\gamma$ or $r_\mathrm{s}$. For
$r_\mathrm{s}=0.1$~kpc, $\rho_\mathrm{s}\sim {\cal
  O}(10^9M_\odot$~kpc$^{-3})$, whereas for $r_\mathrm{s}=1$~kpc,
$\rho_\mathrm{s}\sim{\cal O}(10^7M_\odot$~kpc$^{-3})$. As $J_{\rm sm}
\propto \rho_\mathrm{s}^2$ whereas $J_{\rm sub} \propto
\rho_\mathrm{s}$, the relative amount of $J_{\rm sub}$ with respect to
$J_{\rm sm}$ is expected to decrease with smaller $r_\mathrm{s}$. This
is indeed what we observe in the figure (solid vs dashed lines).}
For small enough $\alpha_{\rm int}$, $B$ is smaller than unity,
and if $\gamma$ is steep enough, $B\approx(1-f)^2$. For large values, a
plateau is reached as soon as $\alpha_{\rm int}d\gtrsim R_{\rm vir}$
(taken to be 3 kpc here). In between, the value of the boost depends on
$r_\mathrm{s}$ and $\gamma$ of the smooth component.
Going beyond this qualitative description is difficult, as the
toy model formulae of Appendix~\ref{sec:sub-clumps} gives results correct to only a factor
of $\sim\!2$ (which is inadequate to evaluate the boost properly). 

To conclude, the maximum value for sub-clump {\em follows} smooth is
$\lesssim 2$, and this value is reached only when integrating the signal out to $R_{\rm vir}/d$.
The boost could still be increased by varying the sub-clump properties
(e.g., taking a higher concentration). Conversely, if dynamical friction
has caused the sub-clump population to become much more centrally
concentrated than the smooth component, then the boost is
decreased. This is detailed in Appendix~\ref{app:boost}. 
For the most realistic configurations, there is \emph{no} significant
boost when a clump mass fraction f = 20\% is used. 
Naturally this result is even more true for the smaller $f$ found in
N-body simulations so we disregard the boost for the rest of this paper and 
consider only the smooth contribution.

\section{Sensitivity of present/future $\gamma$-ray observatories}
\label{sec:detectability}

Major new ground-based $\gamma$-ray observatories are in the planning
stage, with CTA \citep{2010arXiv1008.3703C} and AGIS
\citep{agis:website} as the main concepts. As the designs of these
instruments are still evolving, we adopt here generic performance
curves (described below), close to the stated goals of these
projects. For the Large Area Telescope (LAT) of the Fermi $\gamma$-ray
satellite, the performance for 1 year observations of
point-like, high Galactic latitude sources is known
\citep{fermi:website}, but no information is yet available for longer
exposures or for extended objects. We therefore adopt a toy
likelihood-based model for the Fermi sensitivity, tuned to reproduce
the 1 year point-source curves.  We note that whilst this approach
results in approximate performance curves for both the ground- and
space-based instruments, it captures the key differences (in
particular the differences in collection area and angular resolution)
and illustrates the advantages and limitations of the two instrument
types, as well as the prospects for the discovery of DM annihilation
in dSphs within the next decade.

\subsection{Detector models}

The sensitivity of a major future $\gamma$-ray observatory based on an
array of Cherenkov Telescopes (FCA in the following, for `Future
Cherenkov Array') is approximated based on the point-source
differential sensitivity curve (for a $5\sigma$ detection
in 50 hours of observations) presented by
\citet{2008ICRC....3.1469B}. Under the assumption that the angular
resolution of such a detector is a factor 2 better than HESS
\citep{2008ApJ...679.1299F} and has the same energy-dependence, and
that the effective collection area for $\gamma$-rays grows from
$10^{4}$m$^2$ at 30~GeV to 1~km$^2$ at 1 TeV, the implied cosmic-ray
(hadron and electron) background rate per square degree can be
inferred and the sensitivity thus adapted to different observation
times, spectral shapes and source extensions. Given that the design of
instruments such as CTA are not yet fixed, we consider that such a
simplified response, characterised by the following functions is a
useful tool to explore the capabilities of a generic next-generation
instrument:

\begin{equation}
{\rm LS} = -13.1-0.33\,X+0.72\,X^{2},
\end{equation}
\begin{equation}
{\rm LA} = 6 + 0.46\,X - 0.56\,X^{2},
\end{equation}
\begin{equation}
\psi_{68} = 0.038+\exp{-(X+2.9)/0.61},
\end{equation}  
where 
\begin{equation}
X = \log_{10} {\rm (Photon Energy / TeV)},
\end{equation}
LS = log$_{10}$(Differential Sensitivity/erg\,cm$^{-2}$\,s$^{-1}$), LA
= log$_{10}(\mathrm{Effective Area}/ {\rm m}^{2})$, and $\psi_{68}$ is
the 68\% containment radius of the point-spread-function (PSF) in
degrees. 

For the Fermi detector a similar simplified approach is taken, the
numbers used below being those provided by \citet{fermi:website}.  The
effective area changes as a function of energy and incident angle to
the detector, reaching a maximum of $\approx 8000\,\mathrm{cm}^2$.
The effective time-averaged area is then $\epsilon A\Omega/4\pi$ and
the data-taking efficiency $\epsilon \approx 0.8$ (due to instrument
dead-time and passages through the South Atlantic Anomaly). The point
spread function again varies as a function of energy (with a much
smaller dependence as a function of incidence angle) varying from 10
degrees to a few tenths of a degree over the LAT energy range.  A rate
of $1.5 \times 10^{-5}$ cm$^{-2}$ s$^{-1}$ sr$^{-1}$ ($>100$ MeV) and
a photon index of 2.1 are assumed for the background.  The sensitivity
is then estimated using a simplified likelihood method which provides
results within 20\% of the sensitivity for a one year observation of a
point-like source given by \citet{fermi:website}.

Whilst both detector responses are approximate, the comparison is still
useful. Our work incorporates several key aspects not 
considered in earlier studies, including the strong energy dependence of
the angular resolution of both ground and space based instruments in
the relevant energy range of 1~GeV to 1~TeV and hence the
energy-dependent impact of the angular size of the target region.

\subsection{Relative performance for generic halos}
\label{sec:RelativePerformance}
\begin{figure}
\includegraphics[width=\linewidth]{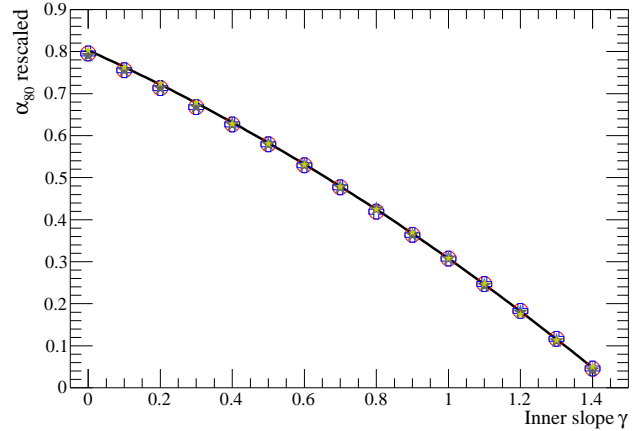}
\caption{The cone angle encompassing 80\% of the annihilation flux at
  as function of the inner slope $\gamma$. Several different values of
  $r_\mathrm{s}$ and distance $d$ are shown for each $\gamma$, all
  scaled by (1 kpc/$r_\mathrm{s}$ and 100 kpc/$d$). The best-fit curve
  is also shown, corresponding to Eq.~(\ref{eq:angsiz}).  }
\label{fig:alpha_80}
\end{figure}
Using the results from Section~\ref{sec:generic} and the detector
performance models defined above we can begin to investigate the
sensitivity of future ACT arrays and the Fermi-LAT detector (over long
observation times) to DM annihilation in dSphs. The
detectability of a source depends primarily on its flux, but also on
its angular extent.  The impact of source extension on detectability
is dealt with approximately (in each energy bin independently) by
assuming that the opening angle of a cone which incorporates 80\% of
the signal is given by
\begin{equation}
\theta_{80} = \sqrt{\psi_{80}^{2}+\alpha_{80}^{2}}\,,
\label{eq:PSF}
\end{equation}
where $\psi_{80}=1.25\,\psi_{68}$ is assumed for the FCA and
interpolated from values given for 68\% and 95\% containment for the
LAT \citet{fermi:website}; here $\alpha_{80}$ is the 80\% containment
angle of the halo emission. The validity of this approximation (at the
level of a few percent) has been tested (see Appendix~\ref{app:PSF})
by convolving realistic halo
profiles with a double Gaussian PSF as found for HESS
\citep{hesspsf}. An 80\% integration circle is close to optimum for a
Gaussian source on a flat background (in the background limited
regime).  Fig.~\ref{fig:alpha_80} shows the 80\% containment radius
of the annihilation flux of generic halos as a function of the inner
slope $\gamma$. This result can be parametrised as:
\begin{equation}
\alpha_{80} = 0.8^{\circ}\,(1-0.48\gamma-0.137\gamma^2) 
    \left(\!\frac{r_\mathrm{s}}{\rm 1\,kpc}\!\right)\left(\!\frac{d}{\rm 100\,kpc}\!\right)^{-1}.
\label{eq:angsiz}
\end{equation}
It is clear that for a broad range of $d$, $\gamma$ and
$r_\mathrm{s}$ the characteristic angular size of the emission region
is {\em larger} than the angular resolution of the instruments under
consideration. It is therefore critical to assess the performance as a
function of the angular size of the dSph as well as the
mass of the annihilating particle.

\begin{figure}
\begin{center}
\includegraphics[width=0.97\linewidth]{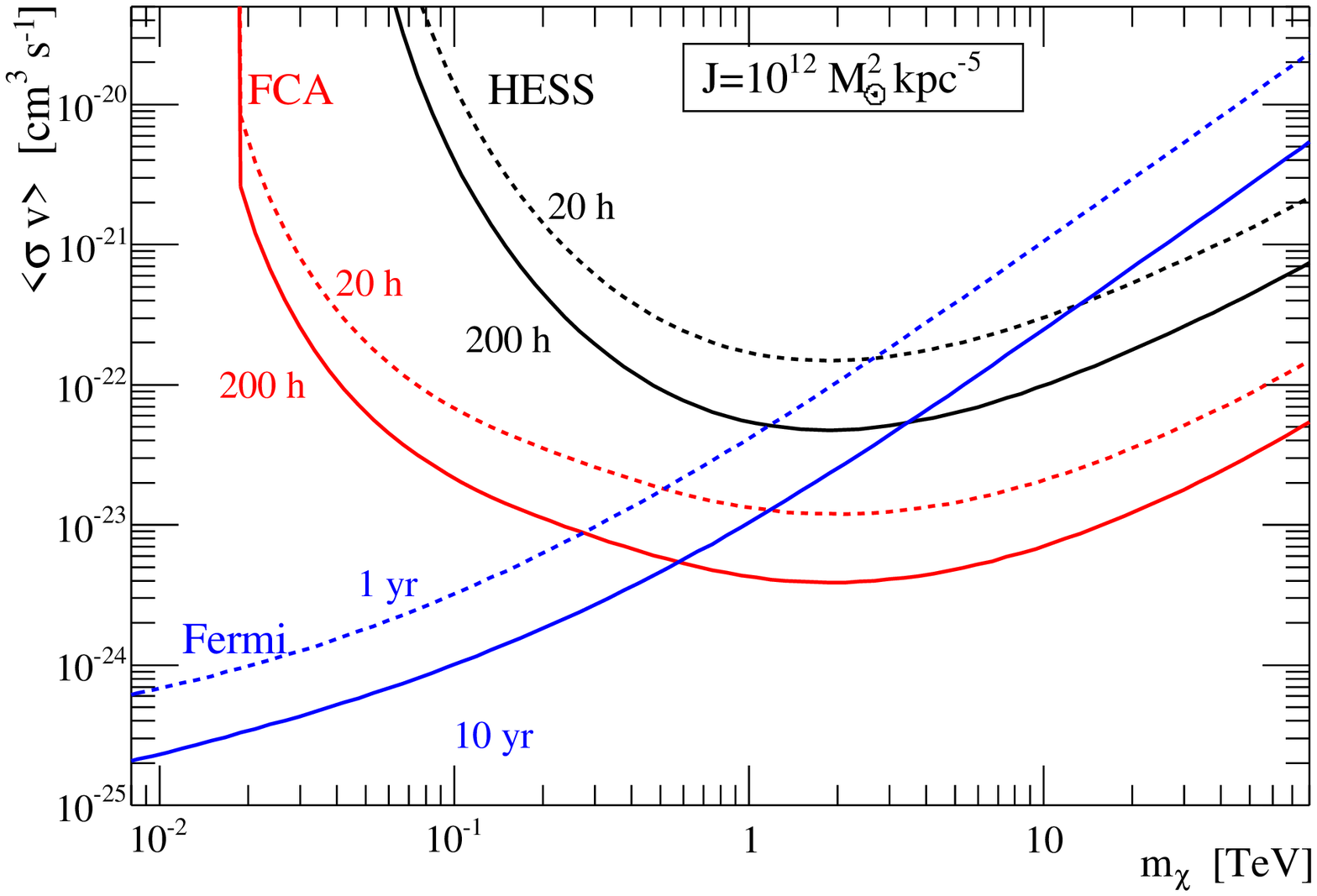}
\includegraphics[width=0.97\linewidth]{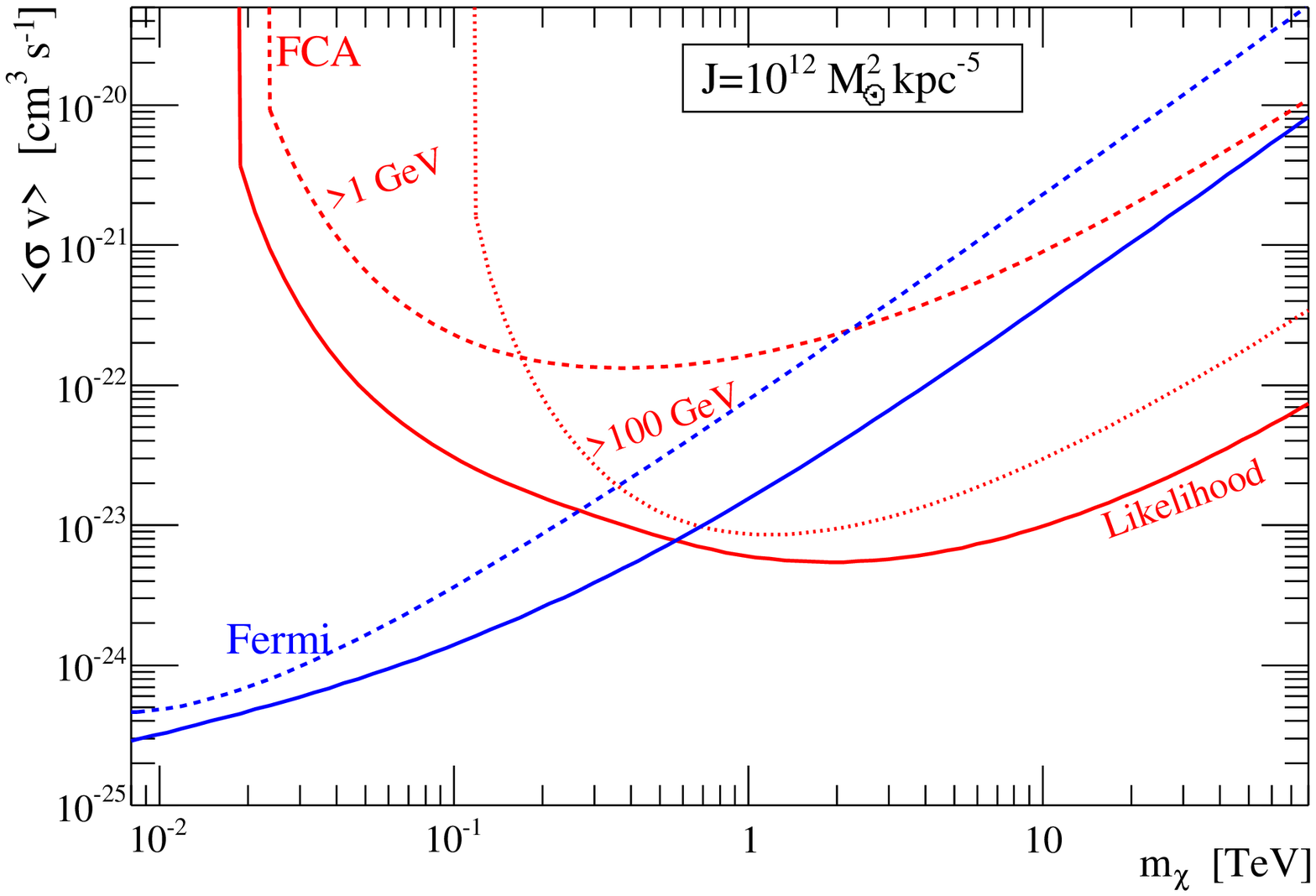}
\includegraphics[width=0.97\linewidth]{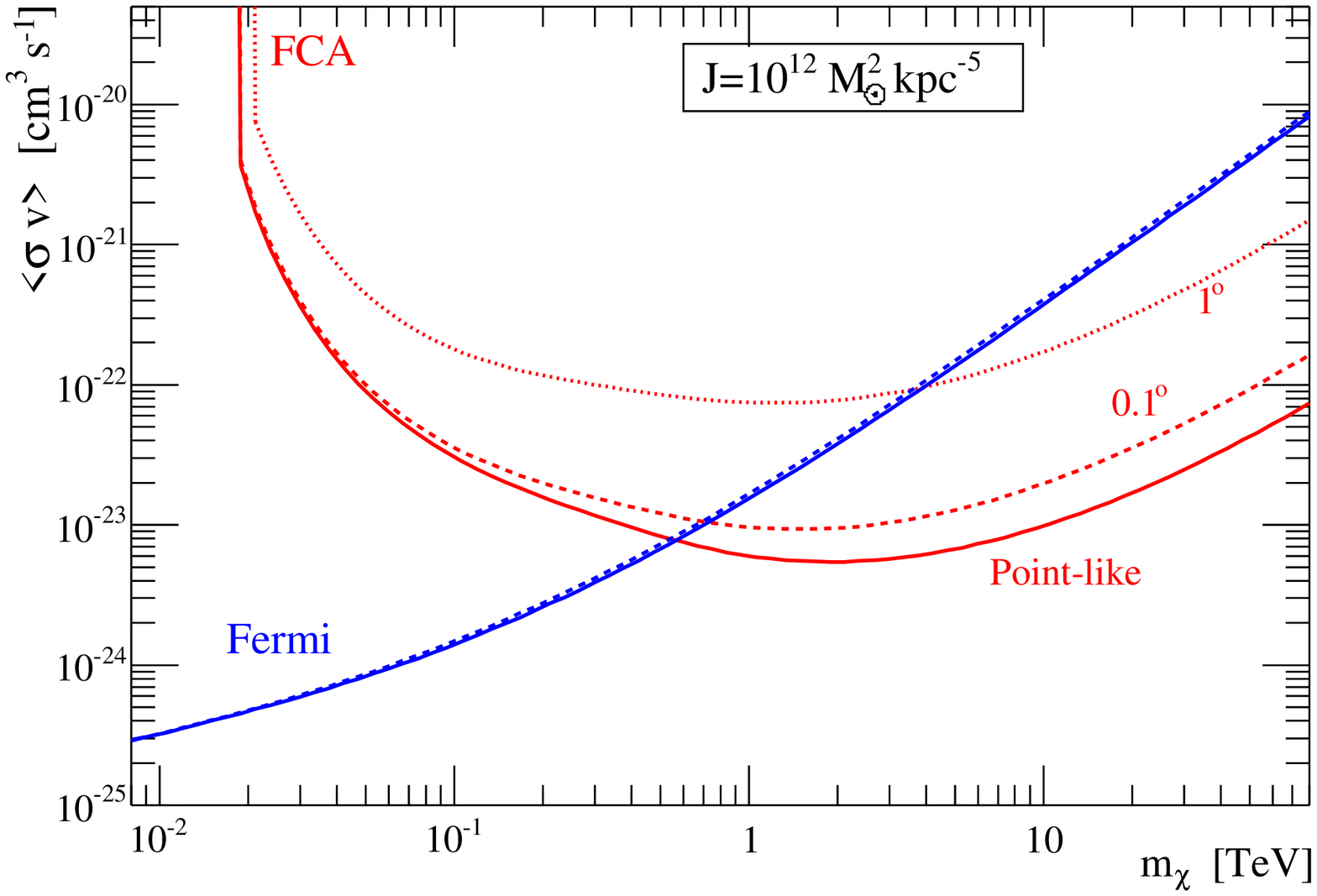}
\caption{ Approximate sensitivities of Fermi-LAT (blue lines), HESS (black
  lines) and the FCA described above (red lines) to a generic halo
  with $J=10^{12}~M_{\odot}^2$~kpc$^{-5}$, as a function of the mass
  of the annihilating particle and for the annihilation spectrum of
  Eq.~(\ref{eq:param-berg}).  {\bf Top}: The impact of observation
  time is illustrated: dashed lines give the 1 year and 20 hour
  sensitivities for Fermi and FCA/HESS respectively while the solid
  lines refer to 10 year (200 hour) observations.  {\bf
    Middle}: the impact of analysis methods is considered for 5 year
  (100 hour) observations using Fermi (FCA). Solid lines show
  likelihood analyses in which the mass and spectrum of the
  annihilating particle are known in advance, while dashed
  and dotted lines show simple integral flux measurements above fixed
  thresholds of 1 GeV (dashed) and 100 GeV (dotted).  Note that the 1
  GeV cut implies accepting all events for the FCA (where the trigger
  threshold is $\approx$20 GeV).  {\bf Bottom}: the impact of the
  angular extension of target sources, as given by the halo profile 
  in Fig.~\ref{fig:alpha_80} is illustrated. The solid lines reproduce 
  the likelihood case from the middle panel for a point-like source, and with
  values of $\alpha_{80}$ of 0.1$^{\circ}$ (dashed) and 1$^{\circ}$
  also shown.}
\label{fig:sensitivity}
\end{center}
\end{figure}
Fig.~\ref{fig:sensitivity} shows the relative sensitivity of Fermi
and an FCA within our framework as a function of the mass of the
annihilating particle, adopting the annihilation spectrum given in
Eq.~(\ref{eq:param-berg}), with the several panels illustrating different
points. From Fig.~\ref{fig:sensitivity} top (the case of a
point-like signal for different observation times) it is clear that
Fermi-LAT has a considerable advantage for lower mass DM particles ($m_\chi\,\ll$
1 TeV) on the timescale for construction of an FCA (i.e. over a 5-10
year mission lifetime) in comparison to a deep ACT observation of 200
hours. Furthermore, Fermi-LAT is less adversely affected by the angular
extent of the target regions (see Fig.~\ref{fig:sensitivity}
bottom), due to its modest angular resolution in the energy range
where it is limited by background, meaning that the source extension is well
matched to the PSF of the instrument. The middle panel of this figure
illustrates the impact of different approaches to the analysis. In the
case that there is a DM candidate inferred from the discovery of supersymmetry 
at the LHC (quite possible on the relevant timescale) a search
optimised on an assumed mass and spectral shape can be made (solid
curves). However, all instruments are less sensitive when
a generic search is undertaken. Simple analyses using all the photon 
flux above a fixed energy threshold (arbitrarily set to reduce background) 
are effective only in a relatively narrow range of particle mass. For example
keeping only $>$100 GeV photons works well for ACTs for 0.3-3 TeV particles;
whereas keeping all photons $>$1 GeV works moderately well in the 0.1-0.2 TeV
range, but is much less sensitive than the higher threshold cut over the rest of
the candidate dark matter particle mass range.
The features of these curves are dictated by the expected shape of the
annihilation spectrum. From Eq.~(\ref{eq:param-berg}) the peak photon
output (adopting the  average spectrum for DM annihilation)
occurs at an energy which is an order of magnitude below the particle
mass -- effective detection requires that this peak occurs within (or
close to) the energy range of the instrument concerned.

\begin{figure}
\includegraphics[width=\linewidth]{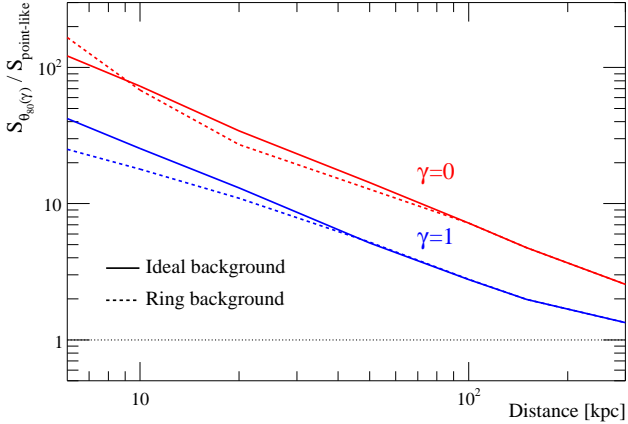}
\caption{Relative DM annihilation detection sensitivity for a
  100~hour FCA observation, as a function of dSph distance for
  different inner slopes $\gamma$ and with $r_{s}$ fixed to 1~kpc. The
  sensitivity for a realistic approach using $\theta_{80}$ is given
  relative to the sensitivity to a point-like source with the same
  flux. Larger values correspond to poorer performance (larger values
  of the minimum detectable flux). The assumed spectral shape is again
  as given by Eq.~(\ref{eq:param-berg}) with $m_{\rm \chi}=300$
  GeV. This sensitivity ratio depends on the strategy used to estimate
  the background level at the dSph position. The dashed lines show the
  impact of using an annulus between 3.5$^{\circ}$ and 4.0$^{\circ}$
  of the dSph centre as a background control region. The solid line
  assumes that the background control region lies completely outside
  the region of emission from the dSph.  }
\label{fig:rel_sensitivity}
\end{figure}
The total annihilation flux from a dSph increases at smaller distances
as $1/d^{2}$ for fixed halo mass, making nearby dSphs attractive for
DM detection. However, as Fig.~\ref{fig:sensitivity} shows, the
increased angular size of such nearby sources raises the required
detection flux.  Fig.~\ref{fig:rel_sensitivity} illustrates the
reduction in sensitivity for an FCA with respect to a point-like
source for generic dSph halos as a function of distance, for inner
slopes, $\gamma$, of zero and one and with $r_\mathrm{s}$ fixed to
1\,kpc, relative to the assumption of the full annihilation signal and
a point-like source. Even for $\gamma=1$, the point-like approximation
leads to an order of magnitude overestimate of the detection sensitivity for nearby
($\sim\!20$ kpc) dSphs. A further complication is how to
establish the level of background emission arising from the
residual non-$\gamma$-ray background.  A common
method in ground-based $\gamma$-ray astronomy is to estimate this
background from an annulus around the target source \citep[see,
e.g.,][]{berge_bg}. The dashed lines in Fig.~\ref{fig:rel_sensitivity}
show the impact of estimating the background using an annulus between
3.5$^{\circ}$ and 4.0$^{\circ}$ from the target. This approach has a
modest impact on sensitivity and is ignored in the following
discussions as it reduces the detectable flux but also $\theta_{80}$ and
leads to a small improvement in some cases.
 
\section{Jeans/MCMC analysis of dSph kinematics}
\label{sec:mcmc}
\subsection{dSph kinematics with the spherical Jeans equation} 
\label{sec:jeans}

\begin{figure*}
\includegraphics[width=0.66\linewidth]{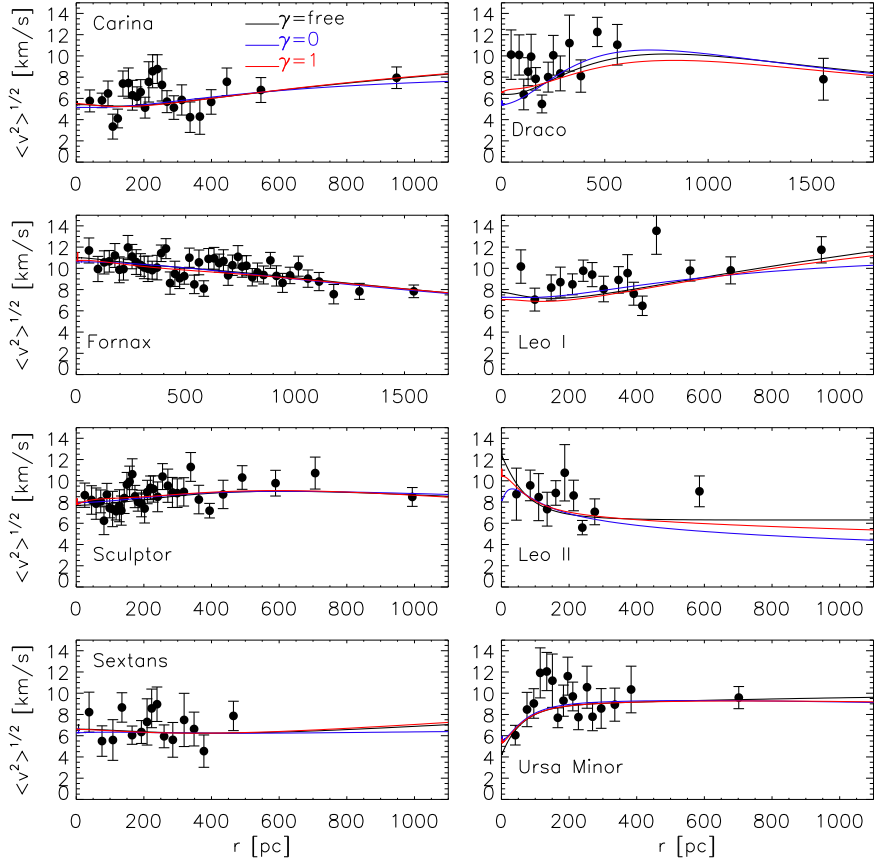}
\caption{Velocity dispersion profile data for the 8 classical dSphs,
  obtained as described in the text (the impact of the binning choice is 
  discussed in Appendix~\ref{app:binning}). The solid lines correspond to
  the best-fit models for the inner slope when $\gamma$ is left free
  (dark), $\gamma$ is fixed to 1 (blue), and $\gamma$ is fixed to 0
  (red). Because of the large degeneracies among the halo parameters
   (see Section~\ref{subsec:gamma-free} for a list), we do not list the corresponding best-fit parameters.
   The motivation for showing these profiles is to illustrate that our halo
   model is capable of describing the kinematic data, and that the inner profile
   is not constrained by the data.}
\label{fig:momentprofiles}
\end{figure*}

Extensive kinematic surveys of the stellar components of dSphs have
shown that these systems have negligible rotational support
\citep[with the possible exception of the Sculptor dSph,
see][]{2008ApJ...681L..13B}.  If we assume that the dSphs are in
virial equilibrium, then their internal gravitational potentials
balance the random motions of their stars.  In order to estimate dSph
masses,  we consider here the behaviour of dSph stellar velocity
dispersion as a function of distance from the dSph centre (analogous
to rotation curves of spiral galaxies).  Specifically, we use the
stellar kinematic data of \citet{2009AJ....137.3100W} for the 
Carina, Fornax, Sculptor and Sextans dSphs, the data of \citet{mateo08} for
the Leo I dSph, and data from Mateo et al. (in preparation) for the
Draco, Leo II and Ursa Minor dSphs.  \citet[W09
hereafter]{2009ApJ...704.1274W} have calculated velocity dispersion
profiles from these same data under the assumption that l.o.s.
velocity distributions are Gaussian.  Here we re-calculate these
profiles without adopting any particular form for the velocity
distributions.  Specifically, for a given dSph we divide the velocity
sample into circular bins containing approximately equal numbers of
member stars,\footnote{Kinematic samples  are often contaminated by
  interlopers from the Milky Way foreground.  Following W09, we
  discard all stars for which the algorithm described by
  \citet{walker09b} returns a membership probability less than
  $0.95$.} and within each bin we estimate the second velocity moment
(squared velocity dispersion) as:
\begin{equation}
  \langle \hat{V}^2\rangle = \frac{1}{N-1}\displaystyle\sum_{i=1}^N[(V_i-\langle \hat{V}\rangle)^2 - \sigma_i^2],
  \label{eq:vdisp}
\end{equation}
where $N$ is the number of member stars in the bin.  We hold $\langle
V\rangle$ fixed for all bins at the median velocity over the entire
sample.  For each bin we use a standard bootstrap re-sampling to
estimate the associated error distribution for $\langle
\hat{V}^2\rangle$, which is approximately Gaussian.  Fig.~\ref{fig:momentprofiles} displays the resulting velocity dispersion
profiles, $\langle \hat{V}^2\rangle^{1/2}(R)$, which are similar to
 previously published profiles.

In order to relate these velocity dispersion profiles to dSph masses,
we follow W09 in assuming that the data sample in each dSph a single,
pressure-supported stellar population that is in dynamical equilibrium
and traces an underlying gravitational potential dominated by dark
matter.  Implicit is the assumption that the orbital motions of
stellar binary systems contribute negligibly to the measured velocity
dispersions.\footnote{\citet{edo96} and \citet{hargreaves96b} conclude
  that this assumption is valid for the classical dSphs studied here,
  which have measured velocity dispersions of $\sim 10$ km s$^{-1}$.
  This conclusion does not necessarily apply to recently-discovered
  `ultra-faint' Milky Way satellites, which have measured velocity
  dispersions as small as $\sim 3$ km s$^{-1}$
  \citep{2010ApJ...722L.209M}.} Furthermore, assuming spherical symmetry,
the mass profile, $M(r)$, of the DM halo relates to (moments
of) the stellar distribution function via the Jeans equation:
\begin{equation}
\frac{1}{\nu}\frac{d}{dr}(\nu \bar{v_r^2})+2\frac{\beta(r)\bar{v_r^2}}{r}=-\frac{GM(r)}{r^2},
\label{eq:jeans}
\end{equation}
where $\nu(r)$, $\bar{v_r^2}(r)$, and $\beta_r\equiv\beta(r)\equiv
1-\bar{v_{\theta}^2}/\bar{v_r^2}$ describe the 3-dimensional density,
radial velocity dispersion, and orbital anisotropy, respectively, of
the stellar component.  Projecting along the l.o.s., the mass
profile relates to observable profiles, the projected stellar density
$I(R)$ and velocity dispersion $\sigma_p(R)$, according to \citep[BT08
hereafter]{bt08}
\begin{equation}
  \sigma_p^2(R)=\frac{2}{I(R)}\displaystyle \int_{R}^{\infty}\biggl (1-\beta_r\frac{R^2}{r^2}\biggr ) \frac{\nu \bar{v_r^2}r}{\sqrt{r^2-R^2}}\mathrm{d}r.
  \label{eq:jeansproject}
\end{equation}
Notice that while we observe the projected velocity dispersion and
stellar density profiles directly, the l.o.s. velocity
dispersion profiles provide {\em no} information about the anisotropy,
$\beta(r)$.  Therefore we require an assumption about $\beta(r)$;
here we assume $\beta = $ constant, allowing for nonzero anisotropy in
the simplest way.  For constant anisotropy, the Jeans equation has the
solution (e.g., \citealt{mamon05}):
\begin{equation}
  \nu\bar{v^2_r}=Gr^{-2\beta_r}\displaystyle\int_r^{\infty}s^{2\beta_r-2}\nu(s)M(s)\mathrm{d}s.
  \label{eq:jeanssolution}
\end{equation}
We shall adopt parametric models for $I(R)$ and $M(r)$ and then find
values of the parameters of $M(r)$ that, via
Eqs.~(\ref{eq:jeansproject}) and (\ref{eq:jeanssolution}), best
reproduce the observed velocity dispersion profiles.

\subsubsection{Stellar Density}
\label{subsec:stellar}
Stellar surface densities of dSphs are typically fit by
\citet{plummer11}, \citet{king62} and/or \citet{sersic68}, profiles
(e.g., \citealt{ih95}).  For simplicity,  we adopt here the Plummer
profile:
\begin{equation}
  I(R)=\frac{L}{\pi r_{\rm half}^2}\frac{1}{[1+R^2/r_{\rm half}^2]^2},
  \label{eq:plummer}
\end{equation}
which has just two free parameters: the total luminosity $L$ and the
projected\footnote{For consistency with \citet{2009ApJ...704.1274W} we
  define $r_\mathrm{half}$ as the radius of the circle enclosing half of the
  dSph stellar light as seen in projection. Elsewhere this radius is
  commonly referred to as the `effective radius'.} half-light radius
$r_{\rm half}$.  Given spherical symmetry, the Plummer profile implies
a 3-dimensional stellar density (BT08) of:
\begin{equation}
  \nu(r)=-\frac{1}{\pi}\int_r^{\infty}\frac{\mathrm{d}I}{\mathrm{d}R}\frac{\mathrm{d}R}{\sqrt{R^2-r^2}}=\frac{3L}{4\pi r_{\rm half}^3}\frac{1}{[1+r^2/r_{\rm half}^2]^{5/2}}.
  \label{eq:abel}
\end{equation}
Since we assume that DM dominates the gravitational potential
at all radii (all measured dSphs have central mass-to-light ratios
$\ga 10$, e.g., \citealt{mateo98}), the value of $L$ has no bearing on
our analysis.  We adopt values of $r_{\rm half}$ (and associated
errors) from Table 1 in the published erratum to W09; these data
originally come from the star count study of \citet{ih95}.
We have checked that a steeper outer slope or a steeper inner slope
for the light profile leaves unchanged the conclusions (see Appendix~\ref{app:light_profile}).

\subsubsection{Dark matter halo}
\label{subsec:halomodel}

For the DM halo we follow W09 in using a generalised
Hernquist profile, as given by Eq.~(\ref{eq:hernquist1}). In terms of
these parameters, i.e, the density $\rho_\mathrm{s}$ at scale radius
$r_\mathrm{s}$, plus the (outer,transition,inner) slopes
$(\alpha,\beta,\gamma)$, the mass profile is:
\begin{eqnarray}
  M(r)=4\pi\displaystyle\int_{0}^{r}s^2\rho(s)\mathrm{d}s=\frac{4\pi \rho_\mathrm{s}r_\mathrm{s}^3}{3-\gamma}\biggl (\frac{r}{r_\mathrm{s}}\biggr )^{3-\gamma}\hspace{0.6in}\\
  _2F_1\biggl [\frac{3-\gamma}{\alpha},\frac{\beta-\gamma}{\alpha};\frac{3-\gamma+\alpha}{\alpha};-\biggl (\frac{r}{r_\mathrm{s}}\biggr )^\alpha\biggr ],\nonumber
  \label{eq:hernquist2}
\end{eqnarray}
where $_2F_1(a,b;c;z)$ is Gauss' hypergeometric function.  

Eq.~(\ref{eq:hernquist1}) includes plausible halo shapes ranging
from the constant-density `cores' ($\gamma=0$) that seem to describe
rotation curves of spiral and low-surface-brightness galaxies (e.g.,
\citealt{deblok10} and references therein) to the centrally divergent
`cusps' ($\gamma>0$) motivated by cosmological N-body simulations
that model only the DM component. For $(\alpha,\beta,\gamma)
= (1,3,1)$ Eq. (\ref{eq:hernquist1}) is just the cuspy NFW
\citep{navarro96,navarro97} profile.

\subsection{Markov-Chain Monte Carlo Method}
\label{subsec:mcmc}

For a given halo model we compare the projected (squared) velocity
dispersion profile $\sigma^2_p(R)$ (obtained from Eq.~\ref{eq:jeansproject}) to the empirical profile $\langle
\hat{V}^2\rangle(R)$ (displayed in Fig.~\ref{fig:momentprofiles})
using the likelihood function
\begin{equation}
  \zeta=\displaystyle \prod_{i=1}^N \frac{1}{\sqrt{2\pi\mathrm{Var}[\langle \hat{V}^2\rangle (R_i)]}}\exp\biggl [-\frac{1}{2}\frac{(\langle \hat{V}^2\rangle (R_i)-\sigma_p^2(R_i))^2}{\mathrm{Var}[\langle \hat{V}^2\rangle (R_i)]}\biggr ],
  \label{eq:likelihood}
\end{equation}
where $\mathrm{Var}[\langle \hat{V}^2\rangle (R_i)]$ is the variance
associated with the empirical mean square velocity, as estimated from
our bootstrap re-sampling.

In order to explore the large parameter space efficiently, we employ
Markov-chain Monte Carlo (MCMC) techniques.  That is, we use the
standard Metropolis-Hastings algorithm \citep{metropolis53,hastings70}
to generate posterior distributions according to the following
prescription: 1) from the current location in parameter space, $S_n$,
draw a prospective new location, $S'$, from a Gaussian probability
density centred on $S_n$; 2) evaluate the ratio of likelihoods at
$S_n$ and $S'$; and 3) if $\zeta(S')/\zeta(S_n)\ge 1$, accept such
that $S_{n+1}=S'$, else accept with probability $\zeta(S')/\zeta(S_n)$
and $S_{n+1}=S_n$ with probability $1-\zeta(S')/\zeta(S_n)$.  In order
to account for the observational uncertainty associated with the
half-light radius adopted from \citet{ih95}, for each new point we
scatter the adopted value of $r_{\rm half}$ by a random deviate drawn
from a Gaussian distribution with standard deviation equal to the
published error. This method effectively propagates the observational
uncertainty associated with the half-light radius to the posterior
distributions for our model parameters.

Solutions of the Jeans equations are not guaranteed to correspond to
physical models, as the associated phase-space distribution functions
may not be everywhere positive. \cite{2006ApJ...642..752A} have
derived a necessary relation between the asymptotic values of the
logarithmic slope of the gravitational potential, the tracer density
distribution and the velocity anisotropy at small radii. Models which
do not satisfy this relation will not give rise to physical
distribution functions. In terms of our parametrisation, this relation
becomes
\begin{equation}
\gamma_{\rm tracer} \gtrsim 2\beta_{\rm aniso}.
\label{eq:anevans}
\end{equation}
We therefore exclude from the
Markov Chain those models which do not satisfy this condition.  Because
the Plummer profiles we use to describe dSph surface brightness profiles
have $\gamma_{\rm tracer}=0$, this restriction implies $\beta_{\rm
aniso}\la 0$. Given our assumption of constant velocity
anisotropy, this disqualifies all radially anisotropic models.
Relaxing this condition affects the results
on the $J$-factors, but the difference is contained within their CLs
(see Appendix~\ref{app:light_profile}).

For this procedure we use the adaptive MCMC engine
CosmoMC \citep{lewis02}. \footnote{available at http://cosmologist.info/cosmomc}
 Although it was developed specifically for analysis
of cosmic microwave background data, CosmoMC provides a generic
sampler that continually updates the probability density according to
the parameter covariances in order to optimise the acceptance rate.
For each galaxy and parametrisation we run four chains simultaneously,
allowing each to proceed until the variances of parameter values
across the four chains become less than 1\% of the mean of the
variances.  Satisfying this convergence criterion typically
requires $\sim 10^4$ steps for our chains.  We then estimate the
posterior distribution in parameter space using the last half of all
accepted points (we discard the first half of points, which we
conservatively assume corresponds to the `burn-in' period).

\section{Detectability of Milky Way dSphs}
\label{sec:results}

This section provides our key results. For the benefit of
readers who start reading here, we summarise our findings so far.

In Section~\ref{sec:method}, we focused on generic $(1,3,\gamma)$
profiles, to show that, most of the time, the sub-structure
contribution is negligible, and to check that the only relevant
dSph halo parameters are the density normalisation $\rho_\mathrm{s}$,
the scale radius $r_\mathrm{s}$, and the inner slope $\gamma$ (because
$J_{\rm dSph}\propto r_\mathrm{s}^{2\gamma}\times (\alpha_{\rm
  int}d)^{3-2\gamma}$, see also Appendix~\ref{app:toyJ}).

In Section~\ref{sec:detectability}, we provided the sensitivity of
present and future $\gamma$-ray observatories, showing how it is
degraded when considering `extended' sources (e.g. a flat profile
for close dSphs), and an instrument response that varies with energy.

In Section~\ref{sec:mcmc}, we presented our method to perform a
Markov-Chain Monte Carlo analysis of the observed stellar kinematics
in the 8 classical Milky Way dSphs under the assumptions of virial
equilibrium, spherical symmetry, constant velocity anisotropy, and a
Plummer light distribution. The analysis uses the observed velocity
dispersion profiles of the dSphs to constrain their underlying dark
matter halo potentials, parametrised using the five parameter models
of Eq.~(\ref{eq:hernquist1}).

\begin{figure*}
\includegraphics[width=0.8\linewidth]{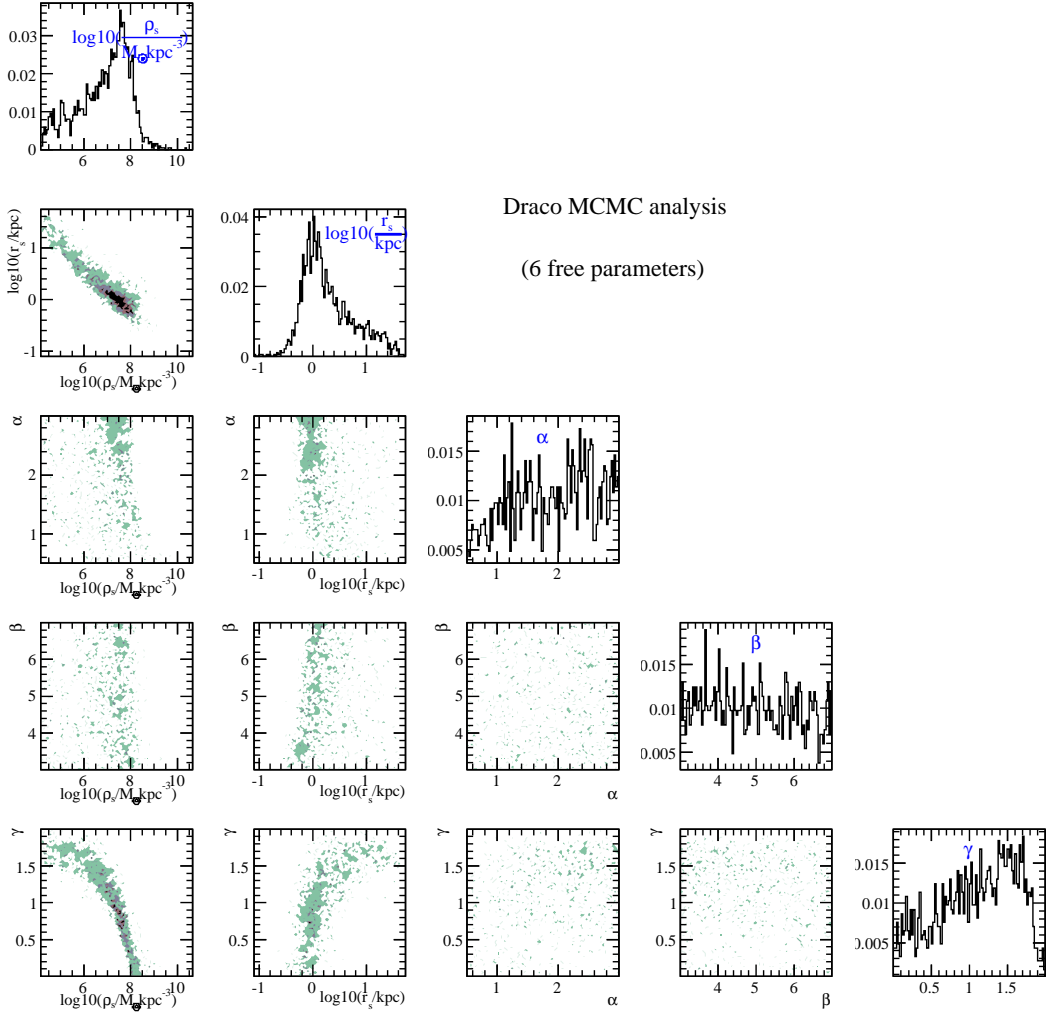}
\caption{Joint distributions and marginalised PDFs of parameters
  entering the MCMC for the Draco dSph. The off-diagonal plots show
  joint distributions that highlight correlations between the
  parameters, while the on-diagonal plots are the marginalised PDFs of
  the parameters. This marginalisation includes the marginalisation
  over the velocity anisotropy parameter $\beta_\mathrm{ani}$. (We do
  not plot a marginalised PDF or correlation for $\beta_\mathrm{ani}$
  since it is a nuisance parameter for our analysis here.)}
\label{fig:cor_par6}
\end{figure*}

\subsection{6-parameter MCMC analysis|varying $\gamma$}
\label{subsec:gamma-free}

Our kinematic models have six free parameters, for which we adopt
uniform priors over the following ranges:
\begin{align*}
  -\log_{10}[1-\beta_{\rm ani}]: [-1,+1];\\
  \log_{10}[\rho_\mathrm{s}/(M_{\odot}\mathrm{pc}^{-3})]: [-10,+4];\\
  \log_{10}[r_\mathrm{s}/\mathrm{pc}]: [0, 4];\\
  \alpha: [0.5, 3];\\
  \beta: [3, 7].\\
  \gamma: [0, 2]~~{\rm or}~~[0, 1];\\
  \label{eq:priors}
\end{align*}
The anisotropy parameter $\beta_{\rm ani}$ does not enter directly the
profile/mass/J calculation, although it is of fundamental importance for the fit
as it can correlate with the DM profile structure parameters (so with the mass and 
the $J$-factor). We have not checked
explicitly the details of these correlations, but we have checked that restricting
the range of possible $\beta_{\rm ani}$ does not significantly impact on the results for
the $J$ calculation. Hence, we do not discuss this parameter further below.

\subsubsection{Parameter correlations}

Fig.~\ref{fig:cor_par6} shows the marginalised probability density
functions (PDFs) of the profile parameters and the joint distributions
of pairs of parameters.  The features of these plots are driven by the
fact that most of the stellar kinematic data lie at radii of up to few
hundred parsecs (see Fig.~\ref{fig:momentprofiles}).  For instance,
the outer slope $\beta$ is not at all constrained (i.e. the fit is
insensitive to the value of $\beta$), because only tracers beyond a
radius of $r\gtrsim 1$~kpc are sensitive to this parameter and these
radii are sparsely sampled by the observations. The transition slope
$\alpha$ and then the inner slope $\gamma$ are the two other least
constrained parameters. In terms of best-fit models, as seen in
Fig.~\ref{fig:momentprofiles}, the match to kinematics data is equally
good for varying $\gamma$ (black) models and models in which we fix
the value to $\gamma=0$ (blue), or $\gamma=1$ (red). In the following,
we will not discuss further the best-fit values. The more meaningful
quantity, in the context of an MCMC analysis providing PDFs, is the
{\em median} of the distribution.

Several groups have  shown recently that in a Jeans analysis, the
observed flatness of dSph velocity dispersion profiles
\citep{walker07b} leads to a constraint on $M(r_{\rm half})$---the mass
enclosed within a sphere of radius $r_{\rm half}$---that is insensitive to
assumptions about either anisotropy or the structural parameters of
the DM halo \citep{2009ApJ...704.1274W,2010MNRAS.406.1220W}.
Using for the appropriate radius the mass estimate Eq.~(\ref{eq:M300})
and the above constraint leads to a relation between the profile
parameters
\[
   \log(\rho_\mathrm{s}) + \gamma \log(r_\mathrm{s}) \approx \mathrm{constant}.
\]
This relation explains the approximately linear correlations between
these parameters seen, for instance, in the bottom left panel of
Fig.~\ref{fig:cor_par6}.

\subsubsection{From $\rho(r)$ to $J(\alpha_{\rm int})$: uncertainty
and impact of $\gamma_{\rm prior}$}

Fig.~\ref{fig:best_CL_profiles} shows the density profile for Draco
as recovered by our MCMC analysis. It is noticeable that the
confidence limits are narrower for radial scales of a few hundreds
pc|this is a common feature of the density profile confidence limits for
all the dSphs we have considered. As discussed above, this is partly
due to the fact that these are the radii at which the majority of the
kinematic data lie. The least constrained $\rho(r)$ (less pronounced
narrowing of the confidence limits) is that of Sextans, for which the
range where useful data can be found is clearly the smallest compared
to other dSphs (see Fig.~\ref{fig:momentprofiles}).

\begin{figure}
\includegraphics[width=\linewidth]{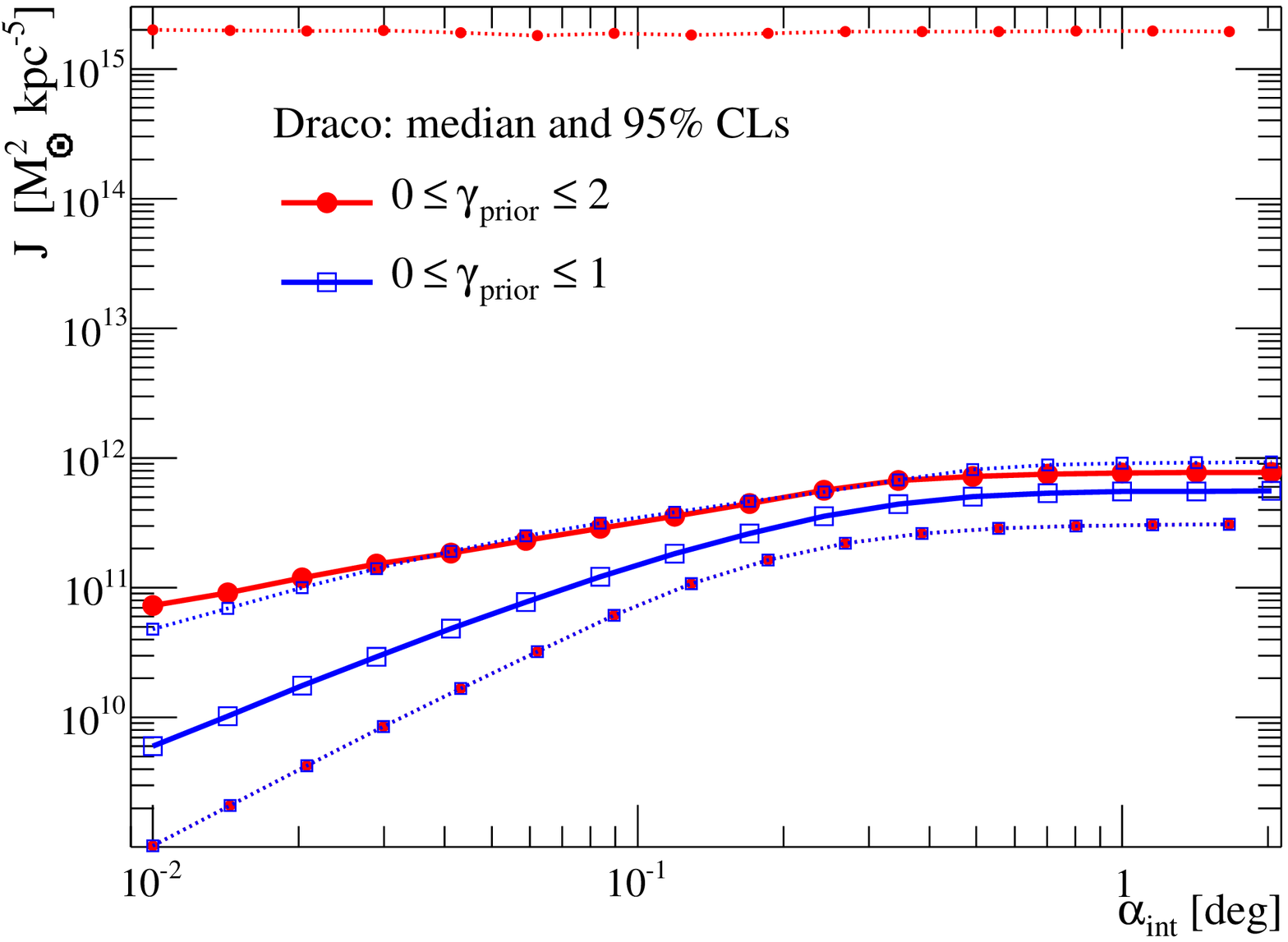}
\includegraphics[width=\linewidth]{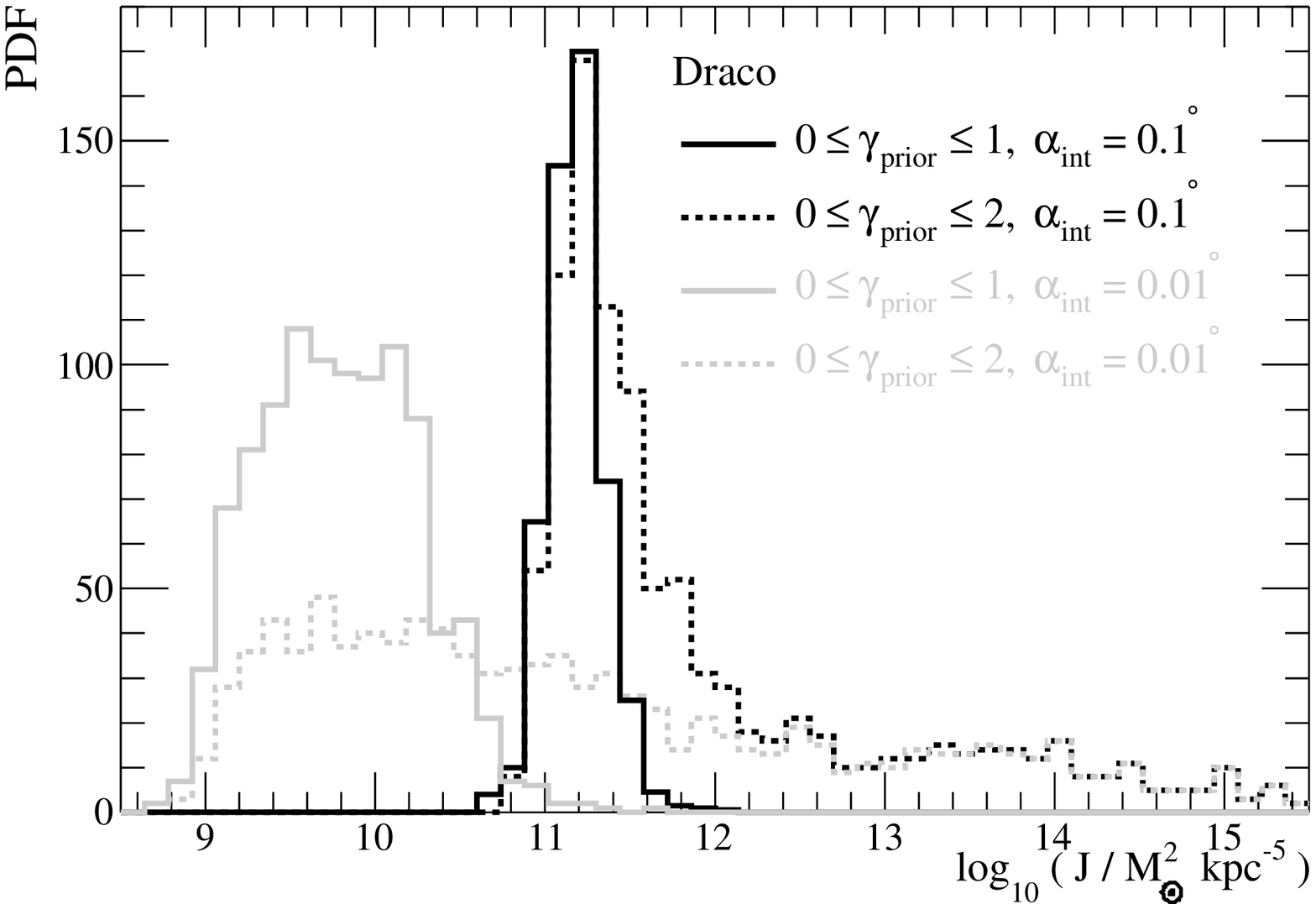}
\caption{{\bf Top}: $J(\alpha_{\rm int})$ for Draco as a function of the
  integration angle.  Solid lines correspond to the median model,
  dotted lines to the 95\% lower and upper CL. The to sets of curves
  correspond to two different $\gamma_{\rm prior}$ for the MCMC
  analysis on the same data. {\bf Bottom:} PDF of the J-factor for $\alpha_{\rm int}=0.01^\circ$ (grey) 
and $\alpha_{\rm int}=0.1^\circ$ (black), when the range of the inner slope prior is $[0-1]$ (solid
lines) or $[0-2]$ (dashed lines).}
\label{fig:cumul_draco}
\end{figure}

The variation of the constraints on $\rho(r)$ as a function of radius
impacts directly on the behaviour of $J$. Complications arise because
it is the profile squared that is now integrated along a l.o.s.
(given the integration angle $\alpha_{\rm int}$, see Eq.~\ref{eq:J}).
The median value and 95\% CL on $J$ as a function
of the integration angle $\alpha_{\rm int}$ is plotted in
Fig.~\ref{fig:cumul_draco} (top), for two different priors on $\gamma_{\rm prior}$. The bottom
panel gives the corresponding PDF for two integration angles. The prior has a strong
impact on the result: the median (thick solid curves and large
symbols -- top panel) is changed by $\sim 50\%$ for $\alpha_{\rm int}\gtrsim
0.1^\circ$, but by a factor of ten for $\alpha_{\rm int}\sim
0.01^\circ$. However, the most striking feature is the difference
between the CLs: for the prior $0\leq\gamma_{\rm prior}\leq2$, the
typical uncertainty is 3 to 4 orders of magnitude (red dotted curves),
whereas it is only $\lesssim$ than one order of magnitude for the
prior $0\leq\gamma_{\rm prior}\leq1$ (blue dotted
curves).\footnote{Note that this behaviour is grossly representative of
  all dSphs, although the integration angle for which the uncertainty
  is the smallest, and the amplitude of this uncertainty depend,
  respectively, on the dSph distance (see Section~\ref{sec:gen_dep}
  for the generic dependence), and on the range/precision of the
  kinematic data (see above).} The bottom panel of Fig.~\ref{fig:cumul_draco}
  shows that $\log_{10}J$ has a long and flat tail (associated to
  large $\gamma$ values). This tail is responsible for the large upper limit of the
  $J$-factor CLs for $0\leq\gamma_{\rm prior}\leq2$.

In Appendix~\ref{app:biases2}, a detailed analysis of the impact of these
two priors is carried on artificial data (for which the true profile
is known). We find that the prior $0\leq \gamma_{\rm prior} \leq 2$
satisfactorily reconstructs $\rho(r)$ and $J(\alpha_{\rm int})$,
i.e. the MCMC CLs bracket the true value. This is also the case when
using the prior $0\leq \gamma_{\rm prior} \leq 1$. However, two
important points are noteworthy:
\begin{itemize}
\item this prior obviously performs better for $0\leq\gamma_{\rm
    true}\leq1$ profiles where it gives much tighter constraints on
  $J$;
\item for cuspier profiles (e.g., $\gamma_{\rm true}=1.5$), this prior
  succeeds slightly less (than the prior $0\leq\gamma_{\rm
    prior}\leq2$) in reconstructing $\rho(r)$, but it does
  surprisingly better on $J$ in terms of providing a value closer to
  the true one (see details and explanations in
  Appendix~\ref{app:biases2}).
\end{itemize}
DM simulations and observations do not favour
$\gamma>1$, although steeper profiles can still fit the kinematic data in a Jeans analysis.
Indeed, the Aquarius simulations indicate values of
$\gamma$ slightly smaller than 1, and although some recent simulations \citep{2010ApJ...723L.195I}
have argued for cuspy profiles, this happens for micro-haloes only.
Given that the $J$-factor for the cuspier profiles are only
marginally more (or even less) reliable when using the prior
$0\leq\gamma_{\rm prior}\leq2$, we restrict ourselves to the $0\leq
\gamma_{\rm prior} \leq 1$ prior below.

Note that other sources of bias exist. First, the reconstruction of
$\rho(r)$ or $J(\alpha_{\rm int})$ is affected by the choice of
binning used in the estimation of the empirical velocity dispersion
profiles.  Appendix~\ref{app:binning} shows that we obtain slightly
different results when we apply our method to empirical velocity
dispersion profiles calculated from the same raw kinematic data, but
using different numbers of bins.  We find that the effects of binning
add an extra factor of a few uncertainty on $J$ for the least well
measured (in terms of radial coverage) dSphs, for which more
measurements are desirable. (On the other hand, Fornax and Sculptor
are found to provide robust results against different binnings.)
Second, we note that the analysis presented here uses a fixed profile
for the light distribution which, when combined with our assumption of
constant velocity anisotropy, restricts the possible halo profiles we
can recover. Our constraints on $\rho(r)$ and
$J(\alpha_{\rm int})$ are therefore sensitive to these
assumptions~\cite[see, e.g., ][for an example of fitting the dSph
kinematic data with cusped profiles when the light profile is also
allowed to be cusped]{2010MNRAS.408.2364S}, although this does
not change our conclusions (see Appendix~\ref{app:light_profile} where
different light profiles are used). This situation is set to
change over the coming years as new distribution function-based models
will permit constraints to be placed on the slope of the DM
density profiles (Wilkinson et al. 2011, in prep.).

\subsubsection{Best constraints on $J$: median value and CLs}

\begin{table*}
  \caption{Positions of the classical dSphs \citep{mateo98} sorted according
    to their distance: longitude, latitude, distance, $2r_{\rm half}$
    (taken from \citealt{ih95}), the galactic angle away from
    centre  $\phi=\cos^{-1}[\cos({\rm long.})\cos({\rm lat.})]$, and
    $\alpha_\mathrm{c} \approx 2r_\mathrm{half}/d$ (see
    \citealt{letter_dsph}). The remaining columns are the
    median value with 68\% (95\%) CLs for $M_{300}$ and 
    $\log_{10}[J(\alpha_{\rm int})]$
    from the six-parameter
    MCMC analysis ($0\leq \gamma_{\rm prior}\leq 1$). For  conversion
    factors to units used in other studies, please refer to numbers given in
    Appendix~\ref{app:defs}.}
\label{tab:res_par6}
\begin{tabular}{@{}lccccccccccc} \hline\hline
  dSph    & long.  & lat.   &   d   & $2r_h$ & $\phi$  & $\alpha_\mathrm{c}$& $M_{300}$ &$\log_{10}[J(0.01^\circ)]$   &  $\log_{10}[J(0.1^\circ)]$     &     $\log_{10}[J^\star(\alpha_c)]$ \\
          &  [deg] & [deg]  & [kpc] &  [kpc] & [deg]  &  [deg]     & $[10^7 M_{\odot}]$   &\multicolumn{3}{c}{$[M_{\odot}^{2}\,$kpc$^{-5}]$}\vspace{0.05cm}\\ 
\hline 
Ursa Minor$\!\!\!\!$& 105.0 &  +44.8 &  66 & 0.56  & 100.6  &  0.49&  $1.54_{-0.21(-0.42)}^{+0.18(+0.33)}$ & $10.5_{-0.6(-1.2)}^{+0.8(+1.5)}$ & $11.7_{-0.3(-0.6)}^{+0.5(+0.8)}$ & $12.0_{-0.1(-0.2)}^{+0.3(+0.5)}$ \vspace{0.2cm}\\
Sculptor  & 287.5  &  -83.2 &   79  &  0.52  & 88.0   &   0.38     &  $1.34_{-0.13(-0.23)}^{+0.12(+0.23)}$ & $10.0_{-0.5(-0.8)}^{+0.5(+0.9)}$ & $11.3_{-0.2(-0.3)}^{+0.2(+0.4)}$ & $11.7_{-0.1(-0.1)}^{+0.1(+0.2)}$ \vspace{0.2cm}\\
Draco     & 86.4   &  +34.7 &   82  &  0.40  & 87.0   &   0.28     &  $1.22_{-0.14(-0.28)}^{+0.15(+0.28)}$ & $9.8_{-0.5(-0.8)}^{+0.5(+0.9)}$  & $11.2_{-0.2(-0.3)}^{+0.2(+0.4)}$ & $11.6_{-0.1(-0.2)}^{+0.1(+0.2)}$ \vspace{0.2cm}\\
Sextans   & 243.5  &  +42.3 &   86  &  1.36  & 109.3  &   0.91     &  $0.61_{-0.31(-0.43)}^{+0.38(+0.96)}$ & $9.4_{-1.2(-1.8)}^{+1.7(+2.9)}$  & $10.7_{-0.8(-1.1)}^{+1.1(+1.9)}$ & $11.1_{-0.4(-0.6)}^{+0.7(+1.5)}$ \vspace{0.2cm}\\
Carina    & 260.1  &  -22.2 &   101 &  0.48  & 99.2   &   0.27     &  $0.59_{-0.07(-0.14)}^{+0.10(+0.60)}$ & $9.3_{-0.4(-0.8)}^{+0.3(+0.8)}$  & $10.5_{-0.1(-0.2)}^{+0.2(+0.4)}$ & $10.9_{-0.1(-0.1)}^{+0.1(+0.1)}$ \vspace{0.2cm}\\
Fornax    & 237.1  &  -65.7 &   138 &  1.34  & 102.9  &   0.56     &  $1.01_{-0.17(-0.28)}^{+0.30(+0.60)}$ & $9.5_{-0.5(-0.8)}^{+0.5(+1.1)}$  & $10.8_{-0.2(-0.3)}^{+0.2(+0.5)}$ & $10.5_{-0.2(-0.4)}^{+0.3(+0.7)}$ \vspace{0.2cm}\\
LeoII     & 220.2  &  +67.2 &   205 &  0.30  & 107.2  &   0.08     &  $0.94_{-0.18(-0.29)}^{+0.26(+0.50)}$ & $11.6_{-0.8(-1.5)}^{+0.8(+1.7)}$ & $11.7_{-0.6(-0.9)}^{+0.7(+1.6)}$ & $11.7_{-0.6(-0.9)}^{+0.7(+1.6)}$ \vspace{0.2cm}\\
LeoI      & 226.0  &  +49.1 &   250 &  0.50  & 117.1  &   0.11     &  $1.22_{-0.21(-0.36)}^{+0.24(+2.52)}$ & $9.7_{-0.2(-0.5)}^{+0.3(+1.0)}$  & $10.7_{-0.1(-0.2)}^{+0.1(+0.3)}$ & $10.7_{-0.1(-0.2)}^{+0.1(+0.3)}$ \\
\hline
\end{tabular}
\\$^\star$ Note that the values for $\log_{10}[J(\alpha_c)]$ differ from those quoted
in \citet{letter_dsph} as the MCMC analysis is slightly different here.
\end{table*}

As validated by the simulated data, we are now able to provide robust
(although possibly not the best achievable with current data) and
model-independent constraints on $J(\alpha_{\rm int})$ for the
8 classical dSphs.  The results are summarised in
Table~\ref{tab:res_par6} in terms of the median, and 68\% and 95\%
CLs. The $J$-factor is  calculated for 
$\alpha_{\rm int}=0.01^{\circ}$ (an angle
slightly better than what can be achieved with FCA),
$\alpha_{\rm int}=0.1^{\circ}$ (typical of the angular
resolution of existing GeV and TeV $\gamma$-ray instruments),
and for $\alpha_{c}=2r_\mathrm{half}/d$
(as proposed in \citealt{letter_dsph}). We do
not report the values of $\rho_\mathrm{s}$ and $r_\mathrm{s}$ as these
vary across a large range|and therefore do not give additional useful
information|nor the value of $\gamma$ as it is forced in the range
$0\leq\gamma_{\rm prior}\leq1$ to give the least biased $J$ value.

There is no simple
way to provide unambiguously the best target, as their relative merit
depends non trivially on their distance, their mass and the
integration angle selected. As proposed in \citet{letter_dsph}, since the
most robust constraint on $J$ is obtained for $\alpha_{\rm
  int}=\alpha_c$, having different integration
angles for each dSph can be a good starting point to establish a relative ranking. 
The situation is  complicated further for
background-limited instruments such as CTA, as some loss of
sensitivity can occur (see, e.g. Figure 4 of \citealt{letter_dsph}).
This is discussed, taking into account the full detail of the instruments,
in Section~\ref{sec:5.3}. However, in this respect, the best target for future
instrument may eventually become Leo II, which despite a quite large
uncertainty outshines all other dSphs at $\alpha_{\rm int}=0.01^\circ$
(see also Fig.~\ref{fig:JdSph_Jdm_bkgd}).  We note
however that it is the dsph with the smallest amount of
kinematic data at present (so it has the most uncertain $J$-factor).

\subsubsection{dSphs in the diffuse galactic DM signal: contrast}

\begin{figure}
\includegraphics[width=\linewidth]{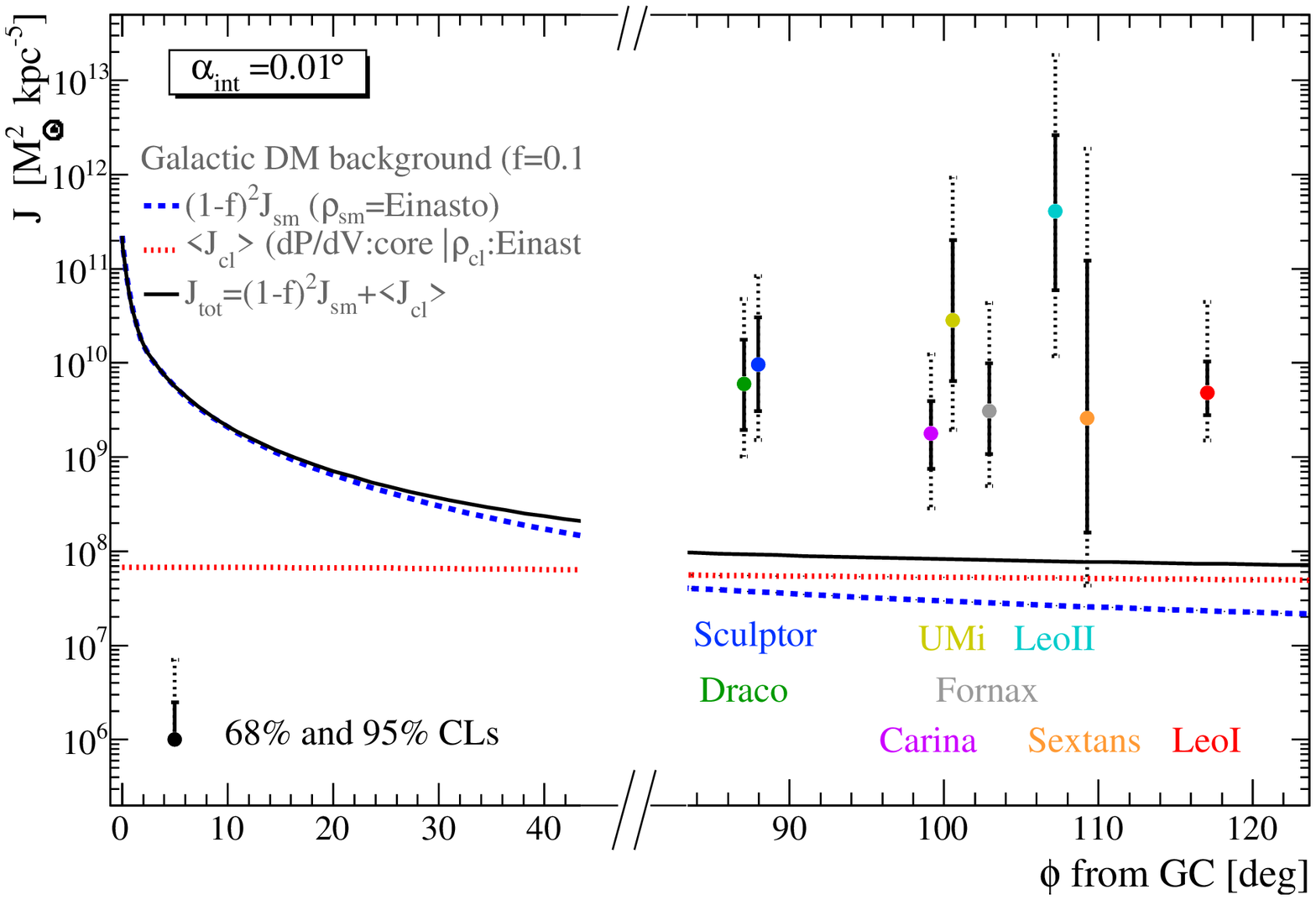}
\includegraphics[width=\linewidth]{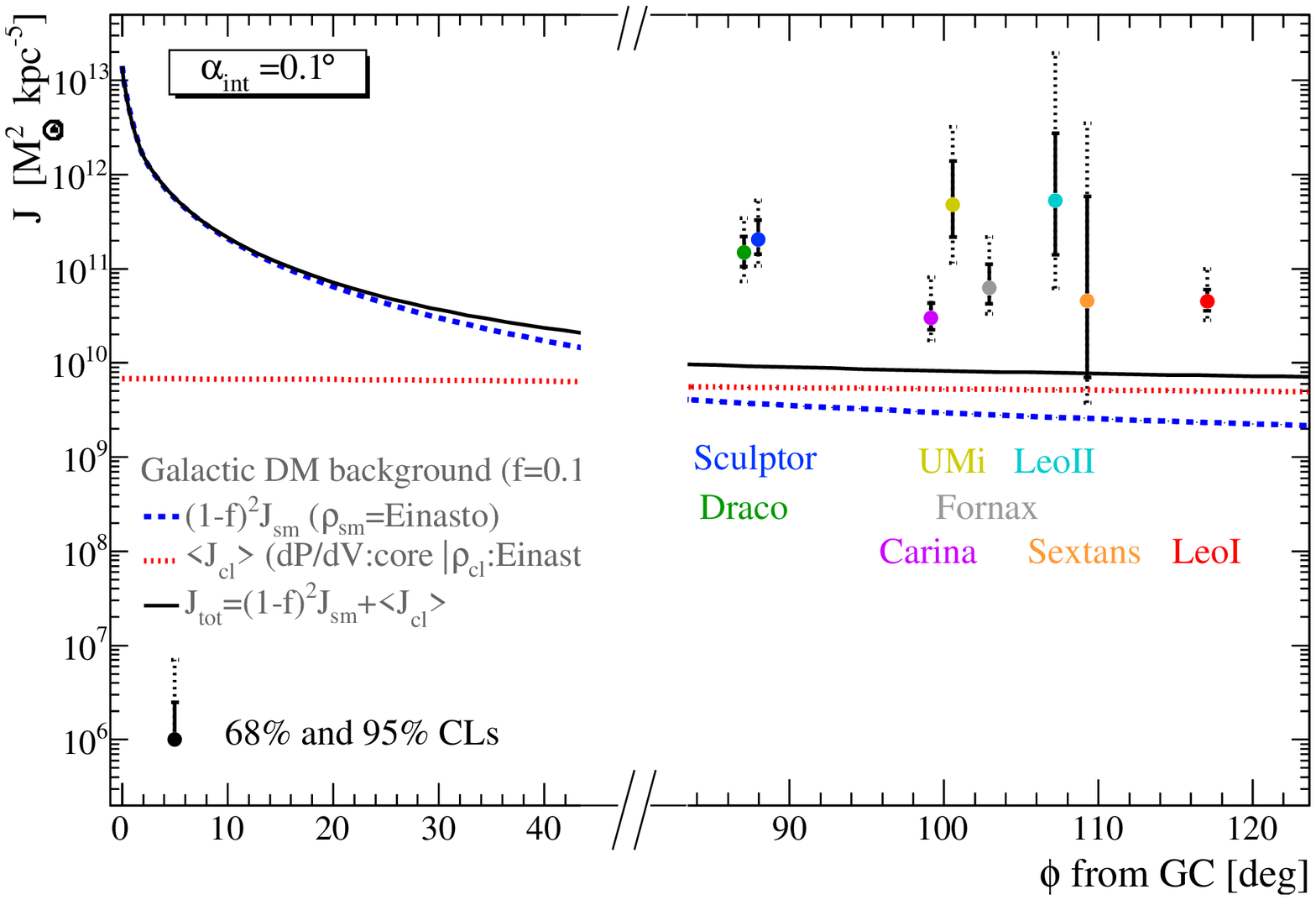}
\includegraphics[width=\linewidth]{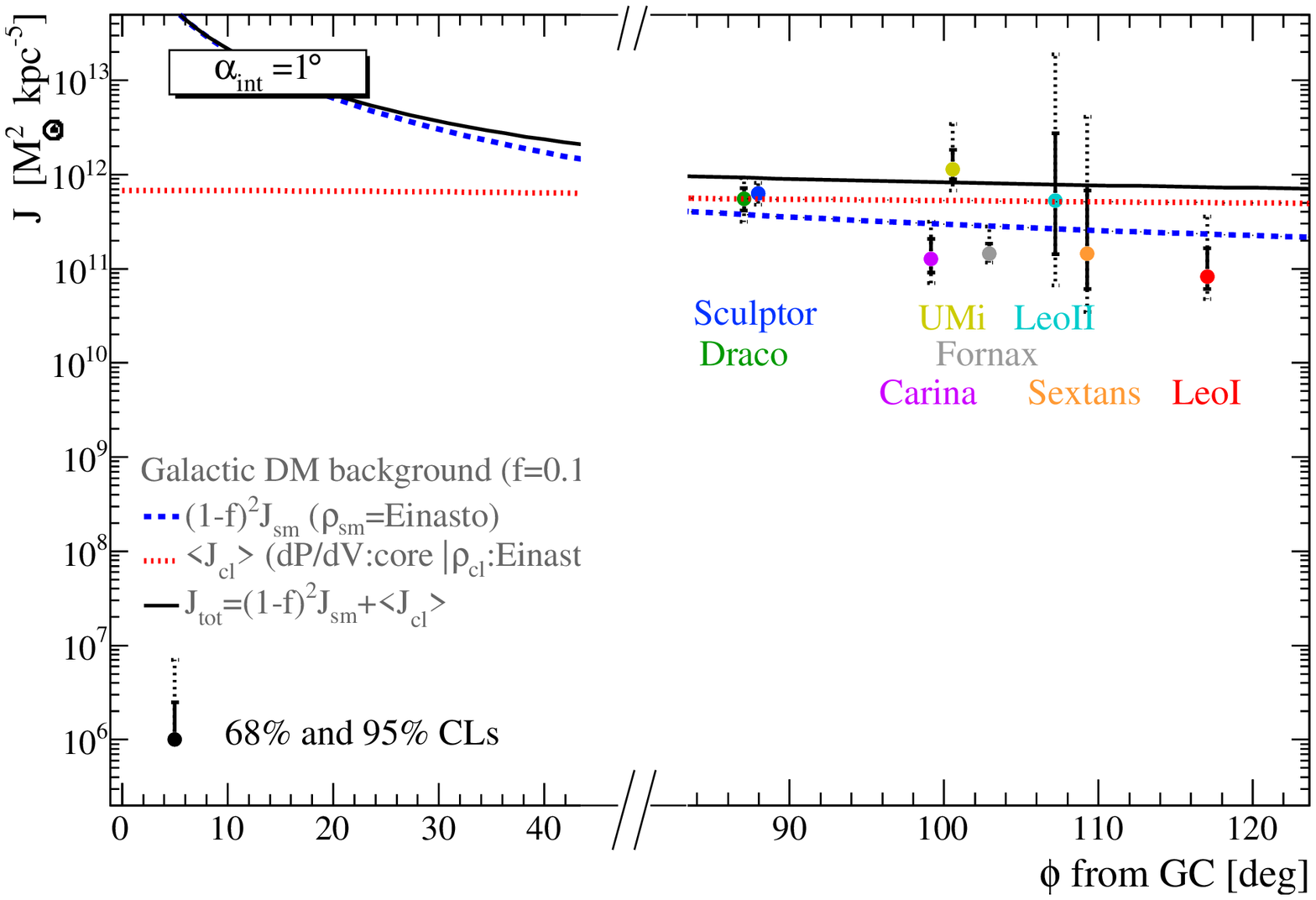}
\caption{Galactic contributions to $J$ for the smooth (blue-dashed
  line), mean clump (red-dotted line) and sum (black-solid line) vs
  the angle from the Galactic centre. The symbols show $J$ for the
  dSphs, assuming a prior of $0 \leq \gamma_{\rm prior} \leq 1$ on the
  central DM slope. The central point corresponds to the
  median values, the solid bars to the 68\% CLs, and the dotted bars
  to the 95\% CLs.  The integration angle is, from top to bottom
  $0.01^\circ$, $0.1^\circ$, and $1^\circ$. The Galactic contributions
  $J_{\rm sm}$ and $\langle J_{\rm subcl}\rangle$ scale as
  $\alpha_{\rm int}^2$, but $J_{\rm dSph}$ does not, changing the
  contrast of the dSphs w.r.t.  to the DM Galactic background (see
  text for details).}
\label{fig:JdSph_Jdm_bkgd}
\end{figure}

The uncertainties in $J$ are illustrated from a different viewpoint in
Fig.\ref{fig:JdSph_Jdm_bkgd}.  It shows, in addition to the mean,
68\% and 98\% CLs on the  $J$s, the latitudinal dependence of the Galactic DM
background (smooth and galactic clump contribution) for the same
integration angle.\footnote{The smooth profile is taken to be an
  Einasto profile, the clump distribution is a core one, whereas their
  inner profile are Einasto with concentration and parameters {\em \`a
    la} \citet{2001MNRAS.321..559B} Normalising the mass
  distribution to have 100 clumps more massive than $10^8M_\odot$, and
  taking $dP/dM\propto M^{-1.9}$ leads to a DM fraction into clumps of
  $\sim 10\%$ for clumps distributed in the range $10^{-6}-10^{10}
  M_\odot$ \citep[see, e.g.,][and references
  therein]{2008A&A...479..427L}. The local DM distribution is fixed to
  the fiducial value $\rho_\odot=0.3$~GeV~cm$^{-3}$. The exact
  configuration is unimportant here as this plot is mostly used for
  illustration purpose.} For a typical present-day instrument resolution
(integration angle $\alpha_{\rm int}\sim0.1^\circ$), we recover the
standard result that the Galactic Centre outshines all dSphs.

The three panels illustrate the loss of contrast (signal from the dSph
w.r.t. to the diffuse Galactic DM signal) as the integration angle is
increased. This is understood as follows: the integrand appearing in
Eqs.~(\ref{eq:J2}) and~(\ref{eq:J3}) is mostly insensitive to the
l.o.s. direction a few tens of degree away from the Galactic centre,
so that Eq.~(\ref{eq:alphint2_dep}) holds, giving an $\alpha_{\rm
  int}^2$ dependence.

For detectability (see also Sec~\ref{sec:detectability}), the
na\"ive approach of maximising the integration angle (to maximise
$J_{\rm dSph}$) must be weighed against the fact that an increased
integration angle means more astrophysical $\gamma$-ray and cosmic-ray
background.  For large integration angles, dSphs also have poor contrast
against the diffuse Galactic DM annihilation signal, indicating that the Galactic
halo is a better target for any search on angular scales $\gtrsim$1
(see e.g. \citealt{2011PhRvL.106p1301A} for such a  search with
H.E.S.S.).

\subsubsection{Comparison to other works}

Comparison between different works can be difficult as every author
uses different definition, notations and units for the astrophysical
factor. To ease the comparison, we provide in Appendix~\ref{app:defs}
conversion factors between standard units (we also point out issues to
be aware of when performing such comparisons).

Below is a comparison to just a few of the works published on the
subject, and only for the objects that these studies and the present one have
in common:
\begin{itemize}
\item The \citet{Evans:2003sc} values of $J/\Delta\Omega$ for Draco (with
  $\Delta\Omega=10^{-5}$ i.e. $\alpha_{\rm int}=0.1^\circ$) for
  all the profiles they explored (cored, $\gamma=0.5$, $\gamma=1$,
  $\gamma=1.5$) are larger  (after correction by
  $\Delta\Omega$, given their definition of the astrophysical factor) than our  95\% CL upper limit for
  this object shown in Table~\ref{tab:res_par6}. The difference is probably related to
  our data set which is about twice as large as that used by \citet{Evans:2003sc}.
  
\item \citet{2007PhRvD..75h3526S} provide directly the $\gamma$-ray
  flux (i.e. including the particle physics term), so that we can only
  compare our respective rankings. These agree in general but for Sculptor  we
  find  a larger flux than Draco, conversely to these authors.

\item \citet{2009A&A...496..351P} focused on Sextans, Carina, Draco
  and Ursa Minor. They found the latter to have the largest $J$
  ($\Phi_{\rm cosmo}$ in their notation) of these 4 objects,
  followed by Draco, Carina and Sextans. But for the last two, this
  ranking is similar to ours. However, while their values of $J$ fall
  within our 68\% (UMi, Sextans) or 95\% (Carina) CL, their
  value for Draco is above our 95\% CL upper limit.

\item \citet{2009PhRvD..80b3506E} also performed a statistical study on Draco and Ursa Minor, to determine their profiles from kinematic data and to derive the confidence levels on the J-factor.
Given that their integration is performed on a slightly larger opening angle ($0.14^\circ$), our results appear 
to be in agreement. Their 90\% CL limits are 2-3 times larger than the 95\% CL limits given in Table~\ref{tab:res_par6} 
but this may be due to the larger range they adopt for the prior on the inner slope (see App.~\ref{app:biases1}).

\item \citet{2010AdAst2010E..45K} gives the astrophysical factors of
  all the dSphs using a point-like
  approximation and a NFW DM profile, and integrated with a
  $\alpha_{\rm int}=0.15^\circ$ angular resolution. These can be
   compared to the median and confidence levels we derived in
  Table~\ref{tab:res_par6} for $\alpha_{\rm int}=0.1^\circ$. The
  values of \citet{2010AdAst2010E..45K} (multiplied by $4\pi$ to match
  our definition of $J$) generally fall inside our 68\% CL
  intervals, but for Leo~II his value is just within our 95\% CI while Draco and
  Carina  cannot be accommodated at all. For these two objects, the values
  of \citet{2010AdAst2010E..45K} are much larger than the ones we find, and this is
  unlikely to be explained by the $0.05^\circ$ difference in
  integration angles. A simple explanation is that \citet{2010AdAst2010E..45K} does not use
  stellar kinematic data directly in his analysis, but stacks suitable 
  Via Lactea halos ($M_{300} \approx 10^7 M_\odot$ and appropriate distances)
  and uses those averages to estimate $J$. Focusing on the ranking 
  (without worrying about contrast to the background and the other instrumental
  constraints), both we and \citet{2010AdAst2010E..45K} agree that
  among the classical dSph UMi is a most promising target. However,
  while we find Sculptor and Draco to be the next most favorable
  targets, \citet{2010AdAst2010E..45K} names Draco and Carina from
  his 'simulation-based' approach.
\end{itemize}

 For completness, we also compare our median values with the $J$ values used
by different experimental groups:
\begin{itemize}
\item The MAGIC collaboration published point source limits for Draco \citep{Albert2008} 
adopting the scheme of \citet{2007PhRvD..76l3509S} of a power law density profile,
with an exponential cut-off. They examine two scenarios, a cored and a 
cusped model, but find no discernable difference when calculating $J$ for 
integration angles $<0.4^\circ$, i.e. larger than the MAGIC PSF. The value of
$J$ they calculate for Draco is a higher than ours (after appropriate scaling
of the integration region and unit conversion) by about a factor of 2.

\item The VERITAS collaboration also published limits on Draco and Ursa Minor
\citep{2010ApJ...720.1174A}. They assume a NFW profile, take the
density profiles from \citet{2007PhRvD..75h3526S} and follow 
\citet{1998APh.....9..137B} for the calculation of J. Whilst the range of density values in 
\citet{2007PhRvD..75h3526S} have a physical motivation the values used
in \citet{2010ApJ...720.1174A} are rather arbitrarily chosen to be the midpoint of that 
range, which leads to consistently higher $J$-values than ours (by a factor of 3 for Draco
and a factor of 1.2 for UMi).

\item The H.E.S.S. collaboration \citep{2011APh....34..608H}
published limits on the southern sources Sculptor and 
Carina using NFW and isothermal profiles with a number of varying 
assumptions. This leads to a range of calculated $J$-values (rather than a 
single solution) that are consistent with our median value and estimated
uncertainties.

\item The Fermi collaboration \citep{2010ApJ...712..147A} has published limits
for a number of the sources studied here. They adopted a NFW profile within the
tidal radius and following \citet{2009JCAP...06..014M} they calculated the $J$-value
(using an MCMC approach on the observed stellar velocities) for a $1^\circ$
integration angle which is compatible with their high energy PSF. From this
they find Draco to have a larger $J$ compared to the other dwarfs (a
factor of $\sim 2$ higher than the next dwarf which is Ursa Minor),
contrary to what we find in this study.

\end{itemize}

\subsection{5-parameter MCMC analysis: $\gamma_{\rm prior}$ fixed}
\label{sec:gamma_fixed_analysis}

Higher resolution numerical simulations following both DM and gas,
additional kinematic data and new modelling techniques may help
constraining the value of $\gamma$ in the near future. With the
knowledge of $\gamma$, we should better constrain the
radial-dependence of $J$, which is crucial to disentangle, e.g. dark
matter annihilation from DM decay
\citep{boyarsky06,2010JCAP...07..023P}. The topic of decaying dark
matter goes beyond the scope of this paper, and it will be discussed
elsewhere. Below, we merely inspect the gain obtained on the $J$
prediction when having a strong prior on $\gamma$, and briefly comment
on the possibility to disentangle $\gamma=0$ profiles from
$\gamma=1.0$ profile in the case of annihilation (if this cannot be
achieved, hopes for disentangling decay from annihilation would be
quite low on a single object).
\subsubsection{Parameter correlations}

We repeat the MCMC analysis for fixed value of the inner slope
$\gamma_{\rm prior}=0.$, 0.5, 1., and 1.5. The priors for the five
other parameters are as given in Section~\ref{subsec:gamma-free}.

\begin{figure}
\begin{center}
\includegraphics[width=0.81\linewidth]{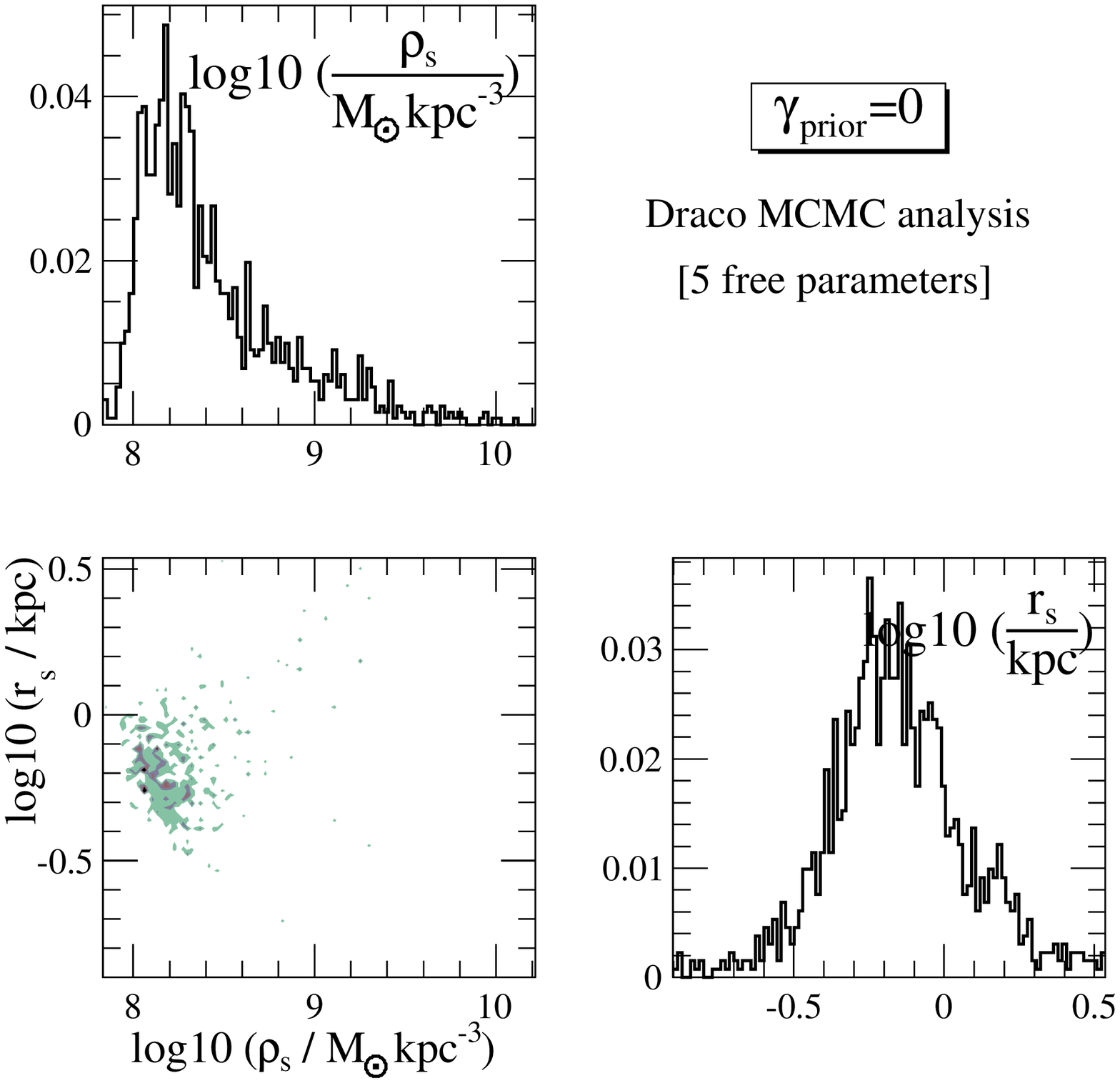}
\includegraphics[width=0.81\linewidth]{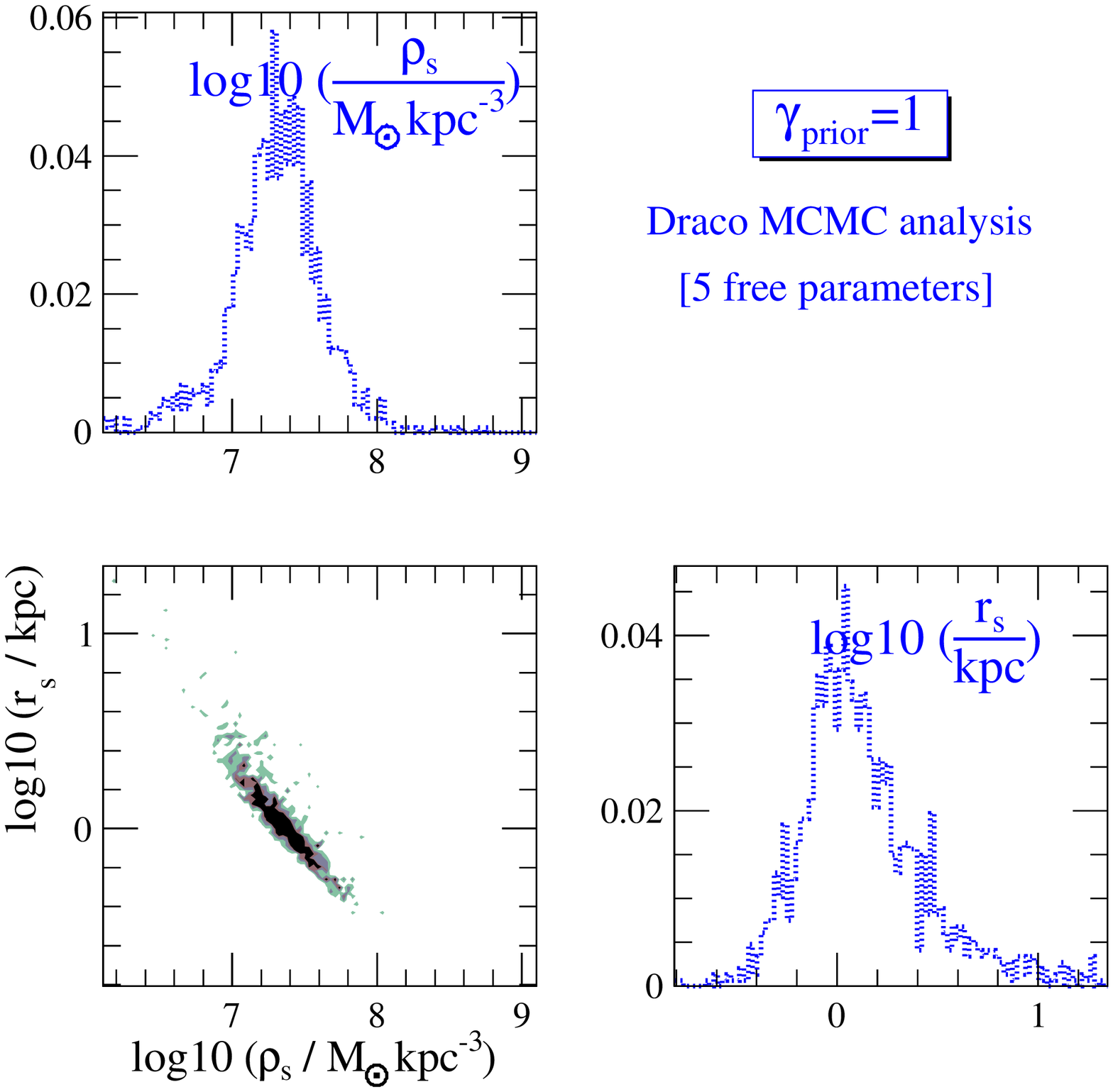}
\includegraphics[width=0.81\linewidth]{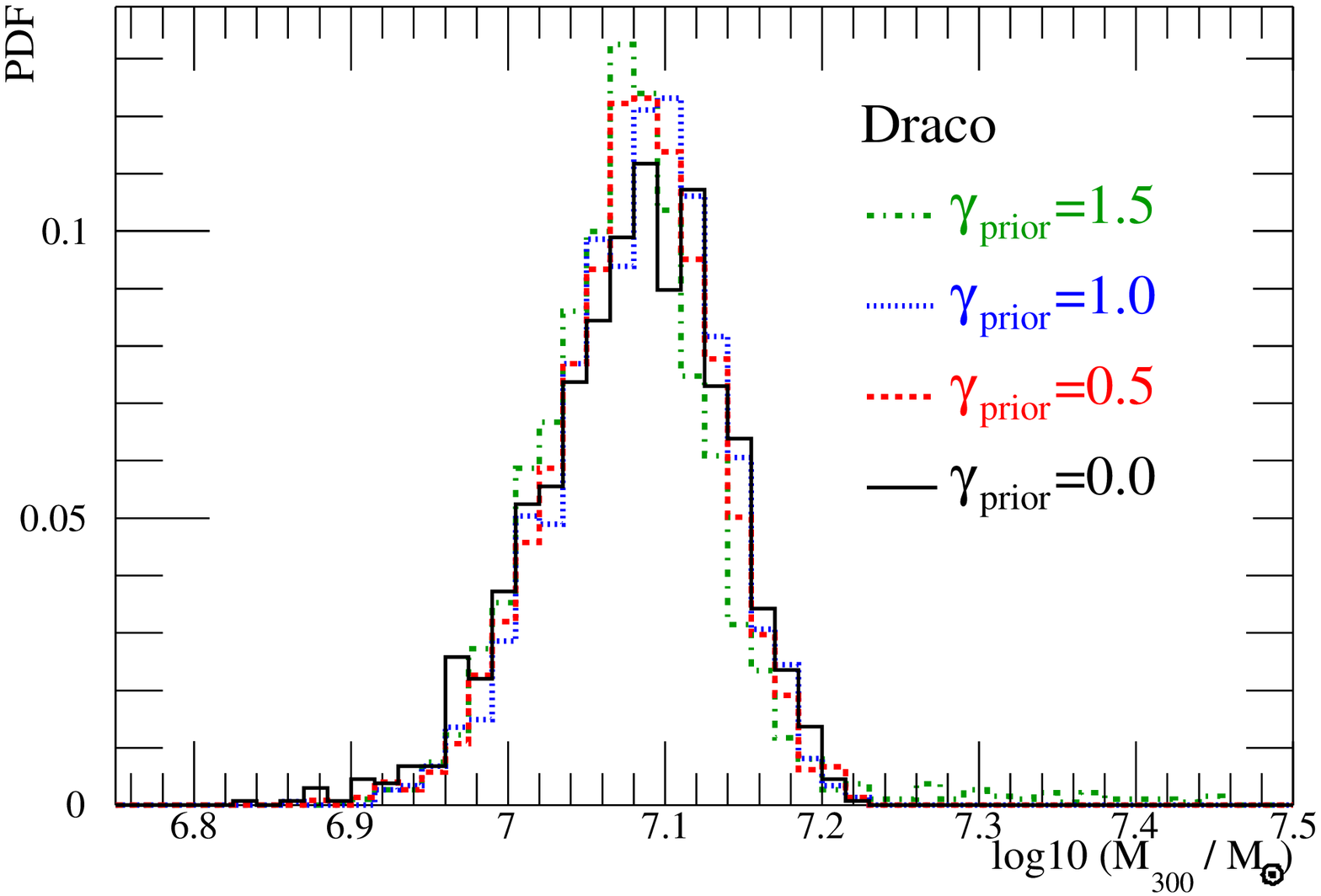}
\caption{{\bf Top:} correlation and PDF of the profile parameters
  $\rho_\mathrm{s}$ and $r_\mathrm{s}$ from the 5-parameter MCMC
  analysis $\gamma_{\rm prior}=0.0$. {\bf Middle:} same, but for
  $\gamma_{\rm prior}=1.0$.  {\bf Bottom:} PDF of $M_{300}$, the mass
  at 300 pc.}
\label{fig:cor_par5}
\end{center}
\end{figure}

Using Eq.~(\ref{eq:M300}) for the mass having a robust estimate of
$M(r_\mathrm{half})$
\citep{2009ApJ...704.1274W,2010MNRAS.406.1220W,amorisco10} gives
$\log(\rho_\mathrm{s}) + \gamma \log(r_\mathrm{s}) \approx \mathrm{constant}$ which
reduces to $\log(\rho_\mathrm{s})\approx \mathrm{constant}$ for $\gamma=0$. As a
result, we expect a strong correlation between $\rho_\mathrm{s}$ and
$r_\mathrm{s}$ when $\gamma_{\rm prior}=1$ and none when $\gamma_{\rm
  prior}=0$. This is confirmed by the result of our MCMC analysis
shown in Fig.~\ref{fig:cor_par5} (here, for the Draco case).  The half-light
radius $r_\mathrm{half}$ for Draco is $\sim 200$~pc, but we choose to
show the PDF for $M_{300}$ in the bottom panel of
Fig.~\ref{fig:cor_par5} as we wish to compare the mass of the dSphs
among themselves (see Table \ref{tab:res_par6}). It confirms that the
mass within an appropriate radius can be reliably constrained by the
data regardless of the value of $\gamma$.
%

\subsubsection{Uncertainties on the profile and on $J$}

\begin{figure}
\includegraphics[width=\linewidth]{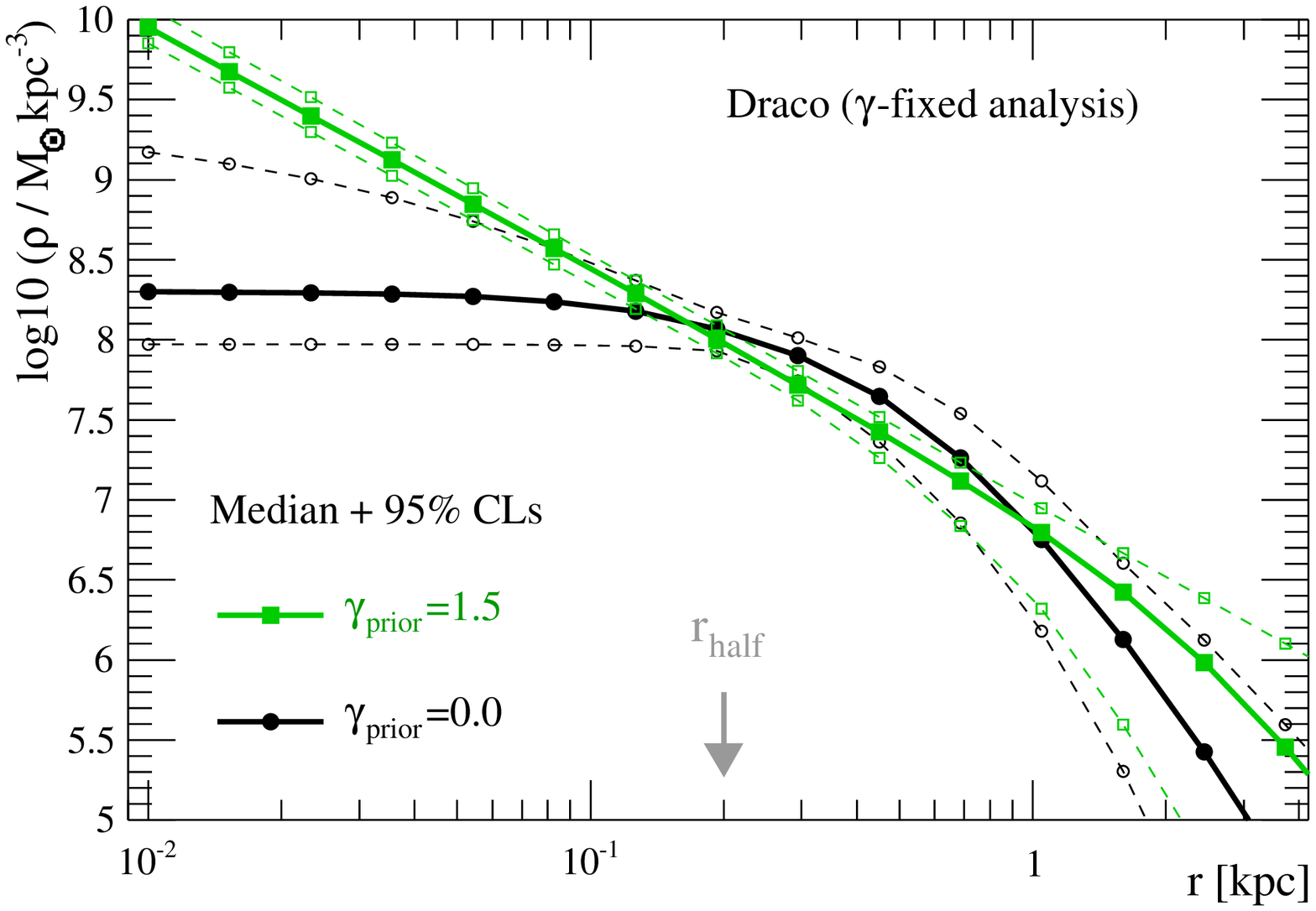}
\includegraphics[width=\linewidth]{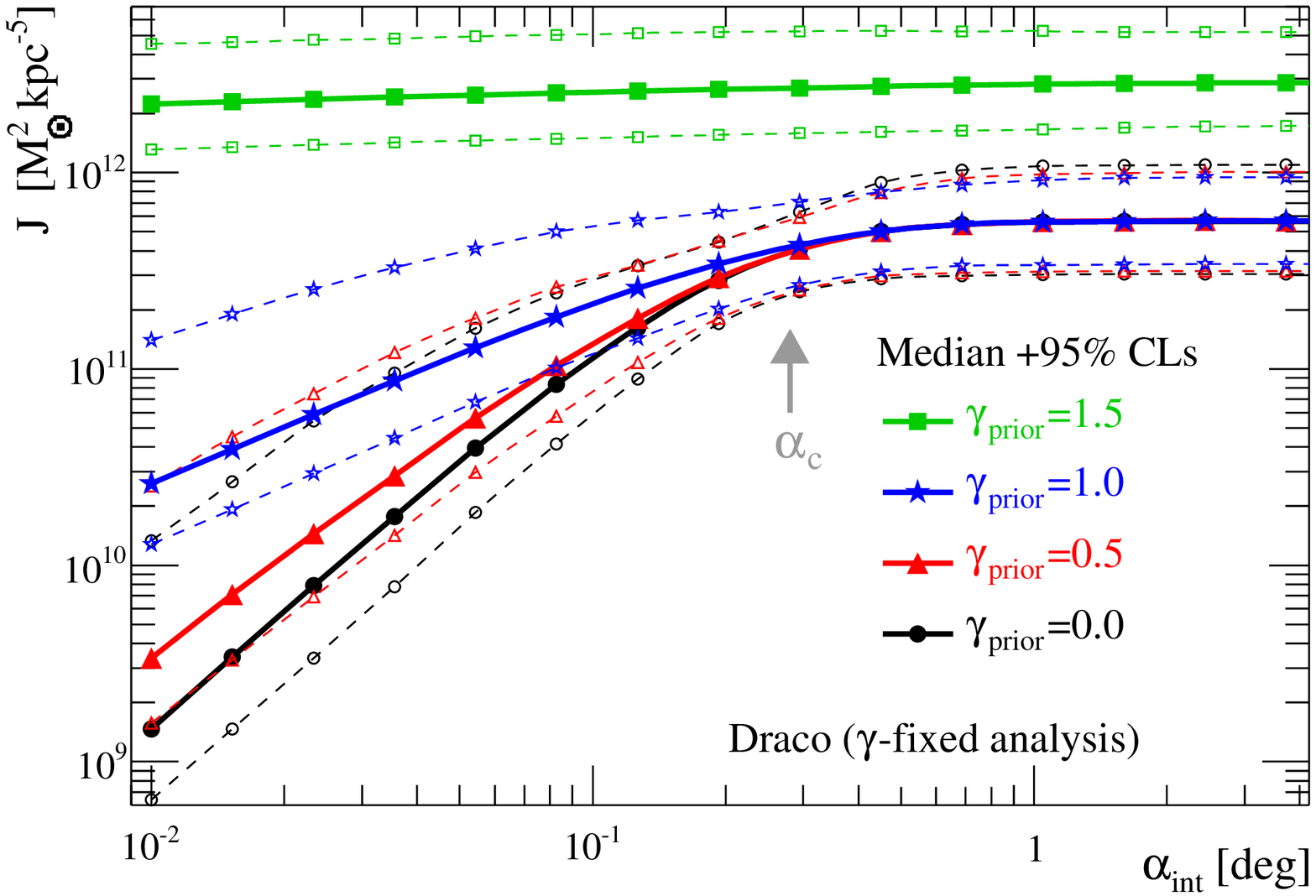}
\caption{Median values (solid lines, filled symbols) and 95\% CLs
  (dashed lines, empty symbols) from the fixed $\gamma_{\rm prior}$
  MCMC analysis on Draco. {\bf Top:} density profiles (the gray arrow
  indicates the value of $r_\mathrm{half}$). {\bf Bottom:} $J$-factor (the gray
  arrow indicates $\alpha_c\approx 2 r_\mathrm{half}/d$).}
\label{fig:rho_gamma_fixed}
\end{figure}

For any given $\gamma$, the uncertainty on $\rho(r)$ at small radii is
related to the range of $r_\mathrm{s}$ values at which the asymptotic
slope is reached (for each profile accepted by the MCMC analysis).
For $\gamma_{\rm prior}=0$, the maximum uncertainty on $\rho(r)$ is directly
related to the maximum uncertainty on $\rho_\mathrm{s}$ (since for $r\ll r_\mathrm{s}$,
$\rho(r)$ is constant) which can be read off the PDF (top-left panel
of Fig.~\ref{fig:cor_par5}). This leads to an order of magnitude uncertainty
on $\rho(r)$ for small $r$, which is consistent with the 95\%
CL shown in top panel of Fig.~\ref{fig:rho_gamma_fixed}.
For $\gamma>0$, the uncertainty has to be read from the dispersion in
the values of $\rho_\mathrm{s} r_\mathrm{s}^{\gamma}$, or
equivalently, the mass $M_{300}$. The bottom panel of
Fig.~\ref{fig:cor_par5} shows that this mass is well-constrained,
independently of $\gamma$ for the case of Draco (see however
in Table~\ref{tab:res_par6} for a larger spread for some dSphs),
resulting in a smaller uncertainty  for $\gamma_{\rm prior}=1.5$ 
than for $\gamma_{\rm prior}=0$
(top panel of Fig.~\ref{fig:cor_par5}).
We checked that the CLs obtained in Fig.~\ref{fig:rho_gamma_fixed}
(in Appendix~\ref{app:biases2}) for the artificial data
enclose correctly the range of reconstructed values: they
are consistent with a larger reconstruction bias
for $\gamma_{\rm prior}=0$ than for $\gamma_{\rm prior}=1.5$ at small radii.

For the uncertainty on $J$, we can obtain a crude estimate by relying
on the approximate formulae given in Appendix~\ref{app:toyJ}.
For $\gamma>0$, $J\propto \rho_\mathrm{s}^2 r_\mathrm{s}^3$,
and substituting the constant $M_{300}$ relationship leads to $J\propto
r_\mathrm{s}^{3-2\gamma}$. The value of
$r_\mathrm{s}$, as seen in its PDF in the top and middle panels of
Fig.~\ref{fig:cor_par5}, varies by roughly a factor of 10. Because of
the weighting power $3-2\gamma$, the uncertainty on $J$ is expected to
be the smallest for $\gamma=1.5$, which is in agreement with the
curves in Fig.~\ref{fig:rho_gamma_fixed} (bottom panel).
However, the analysis of the artificial
data in Appendix~\ref{app:biases2} shows that the typical CL on $J$
obtained in the bottom panel of Fig.~\ref{fig:rho_gamma_fixed} is
likely to be underestimated for $\gamma_{\rm prior}=1.5$ 
(up to factor ${\cal O}(2)$, see Fig.~\ref{fig:fake_J_par5}).
\footnote{This is understood as for
the latter, the inner region ($r\ll r_s$) contribute the most
to $J$, and even small differences for $\rho(r\sim r_s)$ are
bound to translate in sizeable differences for $\rho(r\rightarrow 0)$.
Conversely, similar differences on $\rho$ for shallower profiles
is not an issue as their inner parts do not contribute to $J$.}
This happens for any integration angle. For this reason, we
cannot rely of the $J$ value for $\gamma_{\rm prior}=1.5$
and  focus only on the three cases $\gamma_{\rm prior}=0$,
$\gamma_{\rm prior}=0.5$, and $\gamma_{\rm prior}=1.0$ below.
 
\subsubsection{$J(d)$ and departure from the $1/d^2$ scaling}

\begin{figure*}
\includegraphics[width=\linewidth]{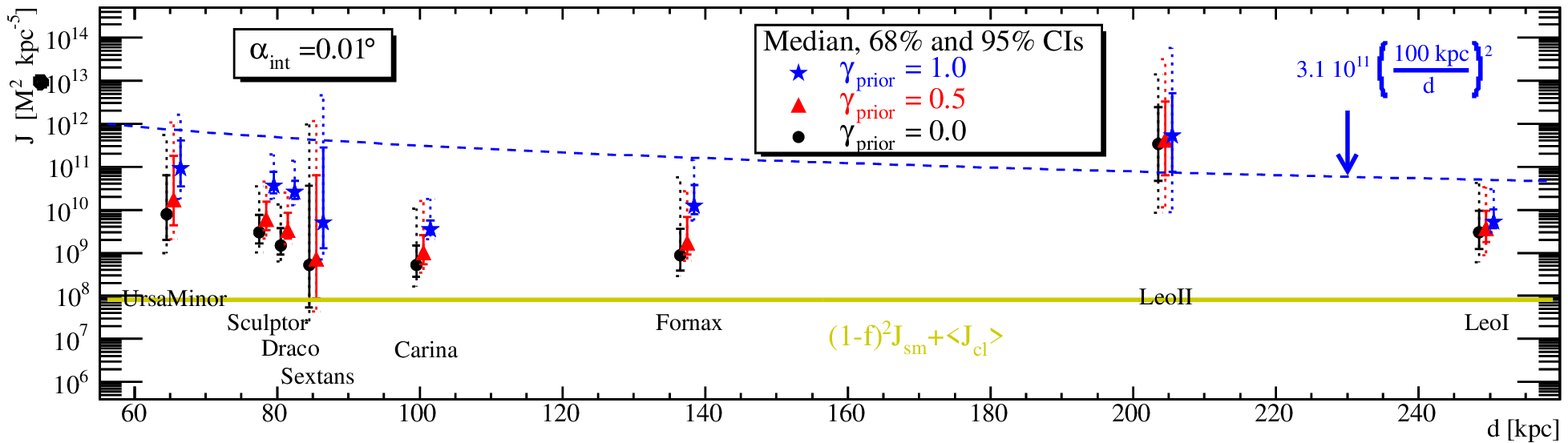}
\includegraphics[width=\linewidth]{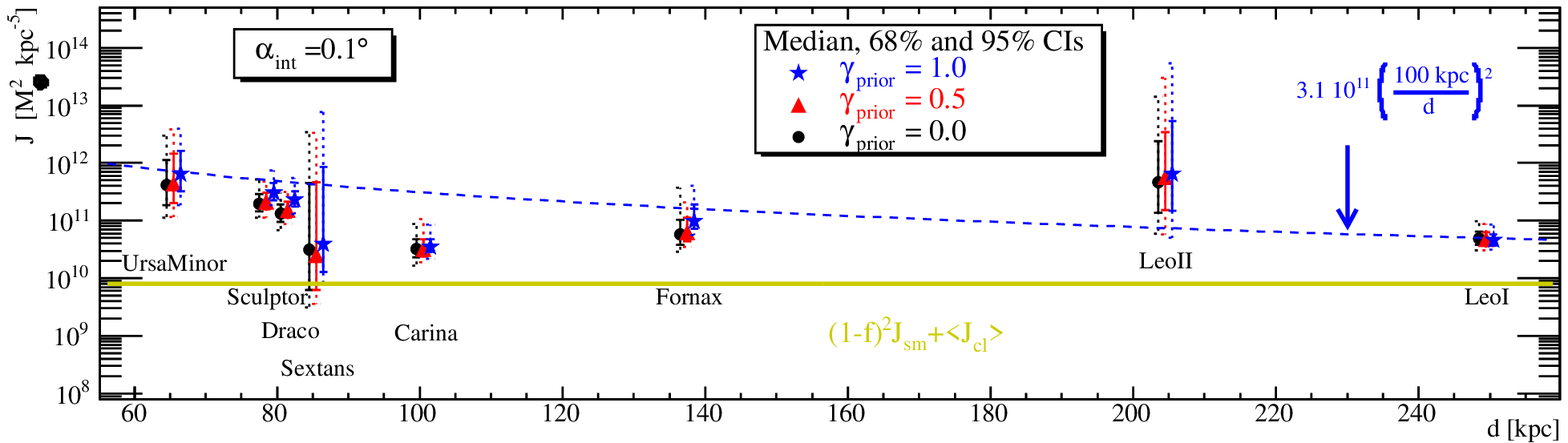}
\includegraphics[width=\linewidth]{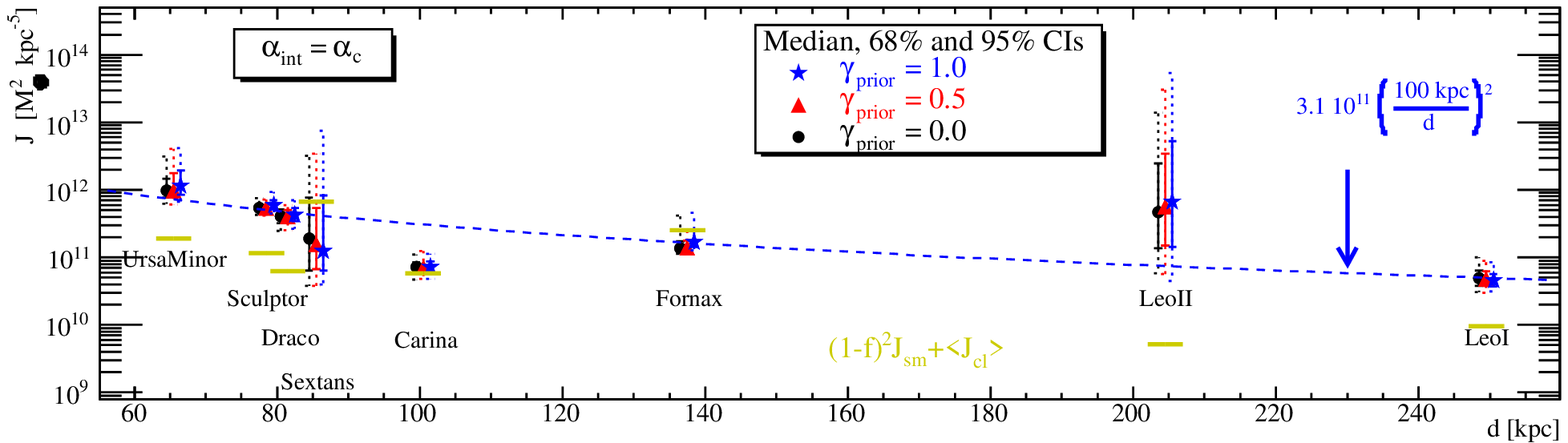}
\caption{Median $J$-factor values (symbols) and 68\%/95\% CLs (solid
  bars; dashed bars) for the fixed $\gamma_{\rm prior}$ analysis (the result for
  $\gamma_{\rm prior}=1.5$ is not shown because it is not reliable, see
  Sect.~\ref{app:biases2}). The blue dashed line shows the expected
  scaling with distance for point sources: $3.1\cdot
  10^{15}d^{-2}~[{\rm M}_\odot^2~{\rm kpc}^{-5}]$. The panels show,
  from top to bottom, three integration angles $\alpha_\mathrm{int}=
  0.01^\circ$, $0.1^\circ$, and $\alpha_c\approx 2r_h/d$ (an angle very similar to the
  angle enclosing 80\% of the flux, see
  Fig.~\ref{fig:min_detect_sigmav})
  that optimises the determination of
  the $J$-factor for a given dSph (hence the error bars are smaller
  in this plot than in the other two). The yellow solid lines (and
  broken lines in the bottom panel) correspond to the Galactic DM
  background including both the smooth and clumpy distributions. For
  the bottom panel, this is not a smooth curve since it depends on the
  integration angle $\alpha_\mathrm{int}$ that varies from dSph to
  dSph in this figure. Note that the choice of using the critical
  angle $\alpha_\mathrm{int} = \alpha_c$  is optimal in the sense that it
  gives the most constrained value for $J$. But where the Galactic
  background annihilation signal approaches that of the dSphs (see
  for example, Sextans and Fornax), the motivation for staring at the
  dSphs rather than simply looking at the Galactic halo is gone.}
\label{fig:Jd_gamma_fixed} 
\end{figure*}

Fig.~\ref{fig:Jd_gamma_fixed} shows the $J$ median values, 65\% and
95\% CIs as symbols, dashed and solid error bars respectively,
for an integration angle of $0.01^\circ$ (top), $0.1^\circ$
(middle), and $\alpha_\mathrm{c} \approx 2 r_\mathrm{half}/d$
\citep{letter_dsph} The $x$-axis is the distance to the dSph (in kpc).
For point-like sources, the $J$-factor of a single dSph scales as
$1/d^2$, as illustrated by the blue-dashed line.  Departure from this
scaling is interpreted as a combination of a mass effect and/or a
profile effect. For instance, Sextans and Carina are  dSphs with
 smaller $M_{300}$ with respect to the other ones (see
Tab.~\ref{tab:res_par6}); consequently they are located below the
dashed blue line in the top panel of
Fig.~\ref{fig:Jd_gamma_fixed}. The exception is Leo II, which has a
`small' mass but is nevertheless above the dashed line.  Although this analysis
cannot constrain $\gamma$, we are tempted to interpret this
oddity  in terms of a `cuspier' profile (w.r.t. those for other
dSphs), which would be consistent with the fact that its $J$ remains
similar in moving from $\alpha_{\rm int}=0.1^\circ$ (middle panel) to
$0.01^\circ$ (top panel).  However, an alternative explanation
(which would be more consistent with the results obtained in this
paper) could be the fact that Leo II has the smallest amount of
kinematic data at present, and that its $J$ is overestimated (see
Appendix~\ref{app:binning} to support this line of argument).
We repeat that the relative brightness of the dSphs is further affected for
background-dominated instruments (as described in Sec.~\ref{sec:detectability}),
so that the ranking has to be based on Fig~\ref{fig:min_detect_sigmav}
discussed in the next section.

The bottom panel of Fig.~\ref{fig:Jd_gamma_fixed} shows the $J$ value
for an `optimal' integration angle $\alpha_\mathrm{c}$ that is twice the
half-light radius divided by the dSph distance\footnote{CLs for
  $J(\alpha_{\rm int})$ are provided along with the paper for readers
  interested in applying our analysis to existing and future
  observatories.} (this corresponds to the integration angle that
minimises the CLs on $J$; see \citealt{letter_dsph}). The
yellow broken solid lines show the expected signal from the diffuse
Galactic DM annihilation background, including a contribution from clumpy
sub-structures (the extragalactic background, which also scales as
$\alpha_{\rm int}^2$, has not been included). The total background may
be uncertain by a factor of a few (depending on the exact Galactic
(smooth) profile and local DM density). Its exact level|which depends on the chosen
integration angle|determines the condition for the loss of contrast of
the dSph signal, i.e. the condition for which looking at the DM halo
(rather than at dSphs) becomes a better strategy.

\subsubsection{Conclusion for the fixed $\gamma_{\rm prior}$ analysis}

The analysis of simulated data shows that the analysis for
$\gamma_{\rm prior}=1.5$ is biased by a factor of ${\cal O}(10)$
and that the CLs obtained on the real data are likely to be
severely under-estimated in that case. But such steeply cusped
profiles are neither supported by observations nor motivated
by current cosmological simulations. For values of
$\gamma_{\rm prior} \leq 1$, this bias is a factor of a
few only, so that it shows that the results from a
fixed $\gamma_{\rm prior}$ analysis of the 8 classical dSphs
are robusts. However, this analysis shows that unless very
small integration angles $\alpha_{\rm int}\lesssim 0.01^\circ$
are chosen (or if $\gamma_{\rm true}\gtrsim 1$), knowing
the exact value of $\gamma$ does not help in improving the
determination of $J$. Indeed,
even using Draco, the stellar population of which is one of the most
studied, the CLs of the three reconstructed fluxes ($\gamma_{\rm prior}=0$ in black full circles,
$\gamma_{\rm prior}=0.5$ in red triangles, and $\gamma_{\rm
  prior}=1.0$ in blue stars) in Fig.~\ref{fig:rho_gamma_fixed}
(bottom), overlap.  Reversing the argument, if we do not know the
inner slope, and if a $\gamma$-ray signal is detected from just one dSph in
future, there will be little hope of recovering the slope of the
DM halo from that measurement only.

 This means that the
best way to improve the prediction of the $J$-factor in the future
relies on obtaining more {\it data} and a more refined MCMC analysis; an
improved prior on the DM distribution makes 
little difference.

\subsection{Sensitivity of $\gamma$-ray observatories to DM annihilation in the dSphs}
\label{sec:5.3}

\begin{figure*}
\includegraphics[width=\linewidth]{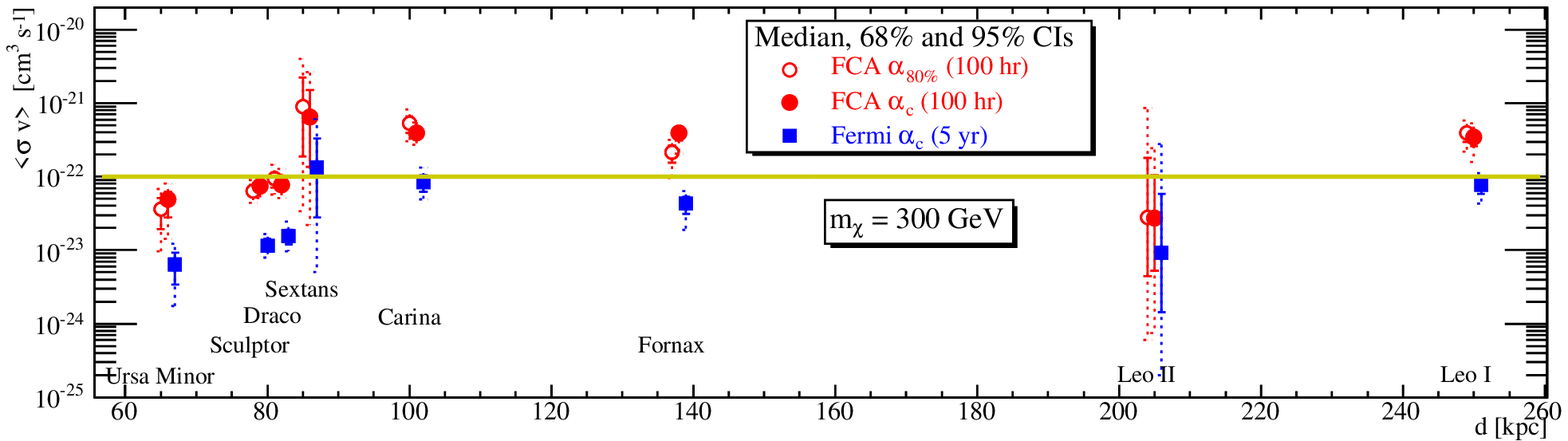}
\includegraphics[width=\linewidth]{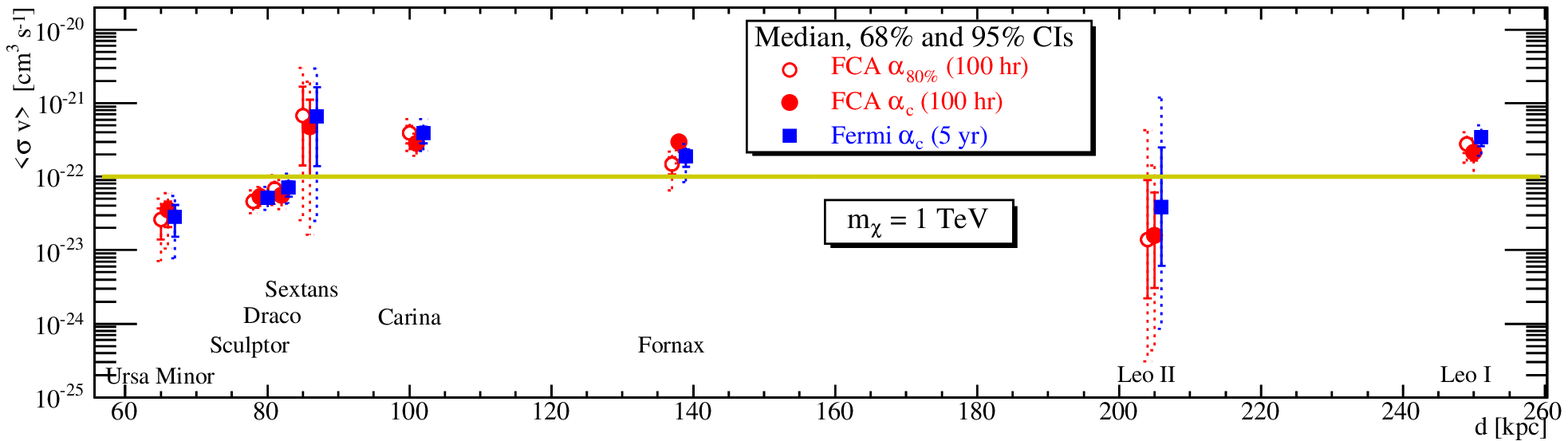}
\includegraphics[width=\linewidth]{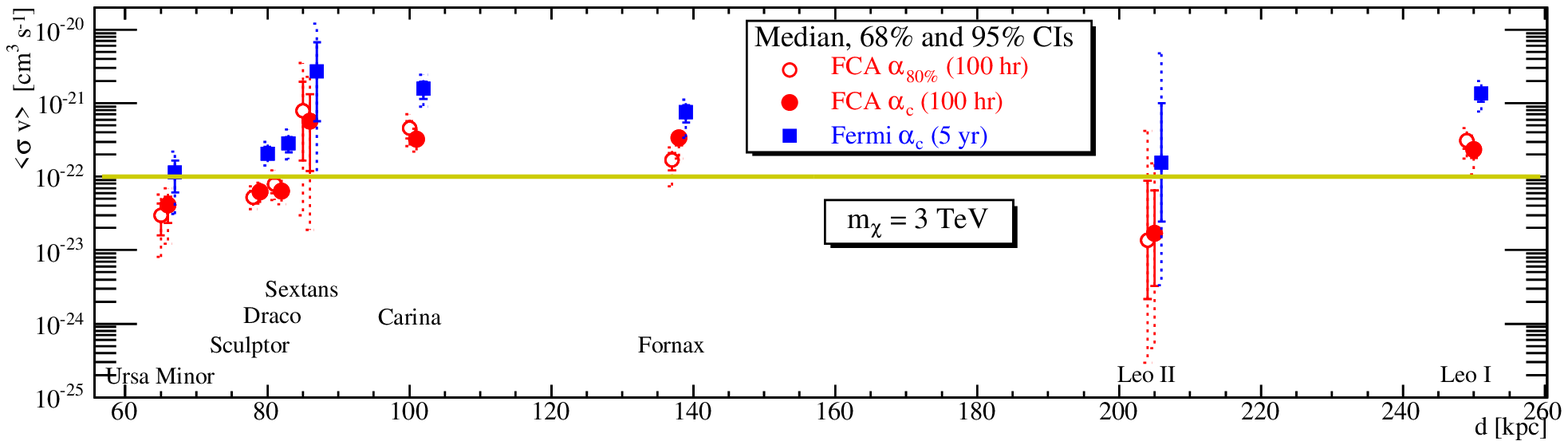}
\caption{Minimum detectable $<\sigma v >$ for known dSphs shown as a
  function of their distance, for different assumed DM masses
  (separate panels). 100 hour observations with an FCA (red circles)
  are compared to 5 years of Fermi observations (blue squares).  Error
  bars indicate 65\% (solid lines) and 95\% (dotted lines) confidence
  limits. The integration angle is adapted to the $\alpha_\mathrm{c}$ of each
  dSph and the energy-dependent PSF of the two instruments. The
  strategy of using $\alpha_{80}$, rather than $\alpha_\mathrm{c}$, is
  indicated with hollow symbols for the FCA case. The line for
  $<\sigma v> = 10^{-22}$~cm$^3$~s$^{-1}$ is drawn for comparison
  purpose between the panels.}
\label{fig:min_detect_sigmav}
\end{figure*}

The potential for using the classical dSph to place constraints on the
DM annihilation cross-section, given the
uncertainties in the astrophysical $J$-factor, can be seen in
Fig.~\ref{fig:min_detect_sigmav}. Previous analyses have adopted
the solid angle for calculation of the $J$-factor to be the angular
resolution of the telescope for a point-like source, typically
assuming a NFW-like profile  \citep{2010ApJ...720.1174A,2010ApJ...712..147A,2011APh....34..608H}.
By contrast our sensitivity plots take into account finite size effects: i)
the $J$ values are based on the MCMC analysis with the prior
$0\leq\gamma_{\rm prior}\leq1$, where the corresponding $J$ are shown in
Fig.~\ref{fig:JdSph_Jdm_bkgd}; ii) the energy dependent angular
resolution has also been taken into account assuming a standard $\gamma$-ray
annihilation spectrum (see Section~\ref{sec:dNdE_xsec}). Moreover for
Fermi-LAT the background level assumed has been increased (resulting
in a 25\% worsening of the sensitivity above 100~MeV) to reflect the
average situation in the directions of the classical dSph (the
variation between the individual dSph is only 7\% rms). A likelihood
based analysis is used for both FCA and Fermi and a nominal
observation zenith angle of 20$^{\circ}$ assumed for the
FCA\footnote{The energy threshold for a ground based instrument is
  dependent on the zenith angle of observation. This means that the
  actual energy threshold for a given object will depend on the
  object's declination and the latitude of the, yet to be determined
  FCA site.}  (see Section~\ref{sec:RelativePerformance}).

The  panels from top to bottom correspond to  increasing
DM (neutralino) masses. At low values, Fermi has a better sensitivity than
FCA; at a mass of about 1 TeV the two are comparable, and
for higher masses the FCA becomes the more sensitive instrument due to
the vastly greater effective area at the photon energies at which the
annihilation spectrum is expected to peak. Note that the precise
value of $\langle\sigma v\rangle$ where
the relative sensitivities of the two instruments cross 
depends on the form of the DM annihilation spectrum. Since
 we are examining the uncertainties in the astrophysical
$J$-factor to the detectability of dSphs, we have used a conservative
spectrum averaged over a number of possible annihilation channels (see
Fig.~\ref{fig:phi-susy}) which results in the  majority
of produced $\gamma$-ray photons having energies $\simeq$10\% of the DM
particle mass. If we were to move from a relatively soft spectrum,
such as $\mathrm{b\overline{b}}$ to a harder one, such as
$\tau^{+}\tau^{-}$, this would benefit both instruments in different
ways. For Fermi-LAT a harder spectrum makes the signal easier to
distinguish above the diffuse $\gamma$-ray background; indeed the
\cite{2010ApJ...712..147A} found that the detectable flux limit from a
potential source could vary by a factor of 2--20 (with lower particle
masses benefiting the most) between these different annihilation spectra.
For the FCA, which has a very large effective area to photons
$\geq100$\,GeV, the
benefits of having more high energy photons is very apparent when it
comes to flux sensitivity. For both observatories, an increased number of
high energy photons needs to be balanced with the correspondingly
better angular resolution, particularly if (e.g. for Fermi-LAT) a point-like
source becomes spatially resolved.

Our analysis places Ursa Minor as the best candidate for the northern
sky (marginally better than Draco, which has long been a
favourite target of northern hemisphere observatories) and Sculptor
for the southern sky, when it comes to a favourable median and
low uncertainty in the $J$-factor. It should be noted, however, that although the closest objects seem to be favoured, Leo II 
has the potential to yield a stronger signal, however more kinematic data
are needed in order to  constrain better its J-factor. In addition, it should be noted that the uneven sensitivity of the Fermi-LAT across the sky, caused in particular by the proximity of bright sources\footnote{In particular there is a bright GeV emitter 1FGL J0058.4-3235 only $\sim 1.1^{\circ}$ away from Sculptor which significantly worsens the upper limit on that object as discussed by \citet{2010ApJ...712..147A}.} as well as the galactic diffuse background can change what is considered the favorite candidate.

We emphasise that in our analysis the inner slope $\gamma$ has not been
constrained, but that a better independent determination of $\gamma$
in  future will not help providing a better determination of $J$
(see Fig.~\ref{fig:Jd_gamma_fixed}); this is discussed further in
the Appendices. Carina, Fornax and Leo I are the targets least
favoured. When compared to existing limits from Fermi-LAT
\citep{2010ApJ...712..147A} or the current generation of ACTs
\citep{2010ApJ...720.1174A,2011APh....34..608H} it can be seen that
our limits are not dissimilar from those that have already been
published. For Fermi this is not surprising, since the source is
unresolved and any difference should relate only to the assumed increase in
exposure from 1 to 5 years, resulting in a factor of a few at
best. The similarity in sensitivity between current and future ACTs
is perhaps more surprising, but this as stated earlier relates to the
na\"ive assumptions made on the form  for the $J$-factor and the 
solid angle integrated over; in order to reach the currently claimed
limits requires a deep exposure with an instrument as sensitive as
CTA.
\begin{figure*}
\includegraphics[width=0.7\linewidth]{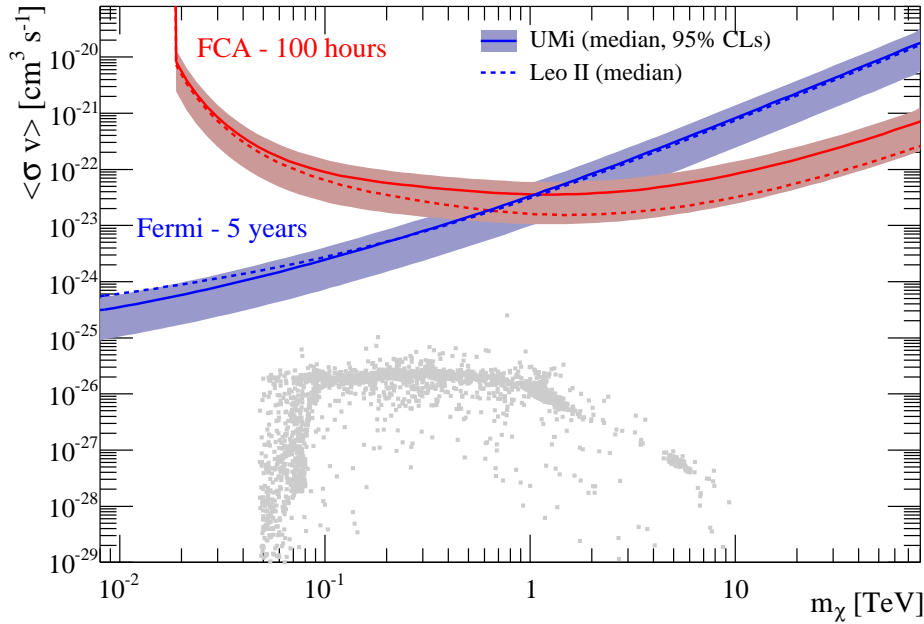}
\caption{Sensitivity reach in the $m_{\chi}-\langle \sigma v\rangle$
  plan for FCA (100 hr) and Fermi (5 yr), for our best candidates UMi
  (median value and 95\% CIs) and Leo II (median only).  Black
  asterisks represent points from MSSM models that fall within 3
  standard deviations of the relic density measured in the 3 year WMAP
  data set (taken from \citealt{2010ApJ...720.1174A}).}
\label{fig:mssm}
\end{figure*} 

One last thing to note is that a common way to synthesise a deeper
exposure is to stack observations of different sources together to
provide an effective long exposure of a generic source. For a common
universal halo profile this may be fine, however any analysis will have to
take into account the different integration angles for each individual
source correctly. If all dSphs do not share a common halo
profile and hence have different $\gamma$ values, we have to rely on
the varying-$\gamma$ analysis presented in the previous section and
the relative ranking of potential targets would then be different.

\section{Discussion and conclusions}
\label{sec:conclusions}

 We have revisited the expected DM annihilation signal from dSph
galaxies for current (Fermi-LAT) and future (e.g. CTA) $\gamma$-ray observatories. The main innovative features of our analysis are that: (i) We have
considered the effect of the {\it angular size} of the dSphs for the
first time. This is important since, while nearby dSphs have higher
$\gamma$ ray flux, their larger angular extent can make them sub-prime
targets if the sensitivity is limited by cosmic ray and $\gamma$-ray
backgrounds. (ii) We determined the astrophysical $J$-factor for the classical dSphs
directly from photometric and kinematic data. We assumed very little
about their underlying DM distribution, modelling the dSph DM
profile as a smooth split-power law, both with and without DM
sub-clumps. (iii) We used a MCMC technique
to marginalise over unknown parameters and determine the sensitivity
of our derived $J$-factors to both model and measurement
uncertainties. (iv) We used simulated DM profiles to demonstrate that
our $J$-factor determinations recover the correct solution within our
quoted uncertainties.
  
Our key findings are as follows: 

\begin{enumerate}

\item Sub-clumps in the dSphs do {\em not} usefully boost the
  signal. For all configurations where the sub-clump distribution
  follows the underlying smooth DM halo, the boost factor is
  at most $\sim\!2-3$. Moreover, to obtain even this mild boost, one
  has to integrate the signal over the whole angular extent of the
  dSph. This is unlikely to be an effective strategy as the diffuse
  Galactic DM signal will dominate for integration angles $\alpha_{\rm
    int}\gtrsim 1^\circ$.

\item Point-like emission from a dSph is a very poor approximation for
  high angular resolution instruments, such as the next-generation CTA.
  For a nearby dSph, using the point-like approximation can lead to an
  order of magnitude overestimate of the detection sensitivity. In 
  the case of a nearby cored profile consisting of very high mass DM 
  particles, a point source approximation can be unsatisfactory even for 
  the modest angular resolution of Fermi-LAT.

\item With the Jeans' analysis, no DM profile can be ruled out by
  current data.  The use of the MCMC technique on artificial data also
  shows that such an analysis is unable to provide reliable values for
  $J$ if the profiles are cuspy ($\gamma=1.5$).  However, using a
  prior on the inner DM cusp slope $0\leq\gamma_{\rm prior}\leq1$
  provides $J$-factor estimates accurate to a factor of a few.

\item The best dSph targets are not simply those closest to us, as
  might na\"ively be expected. A good candidate has to combine high
  mass, close proximity, small angular size ($\lesssim 1^\circ$; i.e.
  not too close); and a well-constrained DM profile. With
  these criteria in mind, we find three categories: well-constrained
  and promising (Ursa Minor, Sculptor and Draco), well-constrained but
  less promising (Carina, Fornax and Leo I), and poorly constrained
  (Sextans and Leo II). Leo II may yet prove to be a viable target
  as it has a larger median $J$-factor than UMi, however more data are
  required to confirm its status.

\item A search based on a known DM candidate (from, e.g., forthcoming discoveries at the LHC)
  will do much to optimise the search strategy and, ultimately,
  the detection sensitivity for all $\gamma$-ray observatories. This is
  because the shape of the annihilation spectrum is a strong driver of
  the photon energy range that can provide the best information on the
  candidate DM particle mass. Fermi-LAT has great potential to
  probe down to the expected annihilation
  cross-section for particles of mass $\ll 700$ GeV, whereas a ground
  based instrument is more suited for probing particle masses above a few
  hundred GeV with a sufficiently deep exposure. However, even for
  5~yr of observation with Fermi-LAT or 100 hrs with FCA, the sensitivity reach (Fig.~\ref{fig:mssm}) remains anywhere
  between 4 to 10 orders of magnitude above the expected
   annihilation cross-section for a cosmological relic (depending on the mass of
  the DM particle candidate). Improving these limits will
  require a harder annihilation spectrum than the conservative average
  we have adopted in this study, or a significant boost (e.g. from the
  Sommerfeld enhancement) to the $\gamma$-ray production.

\end{enumerate}

Finally, the ultra-faint dSphs have received a lot of interest in the
community lately, as they could be the most-DM dominated
systems in the Galaxy. We emphasise that the MCMC analysis we have
performed for the classical dSphs cannot be applied `as is' for these
objects. First, the sample of stars observed is smaller. Second, the
velocity dispersion is smaller and suffers from larger uncertainties
than those for the classical dSphs. The robustness and systematic
biases of the MCMC analysis will be discussed elsewhere (Walker et
al., 2011, in preparation). Results concerning $J$ for the
ultra-faint dSphs will be presented in a companion paper.

\section*{Acknowledgments}

We thank the anonymous referee for their careful reading of the
manuscript and useful comments.
We thank Walter Dehnen for providing his code for use in generating
artificial dSph data sets.
MGW is supported by NASA through Hubble Fellowship grant HST-HF-51283,
awarded by the Space Telescope Science Institute, which is operated by
the Association of Universities for Research in Astronomy, Inc., for
NASA, under contract NAS 5-26555.
CC acknowledges support from an STFC rolling grant at the University
of Leicester.
JAH acknowledges the support of an STFC Advanced Fellowship.
MIW acknowledges the Royal Society for support through a University
Research Fellowship.
JIR acknowledges support from SNF grant PP00P2\_128540/1.
SS acknowledges support by the EU Research \& Training Network
`Unification in the LHC era' (PITN-GA-2009-237920).
Part of this work used the ALICE High Performance Computing Facility at the
University of Leicester.
Some resources on ALICE form part of the DiRAC Facility jointly funded
by STFC and the Large Facilities Capital Fund of BIS.

\appendix

\section{Definitions, notation, conversion factors}
\label{app:defs}
Studies of DM annihilations in the context of dSphs
involves both particle physics and astrophysics. The obvious
difference of scales between the two fields and habits among the two
communities have given rise to a plethora of notations and unit
choices throughout the literature. In this Appendix, we provide some
explanatory elements and conversion factors to ease comparison between
the different works published on the subject.

As mentioned in \S\ref{sec:method}, we define the
differential $\gamma-$ray flux as integrated over the solid angle
$\Delta\Omega$ as
\begin{equation}
     \frac{\mathrm{d}\Phi_{\gamma}}{\mathrm{d}E_{\gamma}}(E_{\gamma},\Delta\Omega)
        = \Phi^{\rm pp}(E_{\gamma}) \times J(\Delta\Omega)\,,
        \nonumber
 \end{equation}
where
\begin{equation}
     \Phi^{\rm pp}(E_{\gamma})\equiv  \frac{\mathrm{d}\Phi_{\gamma}}{\mathrm{d}E_{\gamma}}
        = \frac{1}{4\pi}\frac{\langle\sigma_{\rm ann}v\rangle}{2m_{\chi}^{2}}
          \cdot \frac{dN_{\gamma}}{dE_{\gamma}}\nonumber\;,
 \end{equation}
and
\begin{equation}
      J(\Delta\Omega) = \int_{\Delta\Omega}\int \rho_{\rm DM}^2 (l,\Omega) \,dld\Omega.
     \nonumber
 \end{equation}
 The solid angle is simply related to the integration angle
 $\alpha_{\rm int}$ by
\[
  \Delta\Omega = 2\pi\cdot(1-\cos(\alpha_{\rm int})) \,.
\]
In our work, the units of these quantities are as follows:
\begin{itemize}
\item $\left[\mathrm{d}\Phi_{\gamma}/\mathrm{d}E_{\gamma}\right]
  = {\rm cm}^{-2}{\rm ~s}^{-1} {\rm ~GeV}^{-1}$;
\item $\left[ \Phi^{\rm pp}(E_{\gamma})\right] = {\rm cm}^{3}{\rm ~s}^{-1} {\rm ~GeV}^{-3} ({\rm ~sr}^{-1})$;
\item $[J] = M_\odot^2{\rm ~kpc}^{-5} ({\rm ~sr})$.
\end{itemize}
First of all, note that the location of the $1/4\pi$ factor appearing
in $\Phi^{\rm pp}$ is arbitrary. We followed
\citet{2009A&A...496..351P} and included it in the particle physics
factor. In other works, it can appear in the astrophysical factor $J$
(e.g., \citealt{2009JCAP...01..016B}). Therefore, to compare the
astrophysical factors between several studies, one must first ensure
to correct the value of $J$ by $4\pi$ if needed.  In the text, we did
not explicitly stated the solid angle dependence in the units of $J$
as it is dimensionless quantity. \footnote{Some authors do however
  explicitly express the solid angle dependence in their units,
  e.g. \citet{2009A&A...496..351P}, who express $J$ ($\Phi_{\rm
    cosmo}$ in their notation) in ${\rm GeV}^2{\rm cm}^{-6} {\rm
    kpc}\;{\rm sr}$.  This is completely equivalent to our
  $M_\odot^2{\rm kpc}^{-5}$ but for the unit numerical conversion
  factor.} The conversion factor (once the $4\pi$ issue is resolved)
from our $J$ units to that traditionally found in the literature are:
\begin{itemize}
\item $1\; M_\odot^2{\rm ~kpc}^{-5}=10^{-15}~M_\odot^2{\rm ~pc}^{-5}$
\item $1\; M_\odot^2{\rm ~kpc}^{-5}=4.45\times10^{6}{\rm ~GeV}^2{\rm ~cm}^{-5}$
\item  $1\; M_\odot^2{\rm ~kpc}^{-5} ~(\rm sr) = 1.44\times10^{-15}{\rm ~GeV}^2{\rm ~cm}^{-6} {\rm ~kpc} \;(\rm sr)$ 
\end{itemize}
Before comparing any number, one must also ensure that the solid angle
$\Delta\Omega$ over which the integration is performed is the same. In
most works, a $\alpha_{\rm int}=0.1^\circ$ angular resolution is
chosen, corresponding to $\Delta\Omega=10^{-5}$~sr. However this is
not always the case, as in the present study where we explore several
angular resolutions. Note that the quantity $\bar J\equiv
J/\Delta\Omega$ (in ${\rm GeV}^2{\rm ~cm}^{-5} {\rm ~sr^{-1}}$ for
example) is also in use and the astrophysical factor is can be found
under this form in some articles (e.g., \citealp{Evans:2003sc}).

\section{Toy model for J (in dSphs)}
\label{app:toyJ}
\begin{figure}
\includegraphics[width=\linewidth]{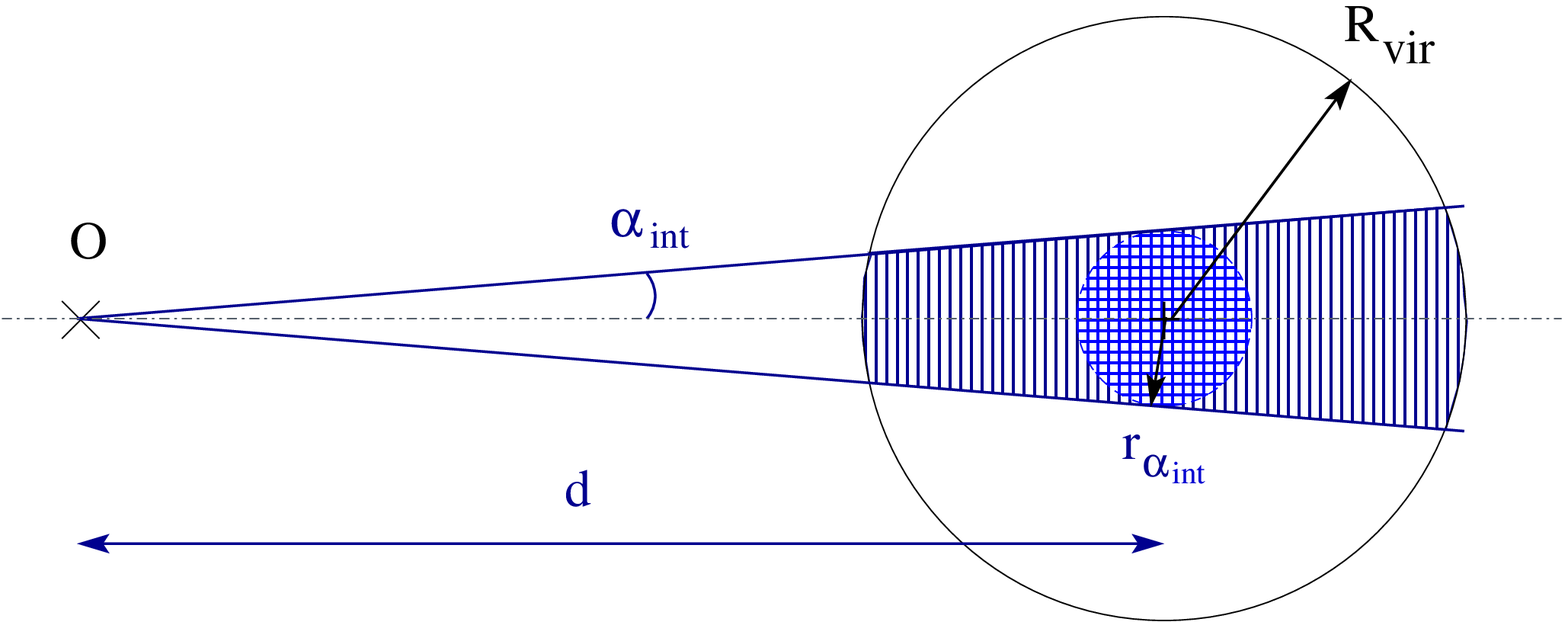}
\caption{Sketch of the integration regions contributing to the J
  factor: shown are the full integration region (vertical hatched) or
  a sub-region (cross-hatched) used for the toy calculations.  The
  letter O shows the observer position, $\alpha_{\rm int}$ is the
  integration angle, $d$ is the distance of the dSph and $R_{\rm vir}$
  its virial radius.}
\label{fig:sketch_dsph}
\end{figure}

The volume of the dSph is not always fully encompassed in the
integration solid angle, as sketched in Fig.~\ref{fig:sketch_dsph}
(vertical hatched region) so that a numerical integration is required
in general.
However, a reasonable approximation for estimating the dependence of
$J$ on the parameters of the problem, i.e. the distance to the dSph
$d$, the integration angle $\alpha_{\rm int}$, and the profile
parameters $\rho_\mathrm{s}$, $r_\mathrm{s}$ and $\gamma$), is to
consider only the volume within the radius
\begin{equation}
 r_{\alpha_{\rm int}} = d\times \sin(\alpha_{\rm int}) \approx d\times \alpha_{\rm int} ,
 \label{eq:Ralpha_alpha}
\end{equation}
where the approximation is valid for typical integration angles
$\alpha_{\rm int} \lesssim 0.1^{\circ}$. This volume corresponds to
the spherical cross-hatched region in Fig.~\ref{fig:sketch_dsph}.

The toy model proposed below to calculate $J$ allows us to cross-check
the results of the numerical integration for both the smooth and
sub-clump contribution. We find that the model is accurate enough up to
a factor of $2$ for $\gamma=0$ and $\gamma>0.5$, so can be used for
gross estimates of any signal from a DM clump.

\subsection{For the smooth distribution}
About 90\% of the clump luminosity is usually contained in a few
$r_\mathrm{s}$, whatever the profile.  The consequences are
twofold. First, as can be read off Table \ref{tab:res_par6},
$r_\mathrm{s}/d\ll 1$, so that the J factor amounts to a point like
contribution
 \begin{equation}
    J_{\rm point-like} = \frac{4\pi}{d^2}\int_{0}^{\min(r_{\alpha_{\rm int}},r_\mathrm{s})} r^2 \rho^2(r) dr.
    \label{eq:Japprox}
\end{equation}
Secondly, it means that Eq.~(\ref{eq:hernquist1}) for the profile can
be simplified into the approximate expression
\begin{equation}
 \rho_{\rm approx}(r) =
   \begin{cases}
      \rho_{\rm sat} &\text{if $r \le r_{\rm sat}$;}\\
      \rho_\mathrm{s} \times \left(\frac{r}{r_\mathrm{s}}\right)^{-\gamma} &\text{if $ r_{\rm sat}<r \le r_\mathrm{s}$;}\\
       0 & \text{otherwise.}
   \end{cases}
   \label{eq:rhoapprox}
\end{equation}
However, for all applications of our toy model, we will keep $\gamma<
3/2$, so that the saturation density above is never reached in the
dSphs considered below.

\paragraph*{Various regimes}
The approximate formulae for $J$ is obtained by combining
Eqs.~(\ref{eq:Japprox}) and (\ref{eq:rhoapprox}):
 \begin{equation}
    J_{\rm approx} =\frac{4\pi}{d^2}\int_{0}^{\min(r_{\alpha_{\rm int}},r_\mathrm{s})} r^2 \rho_{\rm approx}^2(r) dr.
    \label{eq:Japprox_final}
\end{equation}
Using Eq.~(\ref{eq:Ralpha_alpha}), this leads to
\begin{equation}
 J_{\rm approx} = \frac{4\pi}{d^2} \cdot \frac{\rho_\mathrm{s}^2 \,r_\mathrm{s}^{2\gamma}}{3-2\gamma}\cdot
   [\min(r_{\alpha_{\rm int}},r_\mathrm{s})]^{3-2\gamma}\,.
   \label{eq:Japprox2}
\end{equation}
This formula gives satisfactory results for cuspy profiles (see
below), but has to be modified in the following cases:
 \begin{itemize}
 \item If $r_{\alpha_{\rm int}}\gtrsim r_\mathrm{s}$, the integration
   region encompasses $r_\mathrm{s}$. The $(1,3,\gamma)$ profiles
   decrease faster than $r^{-\gamma}$ for $r\sim r_\mathrm{s}$ hence
   integrating the toy model up to $r_\mathrm{s}$ is bound to
   overshoot the true result.  We thus stop the integration at the
   radius $r_x$ such that $\rho_{\rm true}(r_x) = \rho_{\rm
     approx}(r_x)/x$, i.e.
     \[
     r_x = r_\mathrm{s} \cdot [x^{1/(3-\gamma)}-1] \,.
     \]
     Taking $x=2$ gives a satisfactory fit to the full numerical
     calculation (see below).
   \item If $r_{\alpha_{\rm int}}\gtrsim r_\mathrm{s}$ and $\gamma=0$,
     the integration can be performed analytically up to $R_{\rm vir}$
     and is used instead.
   \item If $r_{\alpha_{\rm int}}\lesssim r_\mathrm{s}$ and
     $\gamma=0$, the profile is constant, and integrating on the
     cross-hatched region (instead of the vertical hatched one, see
     Fig.~\ref{fig:sketch_dsph}) undershoots the true result. A better
     approximation is to integrate on a conic section. For the same
     reason as given for the first item, we replace $r_\mathrm{s}$ by
     $r_x$ (with $x=2$) in the calculation of the cone volume.
 \end{itemize}

\paragraph*{Resulting formula}
To summarise, the final toy-model formula proposed for the smooth
contribution of the dSph is:
\begin{equation}
 J_{\rm toy} = \frac{4\pi\rho_\mathrm{s}^2 }{d^2} \times
   \begin{cases}
      \displaystyle r_\mathrm{s}^{2\gamma}\cdot \frac{\min(r_x,r_{\alpha_{\rm int}})^{3-2\gamma}}{3-2\gamma}
             &\text{if $\gamma>0$;}\\
      \displaystyle [I(r_{\alpha_{\rm int}})-I(0)]
             &\text{if $\gamma=$0, $r_{\alpha_{\rm int}}\!>\!r_x$;}\\
      \displaystyle \frac{r_{\alpha_{\rm int}}^2 \cdot r_\mathrm{s}}{2}
             &\text{if $\gamma=$0,  $r_{\alpha_{\rm int}}\!<\!r_x$;}
   \end{cases}
   \label{eq:J_final}
\end{equation}
where
\begin{eqnarray}
  r_{\alpha_{\rm int}} &=& \alpha_{\rm int} \cdot d, \nonumber\\
  r_x &=& r_\mathrm{s} \cdot [x^{1/(3-\gamma)}-1], \nonumber\\
  I(x)&=& -r_\mathrm{s}^6 (r_\mathrm{s}^2+5r_\mathrm{s}x+10x^2)/(30(r_\mathrm{s}+x)^5).
 \label{eq:J_final_def}
\end{eqnarray}

\begin{figure}
\includegraphics[width=\columnwidth]{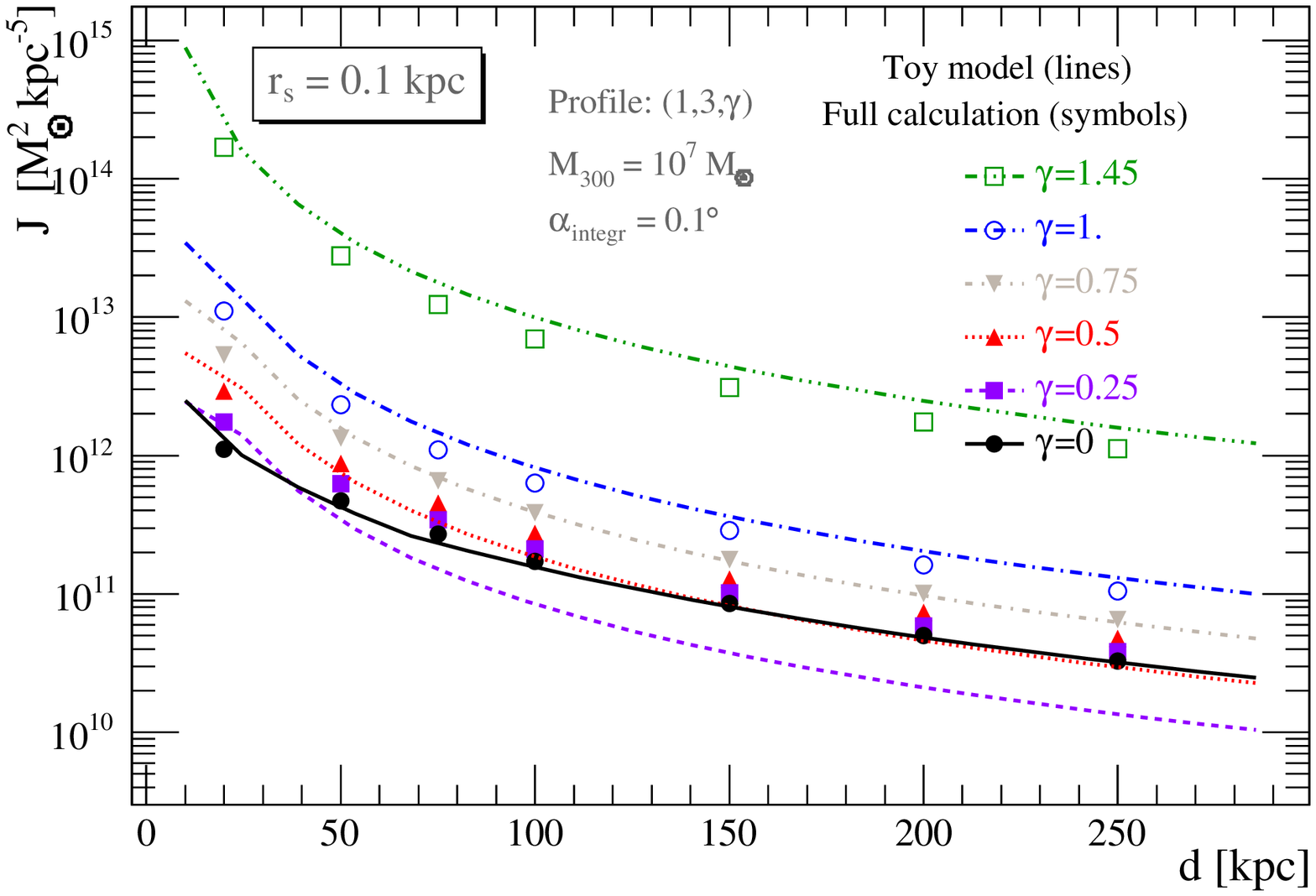}
\includegraphics[width=\columnwidth]{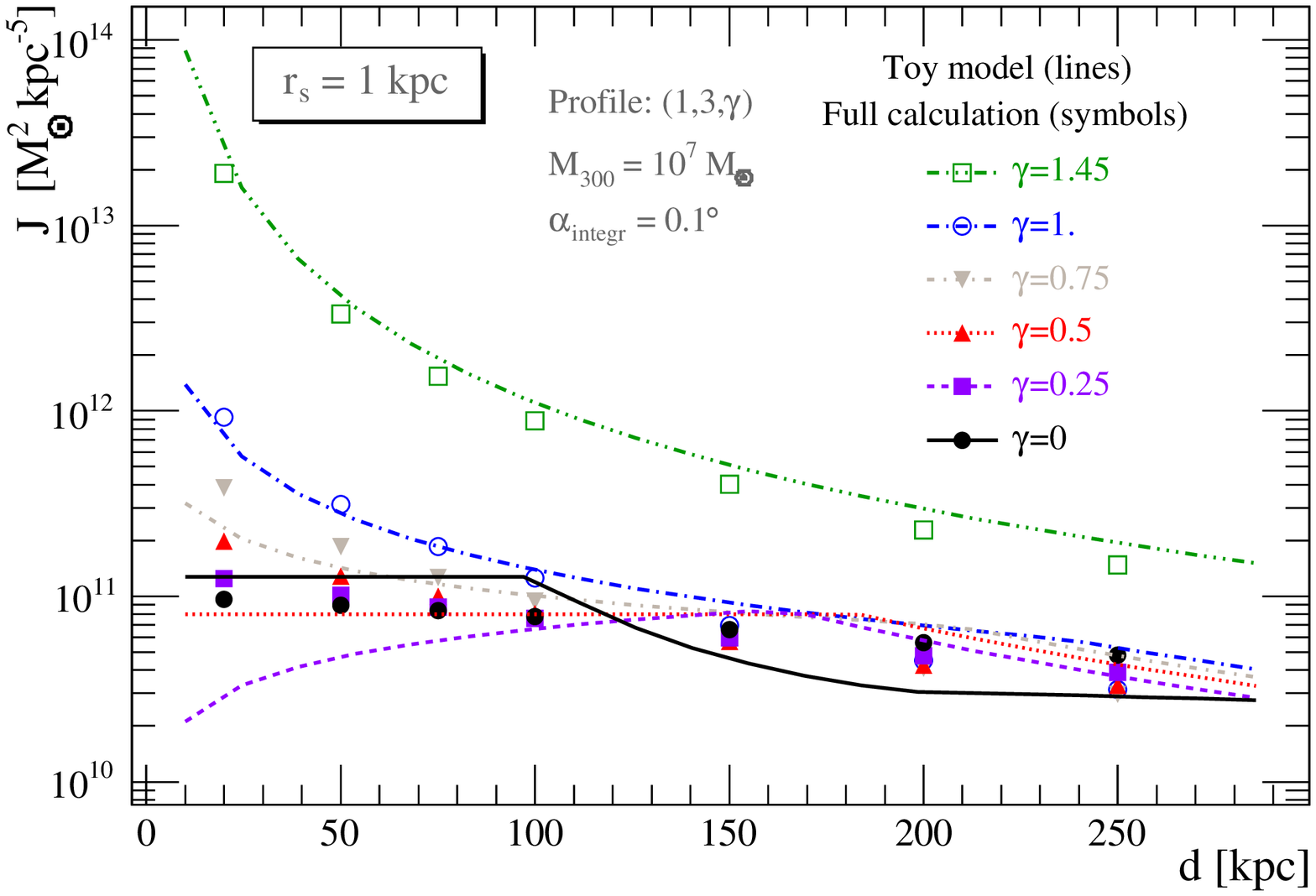}
\caption{Toy-model calculation (lines) vs full numerical integration
  (symbols) of $J$ as a function of the distance to the dSph. The
  integration angle is fixed to $\alpha_{\rm int}=0.1^\circ$ and the
  $(1,3,\gamma)$ profiles are taken to vary from $\gamma=0$ to
  $\gamma=1.45$.  For each model, $\rho_\mathrm{s}$ is calculated such
  as to provide $M_{300}=10^7M_\odot$.  {\bf Top:} dSphs for which
  $r_\mathrm{s}=0.1$~kpc. {\bf Bottom:} dSphs for which
  $r_\mathrm{s}=1$~kpc.  .}
\label{fig:toy_vs_full}
\end{figure}

\paragraph*{Toy model {\em vs} numerical integration}
Finally, we check the validity of this toy model by confronting it
with the full numerical integration.  Various inner slope $\gamma$ of
the profile are considered as provided in
Table~\ref{tab:generic_model_definition}.  Defining the critical
distance $d_{\rm crit}$ for which the dSph is fully encompassed by the
integration region, i.e.,
\[
 d_{\rm crit} = \frac{r_\mathrm{s}}{\alpha_{\rm int}}\,.
\]
we find $d_{\rm crit}\sim$ 50 kpc and 500 kpc for $r_\mathrm{s}=0.1$
and 1~kpc respectively (the integration range is $\alpha_{\rm
  int}=0.1^\circ$). If $r_x$ is used instead of $r_\mathrm{s}$, this
distance is even smaller. This allows us to test the toy model for the
two regimes. The result is shown in Fig.~\ref{fig:toy_vs_full}.  The
symbols show the full numerical integration while the lines show the
toy-model calculations.
For profiles steeper than 0.5, the agreement is better than a factor
of 2 for all distances.  For flatter profiles, the toy model only
gives results within an order of magnitude.  However, for $\gamma=0$,
the fix applied to the toy-model allows to regain the correct results
within a factor of 2.

Hence, given the current uncertainties on the profiles, the set of
formulae (\ref{eq:J_final_def}) and (\ref{eq:J_final_def}) can safely
be used for quick inspection of the $J$ value of any profile with an
inner slope $\gamma$ of 0, or greater than 0.5.

\subsection{For the sub-clumps}
\label{sec:sub-clumps}
The influence of DM sub-structures on the $\gamma$-ray
production has been widely discussed in the literature.  These
sub-structures may enhance the detectability by boosting the
$\gamma$-ray signal. In this appendix, we give an analytical
estimation of the effect of sub-clumps in dSph spheroidal galaxies,
in the same spirit as the toy model developed in the previous section
for the smooth component. For simplicity, we restrict ourselves to one
cored $(\alpha, \beta, \gamma)=(1,3,0)$ and one cusped $(1,3,1)$
profile. To characterise the clump distribution, we use the formalism
given in \citet{2008A&A...479..427L}.

\paragraph*{Sub-structure distribution}
The clump spatial distribution is assumed to follow the dSph DM
profile, namely
\begin{equation}
  \frac{dP(r)}{dV} \propto \left(\frac{r}{r_\mathrm{s}}\right)^{-\gamma}
  \left[1+\left(\frac{r}{r_\mathrm{s}}\right)^{-\alpha}\right]^{\frac{\gamma-\beta}{\alpha}}\,.
\label{eq:distrib_space}
\end{equation}
The mass distribution of the clumps is taken to be independent of the
spatial distribution and takes the usual form,
\begin{equation}
  \frac{dP}{dM} = A M^{-a}\;,
\label{eq:distrib_mass}
\end{equation}
with $M\in[M_{\rm min},M_{\rm max}]$ and $a\sim 1.9$ from cosmological
N-body simulations ($A$ is the normalisation constant for $dP/dM$ to
be a probability).

\paragraph*{Clump luminosity}
Defining $L_i$ the {\em intrinsic luminosity} of the sub-clump $i$ to
be
\begin{equation}
  L_i\equiv\int_{V_{\rm cl}} \rho^2 dV\,,
\end{equation}
the astrophysical contribution to the $\gamma$-ray flux from the
sub-structures of the dSph is
\begin{equation}
  J_{\rm clumps} = \frac{1}{d^2}\cdot \sum_{i=1}^{N^{\rm cl}} L_i\,,
\label{eq:jcl_discreet}
\end{equation}
where $N^{\rm cl}$ is the number of clumps contained within the
integration angle $\alpha$ and $d$ is the distance of dSph. The
luminosity depends only on the mass of the clump, once a
concentration-mass ($c_{\rm vir}-M_{\rm vir}$) relationship is chosen
\citep[see, e.g.,][and references therein]{2008A&A...479..427L}, so
that $ L_i=L (M_i)$.
Moving to the continuous limit, Eq.~(\ref{eq:jcl_discreet}) reads
\begin{equation}
  J_{\rm clumps} = \frac{1}{d^2} \cdot N^{\rm cl} \cdot \int_{M_{\rm min}}^{M_{\rm max}} L(M)\frac{dP}{dM} dM\;.
\end{equation}
Fitting the results from \citet{2008A&A...479..427L}, the intrinsic
luminosity\footnote{In this toy model, we limit ourselves to the NFW
  profiles for the sub-clumps in the dSph, and a $c_{\rm vir}-M_{\rm
    vir}$ relation taken from \citet{2001MNRAS.321..559B}.} varies
almost linearly with the mass of the clump, as
\begin{equation}
  L^{\rm NFW}(M)=1.17\times 10^{8}\;(M/M_\odot)^{0.91}\;M_\odot^2\;{\rm kpc}^{-3},
\end{equation}
so we have
\begin{equation}
  J_{\rm clumps} = \frac{N^{\rm cl}A}{d^2} \left(\frac{1.17\times 10^{8}}{1.91-a}\right)\;\left(M_{\rm max}^{1.91-a}-M_{\rm min}^{1.91-a}\right)\;.
\end{equation}
%

\paragraph*{Number of clumps}
The fraction $F$ of clumps in the spherical integration region
$r_{\alpha_{\rm int}}\approx\alpha_{\rm int} d$ (cross-hatched region
in Fig.~\ref{fig:sketch_dsph}) is given by
\begin{equation}
  F=\frac{N^{\rm cl}}{N_{\rm tot}^{\rm cl}}= \int_0^{r_{\alpha_{\rm int}}} 4\pi r^2
  \frac{dP}{dV} dr\;,
\end{equation}
where $N_{\rm tot}^{\rm cl}$ is the total number of clumps within the
dSph. Upon integration and defining $x_{\rm int}=r_{\alpha_{\rm
    int}}/r_\mathrm{s}$ and $x_{\rm vir}=R_{\rm vir}/r_\mathrm{s}$
this becomes:
\begin{eqnarray}
\label{eq:frac_core}
F_{\rm core}& = &\left[\frac{4x_\alpha+3}{2(x_\alpha+1)^2}+\ln(x_\alpha+1)-\frac{3}{2}\right]\\\nonumber
& \times &\left[\frac{4x_{\rm vir}+3}{2(x_{\rm vir}+1)^2}+\ln(x_{\rm vir}+1)-\frac{3}{2}\right]^{-1} \text{for $(1,3,0)$,}
\end{eqnarray}
and
\begin{eqnarray}
\label{eq:frac_cusp}
F_{\rm cusp}& = &\left[\frac{1}{(x_\alpha+1)}+\ln(x_\alpha+1)-1\right]\\\nonumber
& \times &\left[\frac{1}{(x_{\rm vir}+1)}+\ln(x_{\rm vir}+1)-1\right]^{-1} \text{for NFW.}
\end{eqnarray}
Some care is necessary when evaluating the number of clumps $N^{\rm
  cl}=F\times N_{\rm tot}^{\rm cl}$ in the integration region.
Whatever the profile, most of the clumps are located within
$r_\mathrm{s}$ so when $r_{\alpha_{\rm int}}>r_\mathrm{s}$, the
spherical integration region of our toy model (cross-hatched region in
Fig.~\ref{fig:sketch_dsph}) is a good enough approximation, and
Eq.~(\ref{eq:frac_core}) and (\ref{eq:frac_cusp}) hold. However, if
$r_{\alpha_{\rm int}}<r_\mathrm{s}$ then the remainder of the
intersecting cone (vertically hatched region in
Fig.~\ref{fig:sketch_dsph}) could amount to a significant contribution
to the number of clumps. Cuspy distributions should only be marginally
affected given their high central concentration. However, this effect
may be important for cored profiles. Whenever $r_{\alpha_{\rm int}} <
r_\mathrm{s}$, as for the smooth contribution,
Eq.~(\ref{eq:frac_core}) is therefore multiplied by the ratio of the
intersecting cone volume to the integration sphere volume, in order to
account for that effect.

If the mass of the dSph is $M_{\rm vir}$ and assuming a fraction $f$
of this mass is in the form of clumps, one gets using
Eq.~(\ref{eq:distrib_mass})
\[
N_{\rm tot}^{\rm cl}=f\; \frac{2-a}{A} M_{\rm vir}\left(M_{\rm
    max}^{2-a}-M_{\rm min}^{2-a}\right)^{-1}\;.
\]

\paragraph*{Resulting formulae}
Adding all ingredients together, the contribution of the sub-structures
to the flux is
\begin{eqnarray}
  J_{\rm clumps} & = & 1.17\times 10^{8} \frac{F_{\rm core/cusp}}{d^2} \left(\frac{2-a}{1.91-a}\right)\\\nonumber
  & \times & \left(\frac{M_{\rm max}^{1.91-a}-M_{\rm min}^{1.91-a}}{M_{\rm max}^{2-a}-M_{\rm min}^{2-a}}\right) f\;M_{\rm vir}\;.
\label{eq:toy_cl_final}
\end{eqnarray}

\begin{figure}
\includegraphics[width=\columnwidth]{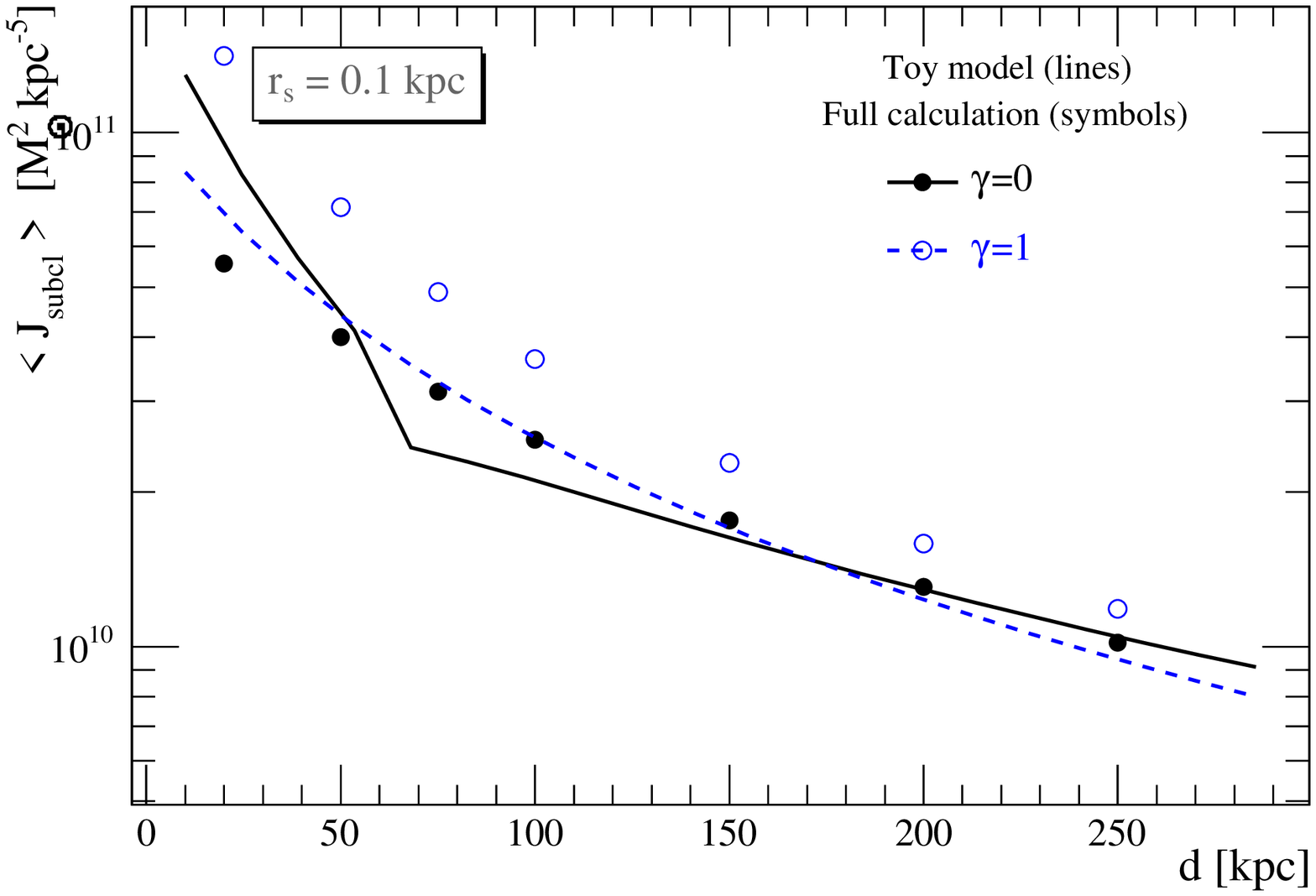}
\includegraphics[width=\columnwidth]{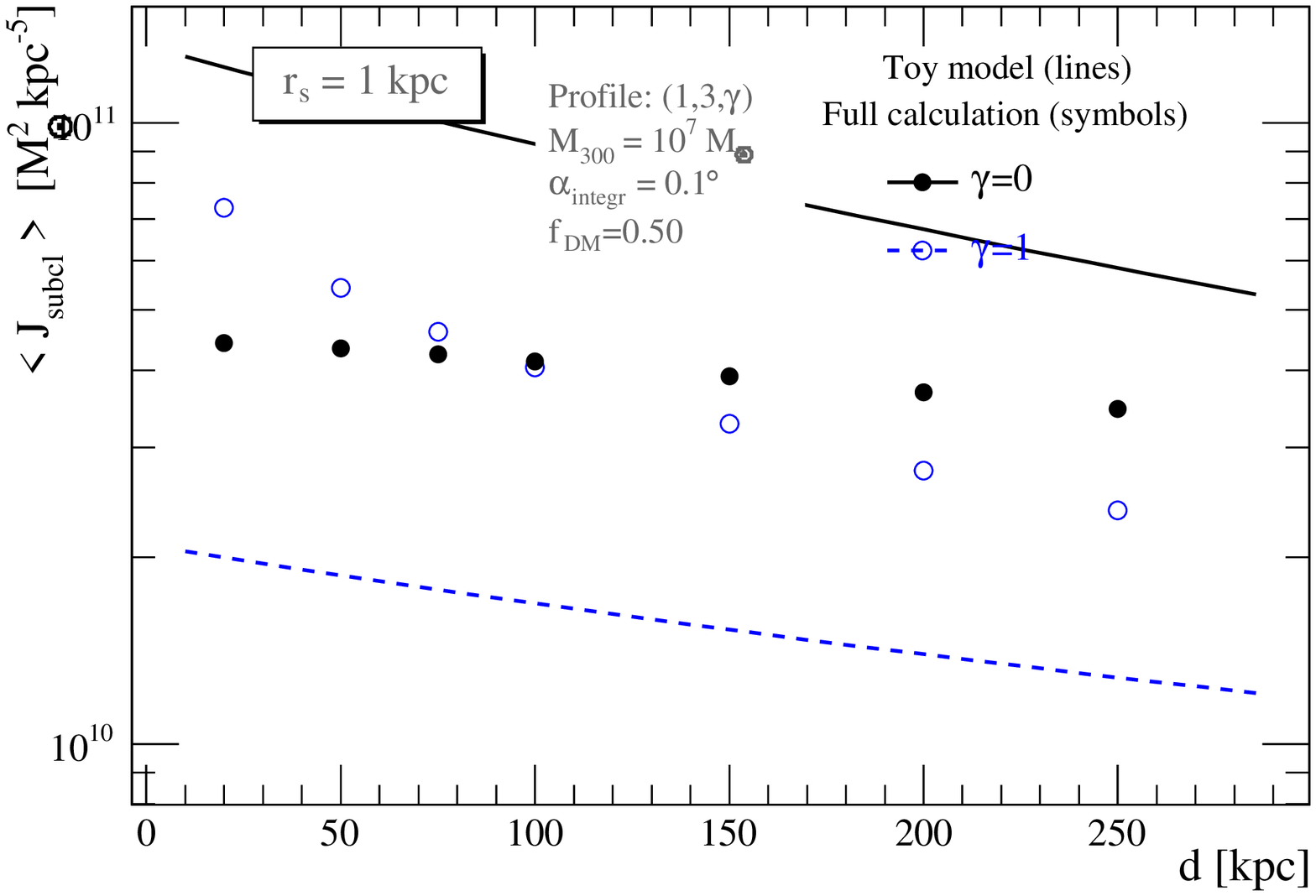}
\caption{Toy-model calculation (lines) vs full numerical integration
  (symbols) of $J$ as a function of the distance to the dSph. The
  integration angle is fixed to $\alpha_{\rm int}=0.1^\circ$ and the
  two $(1,3,\gamma)$ sub-clump spatial distribution are $\gamma=0$ and
  $\gamma=1$ (their inner profile is a NFW with a $c_{\rm vir}-M_{\rm
    vir}$ relation taken from \citet{2001MNRAS.321..559B}).  The
  calculations assume the fraction of DM in sub-clumps to be $f=50\%$
  of the total mass of the dSphs, where the smooth profile is taken as
  in Fig.~\ref{fig:toy_vs_full}. {\bf Top:} $r_\mathrm{s}=0.1$~kpc.
  {\bf Bottom:} $r_\mathrm{s}=1$~kpc.}
\label{fig:toy_vs_full_cl}
\end{figure}

\paragraph*{Toy model {\em vs} numerical integration}
The comparison between the two is shown in
Fig.~\ref{fig:toy_vs_full_cl}.  The symbols show the full numerical
integration while the lines show the toy-model calculations.
For $r_\mathrm{s}=100$~pc, the agreement is better than a factor of 2
for all distances. For $r_\mathrm{s}=1$~kpc, the toy model only gives
results correct to within a factor of 4 for $\gamma=1$.

Hence, given the current uncertainties on the profiles,
Eq.~(\ref{eq:toy_cl_final}) can be used for quick inspection of the
$J$ value for the sub-clump contribution.

\section{Distance and integration angle dependence on J for generic dSphs}
\label{app:dep_generic}

This Appendix completes the study of the $J$-factor dependences started
in Section~\ref{sec:gen_dep}. All the plots and discussions below rely on the
generic profiles given in Table~\ref{tab:generic_model_definition}, and the sub-structure
reference configuration given in Section~\ref{sec:sub_reference}.

\subsection{Distance dependence $J(d)$}

\begin{figure}
\includegraphics[width=\linewidth]{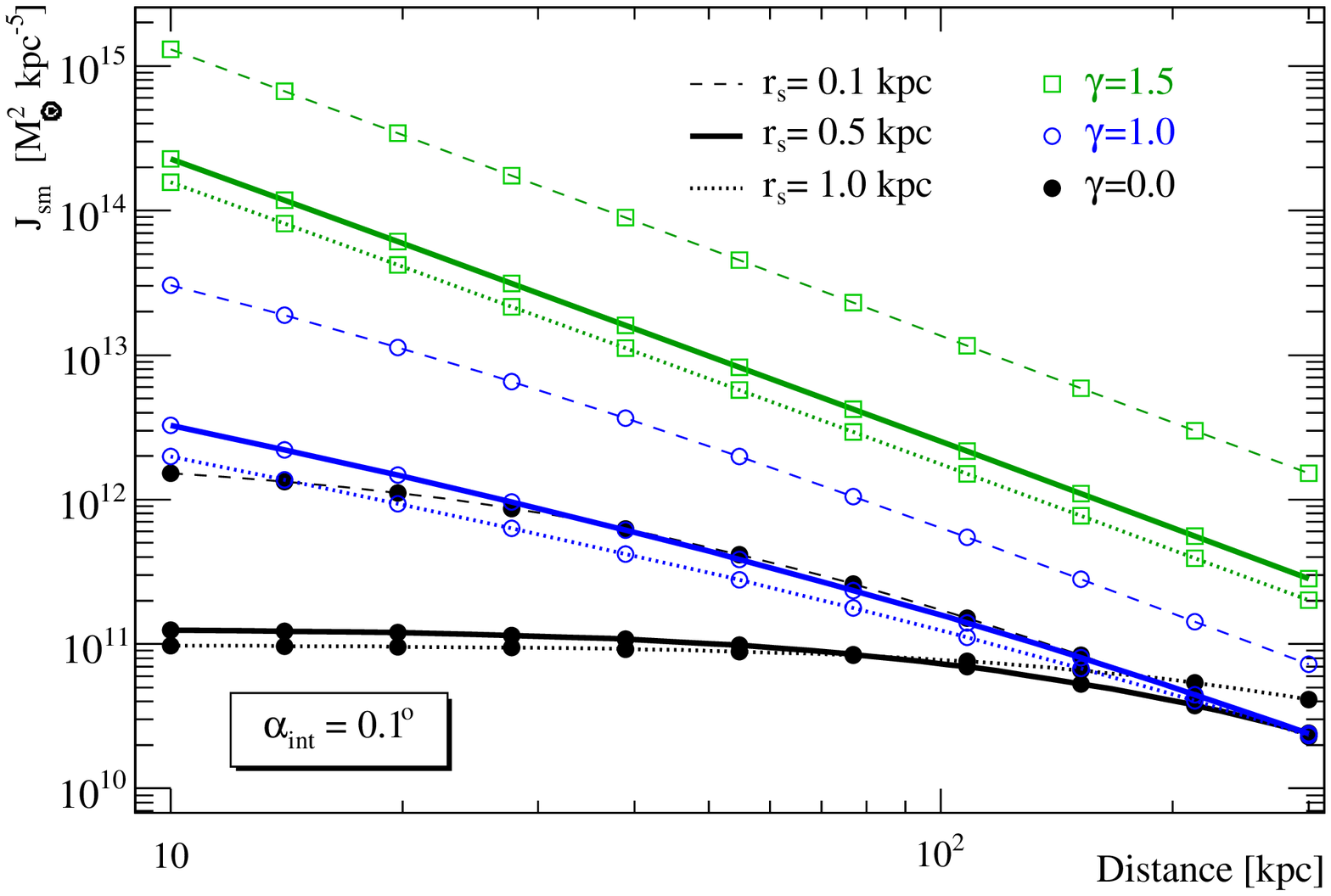}
\caption{$J_{\rm sm}(\theta=0)$ as a function of the distance to the
  dSph for three profiles $\gamma$ and three values of
  $r_\mathrm{s}$. The corresponding values for $\rho_\mathrm{s}$ are
  given in Table~\ref{tab:generic_model_definition}.}
\label{fig:generic_JD_sm_and_subcl}
\end{figure}

Fig.~\ref{fig:generic_JD_sm_and_subcl} shows $J_{\rm sm}$ as a
function of the distance to the dSph (we assume $\alpha_{\rm
  int}=0.1^\circ$ here and that we are pointing towards the dSph
centre, i.e. $\theta=0$). As we have checked earlier, the sub-clump
contribution for the reference model at $\theta=0$ is always
sub-dominant, so for clarity only $J_{\rm sm}$ is displayed ($f=0$) in
the figure.

If the angular size of the signal is smaller than the integration
angle, the distance dependence is expected to be $J_{\rm sm}\propto
d^{~-2}$. This is the case for $\gamma=1.5$ for any value of
$r_s$ (hollow squares curves). Actually, the three curves follow
the point-like source toy formula (\ref{eq:J_final}) appropriate
for steep $\gamma$, i.e.
\begin{equation}
  J(\theta=0) \propto \rho_\mathrm{s}^2 \times \frac{r_\mathrm{s}^3}{d^2}\,.
\end{equation}
However, when the angular size of the emitting region becomes larger
than the integration angle, the above relationship fails. As most of the
flux is emitted within $r_\mathrm{s}$, this happens for a critical
distance
\begin{equation}
d_{\rm crit} \approx \frac{r_\mathrm{s}}{\alpha_{\rm int}}\,.
\label{eq:d_crit}
\end{equation}
For $r_\mathrm{s}=0.1$~kpc, this corresponds to $d_{\rm
  crit}\approx60$ kpc (see the full circles dashed curve
  for $\gamma=0$). Having a
dSph closer than this critical distance does not
increase further the signal (see, e.g., the solid and
dotted full circles curves for $\gamma=0$ and $r_s\gtrsim 0.5$~kpc).
In the latter case, taking a larger integration region
is not always the best strategy as, from an experimental point of view, a larger
integration region increases not only the signal but also the
background. In this case, the gain in sensitivity from having a dSph
close by is not as important as what might na\"ively be expected from
the point-like approximation (see Section \ref{sec:detectability}).

\subsection{Integration angle dependence $J(\alpha_{\rm int})$}

We recall that $ \int_{\Delta\Omega} \mathrm{d}\Omega= \int_{0}^{2\pi}
\mathrm{d}\beta_{\rm int} \int_{0}^{\alpha_{\rm int}} \sin(\alpha_{\rm int})
\mathrm{d}\alpha_{\rm int}$, where $\Delta\Omega = 2\pi(1-\cos(\alpha_{\rm
  int}))$, so that the $J$-factor from Eq.~(\ref{eq:J}) can be rewritten in
the symbolic notation
 \begin{equation}
   J(\psi,\theta, \Delta\Omega) = \int_{0}^{2\pi} F_{[\beta_{\rm int}]} \,\mathrm{d}\beta_{\rm int}
   \label{eq:J1}
 \end{equation}
with
 \begin{equation}
   F_{[\beta_{\rm int}]} = \int_{0}^{\alpha_{\rm int}} F_{[\beta_{\rm int},\alpha_{\rm int}]} \,\mathrm{d}\alpha_{\rm int}' 
   \label{eq:J2}
 \end{equation}
and
 \begin{equation}
   F_{[\beta_{\rm int},\alpha_{\rm int}]} = \sin(\alpha_{\rm int}) \int_{0}^{l_{max}} 
   {\cal F}\large[r(l,\beta_{\rm int},\alpha_{\rm int})\large] \,\mathrm{d}l.
   \label{eq:J3}
 \end{equation}
For small integration angles and the case of a flat enough profile,
the integrand in Eqs.~(\ref{eq:J2}) and~(\ref{eq:J3}) does not vary
much with $\alpha_{\rm int}$, so that for the smooth (${\cal F}\equiv\rho^2$) and the mean
sub-clumps (${\cal F}\equiv\rho$), we have
  \begin{equation}
    J_{\rm sm}\propto \alpha_{\rm int}^2 
       \quad {\rm and} \quad 
       \langle J_{\rm subcl}\rangle\propto \alpha_{\rm int}^2\,.
    \label{eq:alphint2_dep}
  \end{equation}

\begin{figure}
\includegraphics[width=\linewidth]{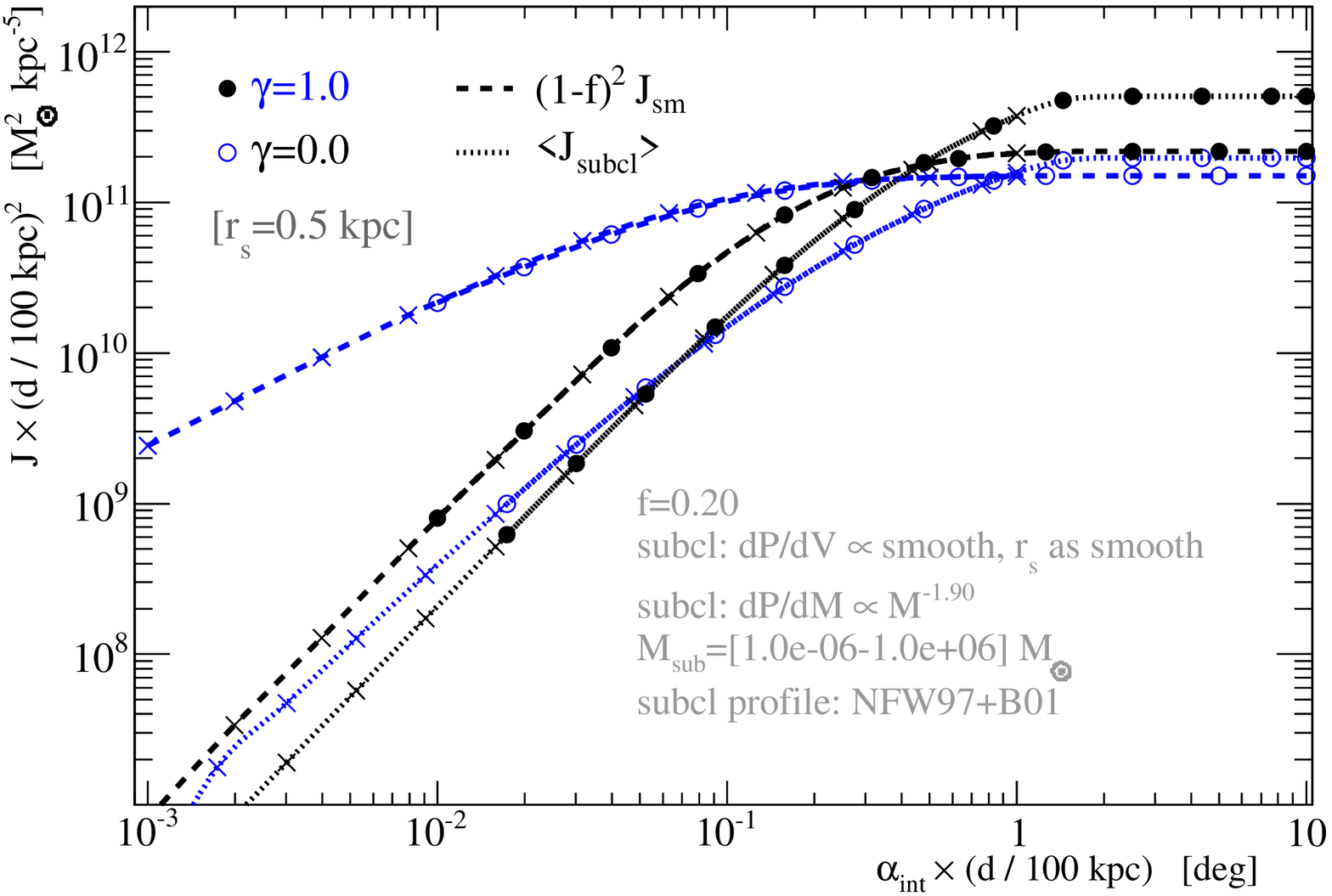}
\caption{$J \times (d/100$~kpc$)^2$ as a function of $\alpha_{\rm int}
  \times (d/100$~kpc) for a generic dSph with $r_\mathrm{s}$=0.5 kpc: smooth
  (thick dashed lines) and sub-clumps (thin dotted lines). With this
  rescaling, the case $d=10$~kpc (stars) superimposes
  on the case $d=100$~kpc (empty and full circles).}
\label{fig:generic_Jalphaint_sm_and_subcl}
\end{figure}

Fig.~\ref{fig:generic_Jalphaint_sm_and_subcl} shows the
integration angle dependence for the smooth $(1-f)^2J_{\rm sm}$
(dashed lines) and the sub-clump mean $\langle J_{\rm subcl}\rangle$ 
(dotted lines) contributions. (The pointing
direction is towards the dSph centre.)
For $\gamma=0$ (solid black circles), the $\alpha_{\rm int}^2$
scaling holds up to $\alpha_{\rm int}^{\rm crit}
\sim\!3^\circ$ if $d=10$~kpc (as given by Eq.~\ref{eq:d_crit}).
A plateau is reached when the entire emitting region of the dSph
is encompassed (i.e. for a few $r_\mathrm{s}/d$).
For $\gamma=1$ (blue empty circles), the curves are
  slightly more difficult to interpret, as the profile is not steep
  enough for it to be considered fully point-like (and thus
  `independent' of $\alpha_{\rm int}$) given the
  integration angles considered. \footnote{The dependence can be
    understood by means of the toy model formulae
    (\ref{eq:J_final}) and (\ref{eq:J_final_def}). For
    $\alpha_{\rm int}<\alpha_{\rm int}^{\rm crit}$, we have
\begin{equation}
  J_{[\gamma \gtrsim 0.5]} \propto r_\mathrm{s}^{2\gamma} \times (\alpha_{\rm int} d)^{3-2\gamma}\,.
  \nonumber
\end{equation}
For $\gamma=1$ (empty blue circles), $J$ is then expected to
scale linearly with $\alpha_{\rm int}$, which is observed for
the smooth (dashed blue line), and to some extent for the
sub-clump contribution (dotted blue line). However, for the latter,
the transition region (around
$r_\mathrm{s}$) falls from a slope $\alpha=1$ towards an outer slope
$\beta=3$ (instead of falling from $\alpha^2=1$ to $\beta^2=6$. Hence,
for $\alpha_{\rm int}>\alpha_{\rm int}^{\rm crit}$, the sub-clump
contribution continues to  build up gradually.}
Finally, the rescaling used in Fig.~\ref{fig:generic_Jalphaint_sm_and_subcl}
implies:
\begin{equation}
   J_{\rm d1} (\alpha_{\rm int}) = 
   J_{\rm d2}\left(\alpha_{\rm int}\frac{d_2}{d_1}\right)
      \times \left(\frac{d_2}{d_1}\right)^2\;.
      \label{eq:rescale_Jalpha}
\end{equation}

\section{Complementary study of the boost factor}
\label{app:boost}
In Section~\ref{sec:boost}, we concluded that the boost could not be
larger than a factor of 2 for all configurations where the sub-clump
spatial distribution follows that of the smooth halo in the dSph. The
calculations were also made for a `reference' configuration of the
sub-clumps. However, the boost can be smaller (or larger) when the
latter parameters are varied.

In Tab.~\ref{tab:boost}, we systematically vary all the parameters
entering the calculation in order to compare with the reference model
case. The two quantities of importance are the maximum boost possible
(which is obtained when $\alpha_{\rm int}$ fully encompasses the
clump), and the transition point $\alpha_{\rm int}d$ for which the
boost equals 1 (the minimum value is always given by $(1-f)^2$). The
{\em reference} results correspond to the numbers obtained from the
dotted lines in Fig.~\ref{fig:generic_boost_default}, i.e. for
$r_\mathrm{s}=1$~kpc. Note that most of the values for $B_{\rm max}$
in the Table would be close to unity if $r_\mathrm{s}=0.1$~kpc were to
be selected.
\begin{table}
\begin{center}
 \caption{Maximum boost and transition regime, i.e. ($\alpha_{\rm
    int}d)_{B=1}$ in deg~kpc, for which $B=1$, for various
  smooth/sub-clump parameters for three inner slope $\gamma$ (for the
  smooth).}
\label{tab:boost}
\begin{tabular}{@{}cccccc} \hline\hline
Config.$^\dagger$         &~~~&   $\gamma=0$      &  $\gamma=0.5$    & $\gamma=1$   \\
\hline
                         &~~~& \multicolumn{3}{c}{~~~~~$B_{\rm max}~|~(\alpha d)_{B=1}$}\\
reference$^{\ddagger}$     &~~~&    $1.9~|~19$     &   $2.2~|~21$     & $2.0~|~30$   \\
\hline
{\bf [global parameters]}  &~~~&                   &                  & \\
$\alpha=1$                 &~~~&    $1.0~|~40$     &   $1.3~|~60$     & $~~1.6~|~160$  \\
$\beta=5$                  &~~~&    $2.3~|~11$     &   $2.0~|~18$     & $1.3~|~36$   \\
$R_{\rm vir}=6$~kpc        &~~~&    $3.0~|~15$     &   $3.5~|~20$     & $2.9~|~29$   \\
$M_{300}=2\cdot10^7M_\odot$&~~~&    $1.3~|~66$     &   $1.4~|~52$     & $~1.3~|~64$   \vspace{0.25cm}\\
{\bf [sub-clump parameters]}&~~~&                   &                  & \\
$dP/dV=$Einasto$^\star$    &~~~&$1.4~|~\dots\!\!$  &$1.7~|~\dots\!\!$ & $1.7~|~22$   \\
$a=1.7$                    &~~~&    $1.3~|~62$     &   $1.5~|~50$     & $1.3~|~61$   \\
$a=2.0$                    &~~~& $2.8~|~0.2\!$     &   $3.4~|~8~\,$   & $2.9~|~16$   \\
$M_{\rm min}=1 M_\odot$    &~~~&    $1.5~|~43$     &   $1.7~|~37$     & $1.5~|~47$   \\
$M_{\rm max}=10^4M_\odot$  &~~~&    $\!\!2.4~|~4~$ &   $2.8~|~14$     & $2.5~|~22$   \\
$f=0.5$                    &~~~&    $3.4~|~10$     &   $4.2~|~16$     & $3.5~|~25$   \\
$\rho_{\rm subcl}=$Einasto &~~~&$~8.7~|~0.05\!\!\!$& $10.6~|~0.35\!\!$& $9.0~|~4~~$   \\
$c_{\rm vir}\times 2$      &~~~&$~7.6~|~0.06\!\!\!$&  $~\,9.3~|~0.4$  & $7.9~|~4.5\!\!$  \\
\hline
\end{tabular}
\begin{flushleft}
 $^\dagger$ All parameters are as for {\em reference}, except those quoted.\\
 $^\ddagger$ Reference configuration ($M_{300}=10^7M_\odot$):\\
 \hspace{0.5cm}   $\cdot$ $\rho_{\rm sm}$ with $(\alpha,\beta,\gamma)=(1,3,\gamma)$ and $dP/dV\propto \rho_{\rm sm}$;\\
 \hspace{0.5cm}   $\cdot$ $R_{\rm vir}=3$ kpc and $r_\mathrm{s}=1$ kpc (for $\rho_{\rm sm}$ and $dP/dV$);\\
 \hspace{0.5cm}   $\cdot$ $dP/dM=M^{-a}$ $(a=1.9)$, and $M_{\rm sub} \in [10^{-6}-10^{6}]~M_\odot$;\\
 \hspace{0.5cm}   $\cdot$ $f=0.2$, $\rho_{\rm subcl}=$NFW, and $c_{\rm vir}-M_{\rm vir}$=B01.\\
 $^\star$ Einasto parameters taken from \citet{2006AJ....132.2685M}.
\end{flushleft}
\end{center}
\end{table}

\subsection{Varying the [global parameters]}
The four lines under '[global parameters]' keeps the recipe of $dP/dV
\propto\rho_{\rm sm}$, but some previously fixed parameters are now
varied. The trend is that a sharper transition zone (larger $\alpha$),
a larger radius of the dSph, or a smaller mass imply a larger $B_{\rm
  max}$. The impact of the outer slope $\beta$ depends on the value of
the inner slope $\gamma$. However, the maximum boost factor reached
for these parameters is never larger than $\sim3$. The typical
transition value lies around $20^\circ$~kpc, which corresponds, for a
dSphs located 100 kpc away, to an integration angle of
$0.2^{\circ}$. Hence, for all these configuration, large integration
angle should be preferred (this is even worse for closer dSph).

\subsection{Varying the [sub-clump parameters]}
The remaining lines under [sub-clump parameters] show the impact of the
choice of the distribution of sub-clumps, the mass distribution
parameters (minimal mass and maximal mass of the sub-clumps, slope $a$
of $dP/dM$), and the density profile of the sub-clumps. Relaxing the
condition $dP/dV \propto\rho_{\rm sm}$ has no major impact. In
\citet{2008MNRAS.391.1685S}, a simple Einasto profile with universal
parameters was found to fit all halos (from the Aquarius simulation)
independently of the halo mass. For that specific case, we use the
values found for the Galaxy in \citet{2006AJ....132.2685M}. The
Einasto profile is steeper than $\gamma=0$ but it decreases
logarithmically inwards. Only for $\gamma\gtrsim1$ (for the smooth
component) such a model is able to marginally increase the maximum
boost w.r.t. the reference model (instead of decreasing it), which is
not unexpected.\footnote{For smaller $\gamma$, the smooth distribution,
  in that case, is flatter than the sub-clump one, so that the boost is
  larger than one for small $\alpha_{\rm int}$ and the transition
  where $B=1$ is ill-defined. However, such a configuration is highly
  unlikely as it is exactly the opposite of what is observed in all
  N-body simulations.} Varying the mass distribution slope $a$ is
understood as follows: for $a\approx1.9$, all decades in mass
contribute about the same amount. When $a$ is decreased, the less
massive sub-halos dominate, whereas for $a\gtrsim1.9$, the most
massive sub-halos dominate the luminosity (e.g. Fig.~4 of
\citealt{2008A&A...479..427L}). This has to be balanced by the fact
that the fraction of DM going into sub-clumps remain the same
($f=0.2$), regardless of the value of $a$, so that the total number of
clumps in a mass decade also changes. The net result is a smaller
boost when $a$ is decreased, and a larger boost from the more massive
sub-structure when $\alpha$ is increased. In a similar way ($a$ is now
fixed to 1.9 again), the mass also impact on $B$, but in a marginal
way.
The only sizeable impact comes from varying the fraction of mass into
clumps, the sub-clump profile or the concentration of sub-clumps. In
the first case, when $f$ increases, the smooth signal decreases by
$(1-f)^2$ whereas the sub-clump signal increases as $f$. Even if $f$ is
increased up to 50\%, which is very unlikely (recent simulations such
as \citet{2008MNRAS.391.1685S} tend to give an upper limit of
$f\lesssim 10 \%$) this gives only a mild enhancement. In the second
configuration, the NFW profile for the sub-clumps is replaced by an
Einasto one. Despite its logarithmic slope decreasing faster than the
NFW slope $\gamma=1$ below some critical radius, the latter profile is
known to give slightly more signal than the NFW one ($\rho_{\rm
  Einasto}(r)>\rho_{\rm NFW}(r)$ for a region that matters for the $J$
calculation). This results in a boost close to 10, regardless of the
dSph's smooth profile. Finally, we recall that the B01 $c_{\rm
  vir}-M_{\rm vir}$ relation is used to calculate the value of the
scale parameter for any sub-clump mass. In the last configuration, the
concentration parameter is simply multiplied by a factor of 2, which
is probably not realistic. Again, the same boost of $\sim10$ is
observed. Accordingly, for these last two cases, the transition angle
is reduced, and corresponds to $\alpha_{\rm int}< 0.01^\circ$ (for a
dSphs at 100 kpc).

To conclude, although boosts by as large as a factor of 10 can be obtained
through suitable combinations of parameters, most of these combinations
are unlikely and require the signal to be integrated on large angles.

\section{Impact of the PSF of the instrument}
 \label{app:PSF}

\begin{figure} 
\includegraphics[width=\linewidth]{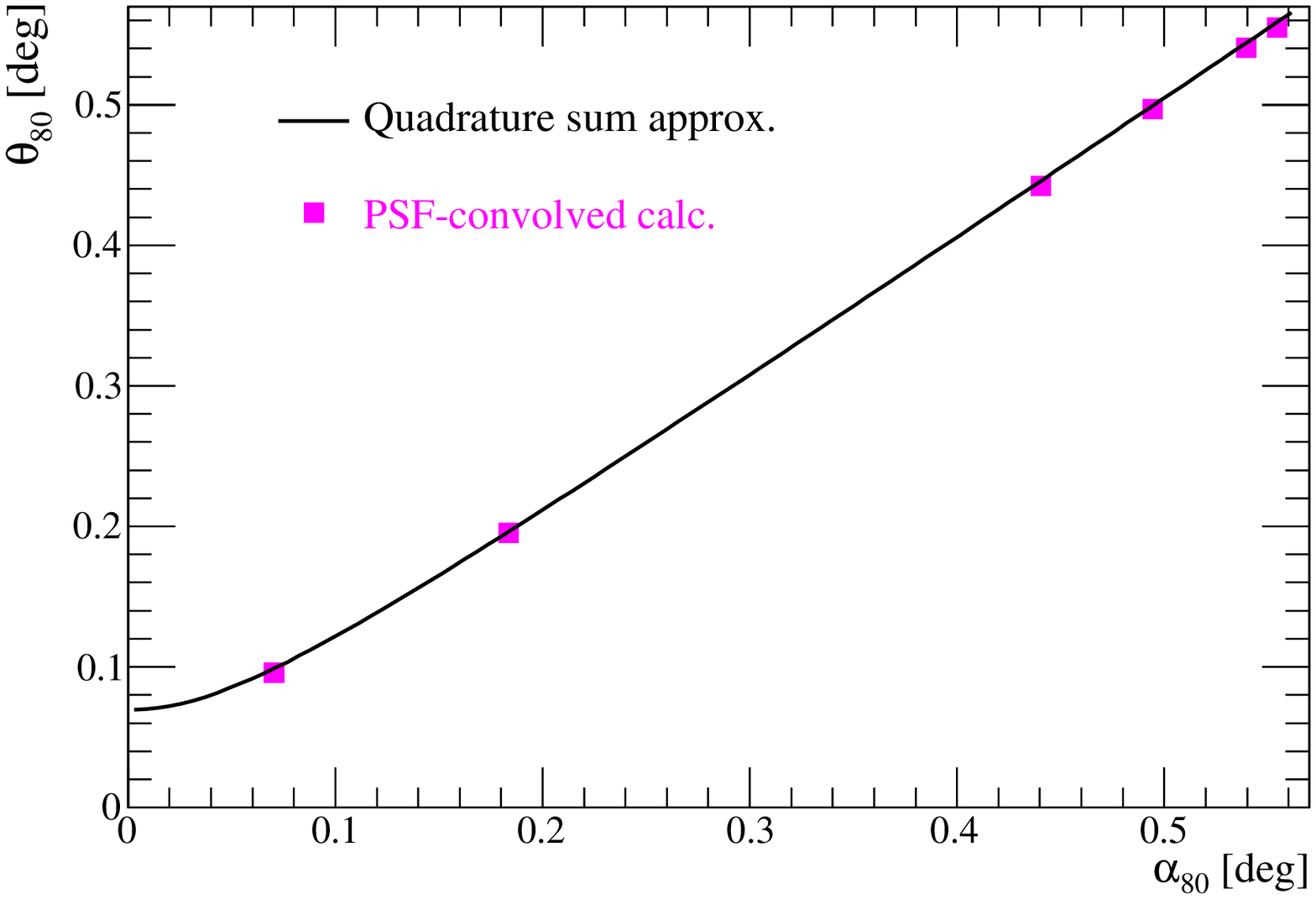}
\caption{80\% containment radius ($\theta_{80}$) of PSF-convolved DM
  annihilation halo models versus $\alpha_{80}$.}
\label{fig:PSFapprox}
\end{figure}

 Fig.~\ref{fig:PSFapprox} shows the impact of the instrument angular
 resolution on the 80\% containement radius for $J$ (for the generic
 dSphs studied in Sec.~\ref{sec:detectability}).
 The solid line corresponds to the quadrature approximation given by
 Eq.~(\ref{eq:PSF}), whereas the symbols correspond to the convolved
 PSF$*$halo profile.  The PSF is described by the sum of two Gaussians
 and is a scaled (factor two improved) version of the PSF appropriate
 for H.E.S.S. at 200 GeV.  Calculated halos for a range of
 $\alpha,\,\beta,\,\gamma$ models consistent with the stellar
 kinematics of the classical dSphs are shown as gray squares. The
 quadrature sum approximation used in this work is shown as a solid
 line.

\section{Confidence levels and priors}
 \label{app:CLs}

 In this Appendix, we describe how confidence intervals for the
 quantities such as $\rho(r)$ or $J$ are chosen.

\subsection{Sensitivity of the result to the choice of prior}
In the Bayesian approach, the PDF of a parameter $x$ is given by the
product of the MCMC output PDF ${\cal P}(x)$ and the prior $p(x)$. The
resulting PDF is therefore subjective, since it depends on the
adoption of a prior. However, whenever the latter are not strongly
dependent on $x$, or if ${\cal P}(x)$ falls in a range where $p(x)$
does not strongly varies, the PDF of the parameter becomes insensitive
to the prior. This happens for instance if the data give tight
constraints on the parameters.

In our MCMC analysis, we assumed a flat prior for all our halo
parameters, as there is no observationally motivated reason for doing
otherwise.  Note, however, that flat priors on the model parameters do
not necessarily translate into flat priors on quantities derived from
those parameters.  Specifically, the flat priors on our model
parameters imply a non-flat prior on the DM density (and also
on its logarithm) at a given radius, and hence a non-flat prior on
$J$.  In principle it is possible to choose a combination of priors
for the parameters that would translate into flat priors on $\rho(r)$,
but we have not done so here.  The general impact of such choices, and
the methodology to study the prior-dependent results, has been
discussed in the context of cosmological studies by
\citet{2008PhRvD..78f3521V}. In this study, we only use a flat prior
on the parameters (or on the log for $r_\mathrm{s}$ and $\rho_s$).
The test with artificial data demonstrate that our reconstructed
$\rho$ and $J$ values are sound.

\subsection{Confidence intervals for $\rho(r)$ and cross-checks}

\subsubsection{Definition}
Confidence intervals $\Delta_x$ (CI), associated with a confidence
level $x\%$ (CL), are constructed from the PDF. The asymmetric
interval $\Delta_x\equiv [\theta^-_x, \theta^+_x]$ such as
\[
{\rm CL}(x)\equiv \int_{\Delta_x}{\cal P}(\theta) {\rm d} \theta= 1 - \gamma,
\]
defines the $1 - \gamma$ confidence level (CL), along with the CI of
the parameter $\theta$.
We rely on two standard practises for the CI selection. The first one
(used only in this Appendix)
is to fix $\theta^-_x$ to be the lowest value of the PDF. The CLs
correspond then to quantiles. This is useful for CI selection of
$\chi^2$ values, to ensure that the best-fit value of a model
(i.e. the lowest $\chi^2$) falls in the CI (see, e.g., Fig.~7 of
\citealt{2009A&A...497..991P}). In the second approach, the CI, i.e
$\theta^-_x$ (resp. $\theta^+_x$), is found by starting from the
median $\theta^{\rm med}$ of the PDF and decrease (resp. increase)
$\theta_x$ until we get $x\%/2$ of the integral of the PDF. This
approach ensures that the median value of the parameters falls in the
CI, any asymmetry in the CI illustrating the departure from a Gaussian
PDF: this is the one used thoughout the paper.

\subsubsection{Comparison of several choices for the PDF of $\rho(r)$}

\begin{figure}
\includegraphics[width=\linewidth]{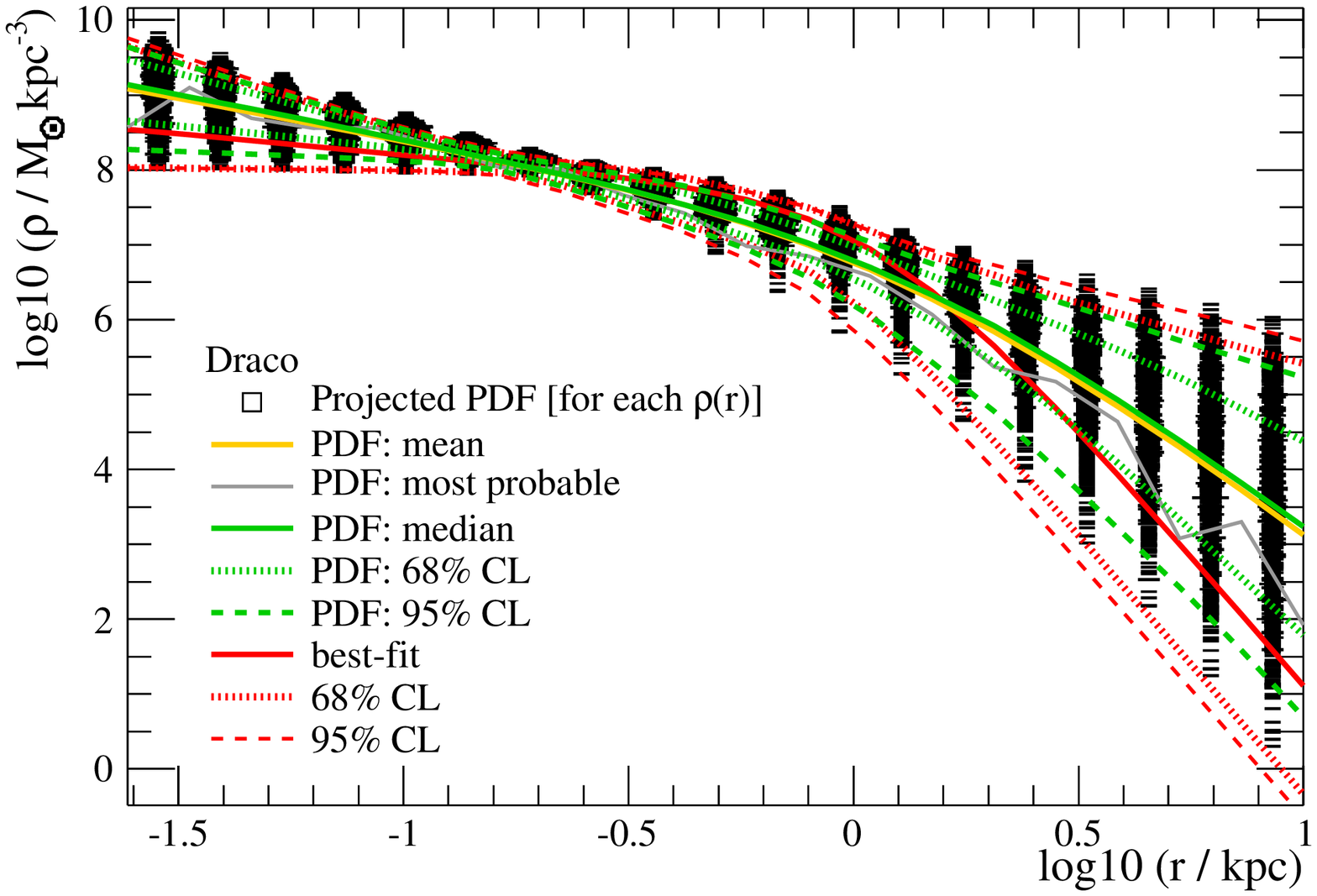}
\caption{Projected distribution of $\log_{10}(\rho)$ along with the
  value of several other estimators for the MCMC analysis of Draco.
  In this box projection, the larger the box, the most
  likely the probability of $\log_{10}(\rho)$. For instance, on the top
  panel, for $\log_{10}(r)=-1.5$ the probability density function of
  $\log_{10}(\rho)$ is distributed in the range $[8-10]$ and
  peaks around $9.5$).}
\label{fig:best_CL_profiles}
\end{figure}

Fig.~\ref{fig:best_CL_profiles} shows the projection for each $r$ of
the PDF calculated from the output MCMC file. To do so, $\rho(r)$ is
calculated for each entry of the thinned chains and then stored as an
histogram. This results in 'boxes': the larger the box, the more
likely the value of $\rho(r)$.  From this distribution, we can
calculate the median (thick solid black line), the most probable value
(thick dotted black line). The thick solid red line correspond to the
model having the smallest $\chi^2$ value. We see that the latter
differ from the median one for this dSph, though they can be close for
other dSph in our sample. In this paper, as our analysis is based on
the Bayesian approach, we disregard the best-fit model and only retain
the median value.

In the first approach, the 68\% and 95\% CLs are calculated from the
distribution $\rho_r$ (at each $r$) They are shown as dashed and
dotted thick black lines. Note that none of all the above lines
corresponds to a {\em physical} configuration of $\rho(r)$.

A second approach is to construct the 68\% CLs from a sampling of the
(still) correlated parameters.  This is achieved by using all sets of
parameters $\{\vec{\theta}\}_{x\% \rm
  CL}=\{\vec{\theta}_i\}_{i=1\cdots p}$, for which
$\chi^2(\vec{\theta}_i)$ falls in the $68\%$ CL of the $\chi^2$ PDF
(see above). Once these sets are found, we calculate $\rho(r)$ for
each of them, and keep the maximum and minimum values for each
position $r$. This defines envelopes of $\rho(r)$ (CIs are found for
each $r$). This is shown as dotted and dashed red lines. Such an
approach was used in \citet{2009A&A...497..991P}.  The CLs obtained
from it are larger than the previous one. In the above paper, the
uncertainties were small even with that method, so that was not an
issue. However, in this study, this makes a huge difference in the
resulting value CL of $J$.

In order to check which approach was the correct one, we bootstrapped
the Draco kinematic data and calculate from the collection of
$\rho(r)$ from each bootstrap sample the median value and the
uncertainty. The first approach, where the CLs are directly calculated
from the full set of MCMC samples was in agreement with the bootstrap
approach, meaning that the second one biases the results toward too
large uncertainties. The results of the paper rely thus on the first
and correct approach.

\section{Artificial data sets: validation of the MCMC analysis}
\label{app:fake}

In this section, we examine the reliability of the Jeans/MCMC analysis
by applying it to artificial data sets of 1000 stellar positions and
velocities drawn directly from distribution functions with constant
velocity anisotropy. We assume the form $L^{-2\beta_{\rm aniso}}
f(\varepsilon)$ for the distribution functions, where the (constant)
velocity anisotropy is given by $\beta_{\rm aniso} = 1 - \sigma_{\rm
  t}^2/\sigma_{\rm r}^2$, with $\sigma_{\rm t}^2$ and $\sigma_{rm
  r}^2$ being the second moments of the velocity distribution in the
radial and tangential directions, respectively. The function
$f(\varepsilon)$ is an unspecified function of energy $\varepsilon$
which we determine numerically using an Abel inversion once the halo
model and stellar density are
specified~\citep{1991MNRAS.253..414C}. We used the same models in
\citet{letter_dsph}, but we present here a more general study. The set
of artificial data covers a grid of models with $\gamma =
{0.1,0.5,1.0}, r_{\rm h}/r_{\rm s} = {0.1, 0.5, 1.0}$ and
$\beta=3.1$. For each halo model, we assume $\beta_{\rm aniso}$ values
of $0$ (isotropic), $0.25$ (radial) or $-0.75$ (tangential): the
$\beta_{\rm aniso}$ values for the anisotropic models are chosen to
give models with roughly equivalent levels of anisotropy (in terms of
the ratios of the velocity dispersions in the radial and tangential
directions). We also generate a grid of models with a steeper inner
slope $\gamma = 1.5$ and $\beta=4.0$. In all cases, the haloes contain
$\sim 10^7$M$_\odot$ within 300pc. We mimic the effects of
observational errors by adding Gaussian noise with a dispersion of
$2\,$km\,s$^{-1}$ to each individual stellar velocity generated from
the distribution function.
The reconstruction depends on the choice of the prior $\gamma_{\rm
  prior}$, and this effect is explored in the two sections below.

\subsection{Prior: $0\leq\gamma_{\rm prior}\leq1$ versus $0\leq\gamma_{\rm prior}\leq2$}
\label{app:biases1}
\begin{figure*}
\includegraphics[width=0.49\linewidth]{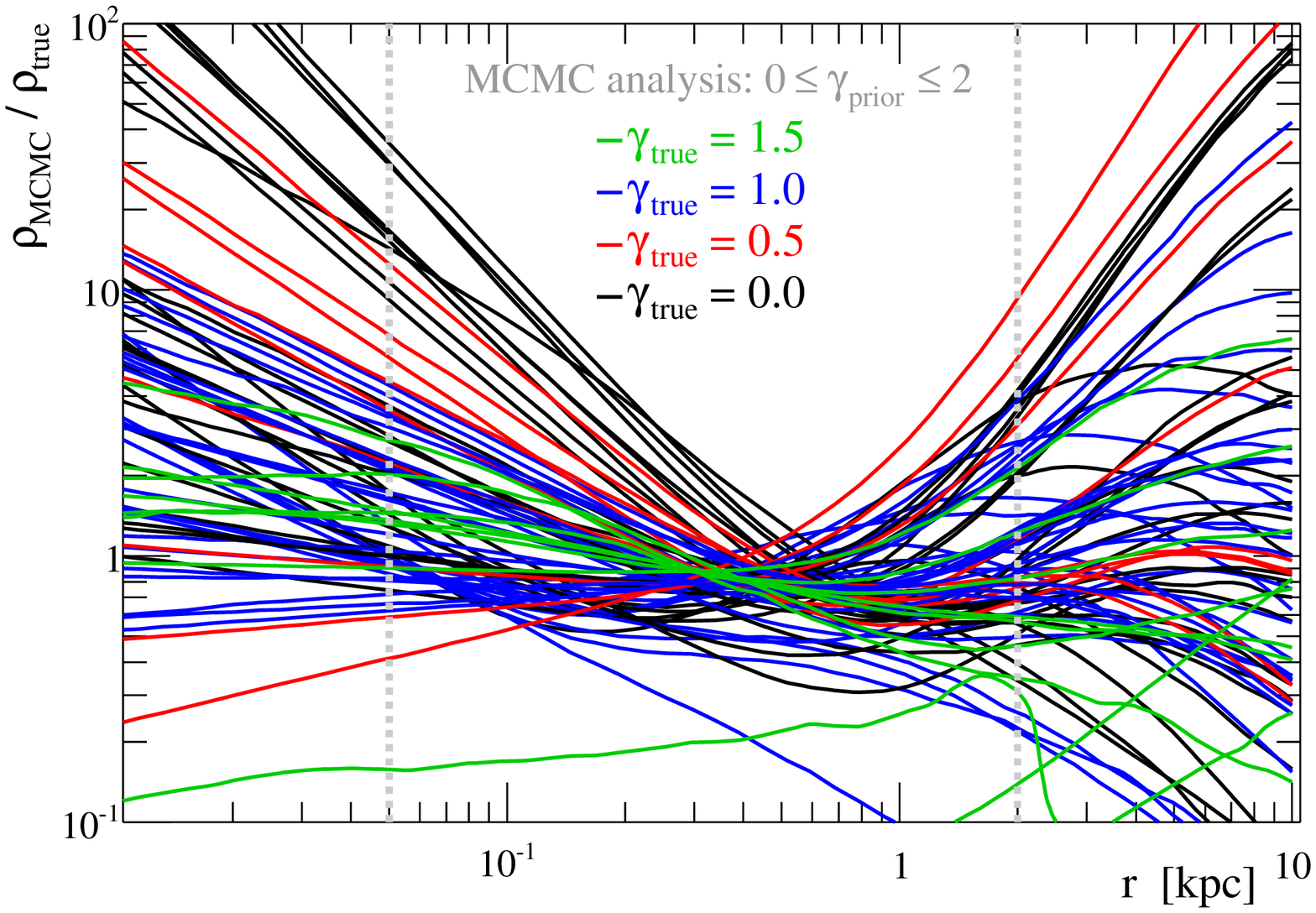}
\includegraphics[width=0.49\linewidth]{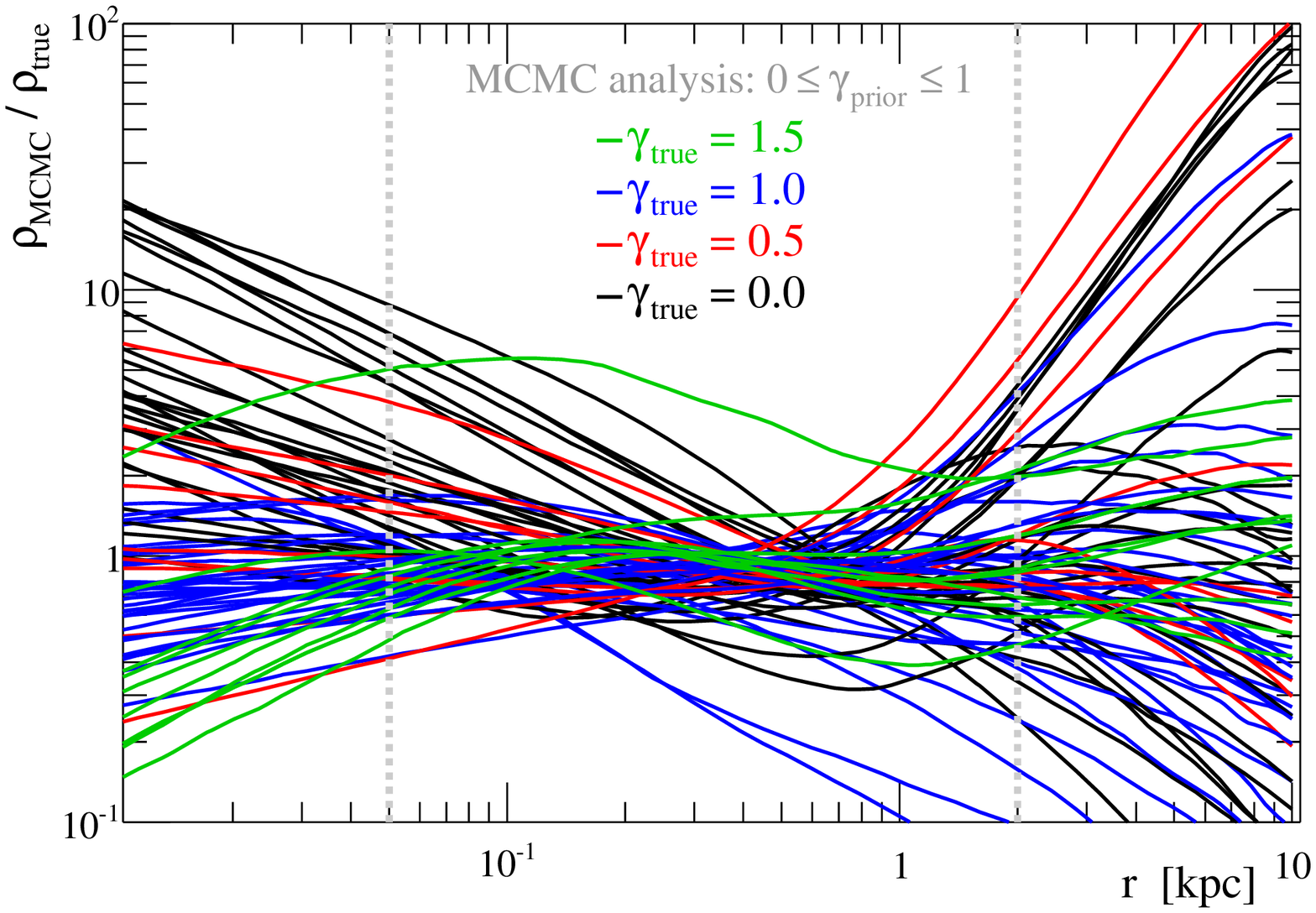}
\includegraphics[width=0.49\linewidth]{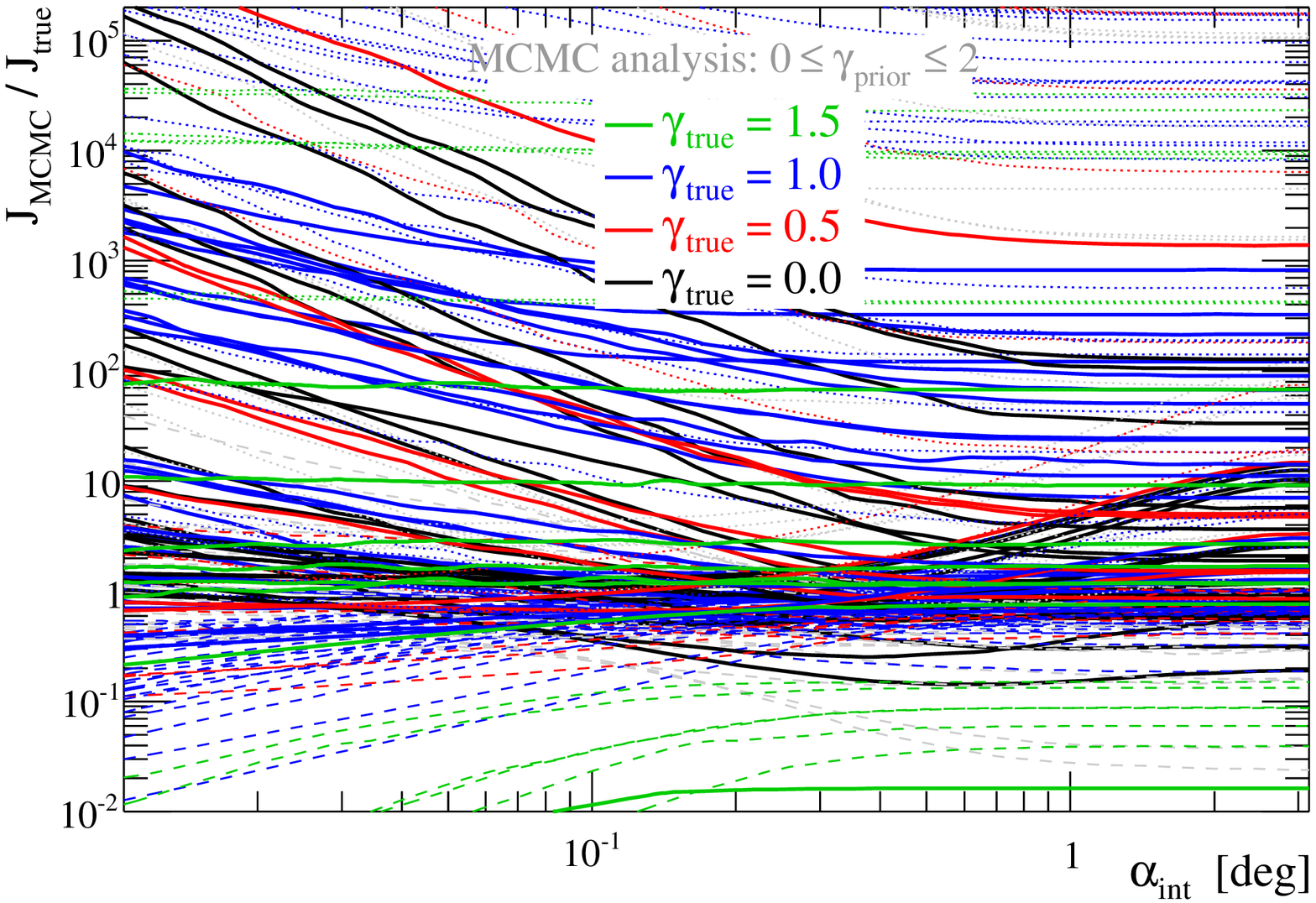}
\includegraphics[width=0.49\linewidth]{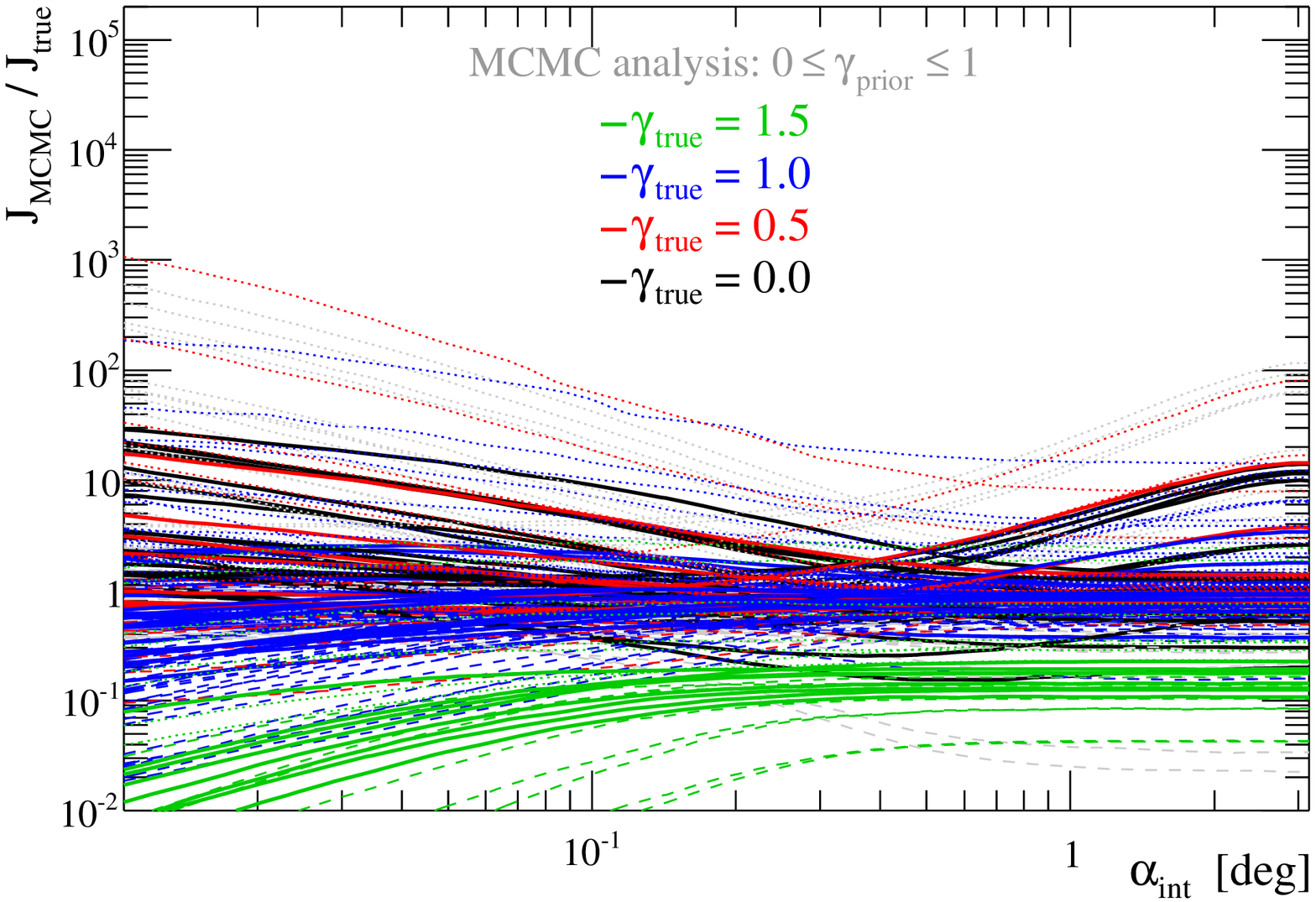}
\caption{Ratio of the MCMC profile to the true profile. The lines are
  colour-coded with respect to the value of the true inner slope
  $\gamma_{\rm true}$ of the artificial data. {\bf Top panels:} ratio
  of the median $\rho(r)$.  The two vertical gray dashed lines
  correspond to the typical range within which the artificial data bin
  are taken. {\bf Bottom panels:} ratio of $J(\alpha_{\rm int})$ for
  the artificial dSphs located at 100 kpc.  {\bf Left panels:} MCMC
  analysis with the prior $0\leq\gamma_{\rm prior}\leq2$.  {\bf Right
    panels:} the prior is $0\leq\gamma_{\rm prior}\leq1$.}
\label{fig:fake_par6}
\end{figure*}
We start with the free $\gamma_{\rm prior}$ analysis (see
Sec.~\ref{subsec:gamma-free}) based on two different priors. Top
panels of Fig.~\ref{fig:fake_par6} show the ratio of the reconstructed
median profile to the true profile.  There is no significant
differences for $\rho(r\gtrsim 1$~kpc) when using the prior
$0\leq\gamma_{\rm prior}\leq2$ (top right) or $0\leq\gamma_{\rm
  prior}\leq1$ (top left): at large radii, the profile does not depend
any longer on the $\gamma$ parameter.  However, it is striking to see
that restricting the prior to $0\leq\gamma_{\rm prior}\leq1$ greatly
improves the determination of the inner regions for the profile,
regardless of the value of $\gamma_{\rm true}$. Even for $\gamma_{\rm
  true}=1.5$ (green curves), using an incorrect prior does not degrade
to much the reconstruction of the profile.

This results is further emphasised when looking at $J$.  
The bottom panels of Fig.~\ref{fig:fake_par6}
are plotted with the same scale to emphasise the difference. As $J$
integrates over the inner parts of the profile, the median MCMC
value can strongly differ from the true value for cuspy profiles.
This difference can reach up to 5 orders of magnitude (over the whole
range of $\alpha_{\rm int}$) for $\gamma_{\rm true}\gtrsim 0.5$ when using
the prior $0\leq\gamma_{\rm prior}\leq2$ . The prior 
$0\leq\gamma_{\rm prior}\leq1$ does generally better, and
accordingly, the confidence intervals are much smaller than for the
other prior (for any integration angle).

The behaviour of the $\gamma_{\rm true}=1.5$ case is unexpected. Using
the prior $0\leq\gamma_{\rm prior}\leq1$ does better than the other
one for any integration angle.  Indeed, even if the reconstructed
median value is shifted by a factor of 10, its CLs correctly encompass
the true value. It does better than the $0\leq\gamma_{\rm prior}\leq1$ prior,
which correctly provides CLs (that bracket the true value), but which are
completely useless as these CLs can vary on $\sim 8$ orders of magnitude.

\subsection{Strong prior: $\gamma_{\rm prior}$ fixed}
\label{app:biases2}

\begin{figure}
\includegraphics[width=\linewidth]{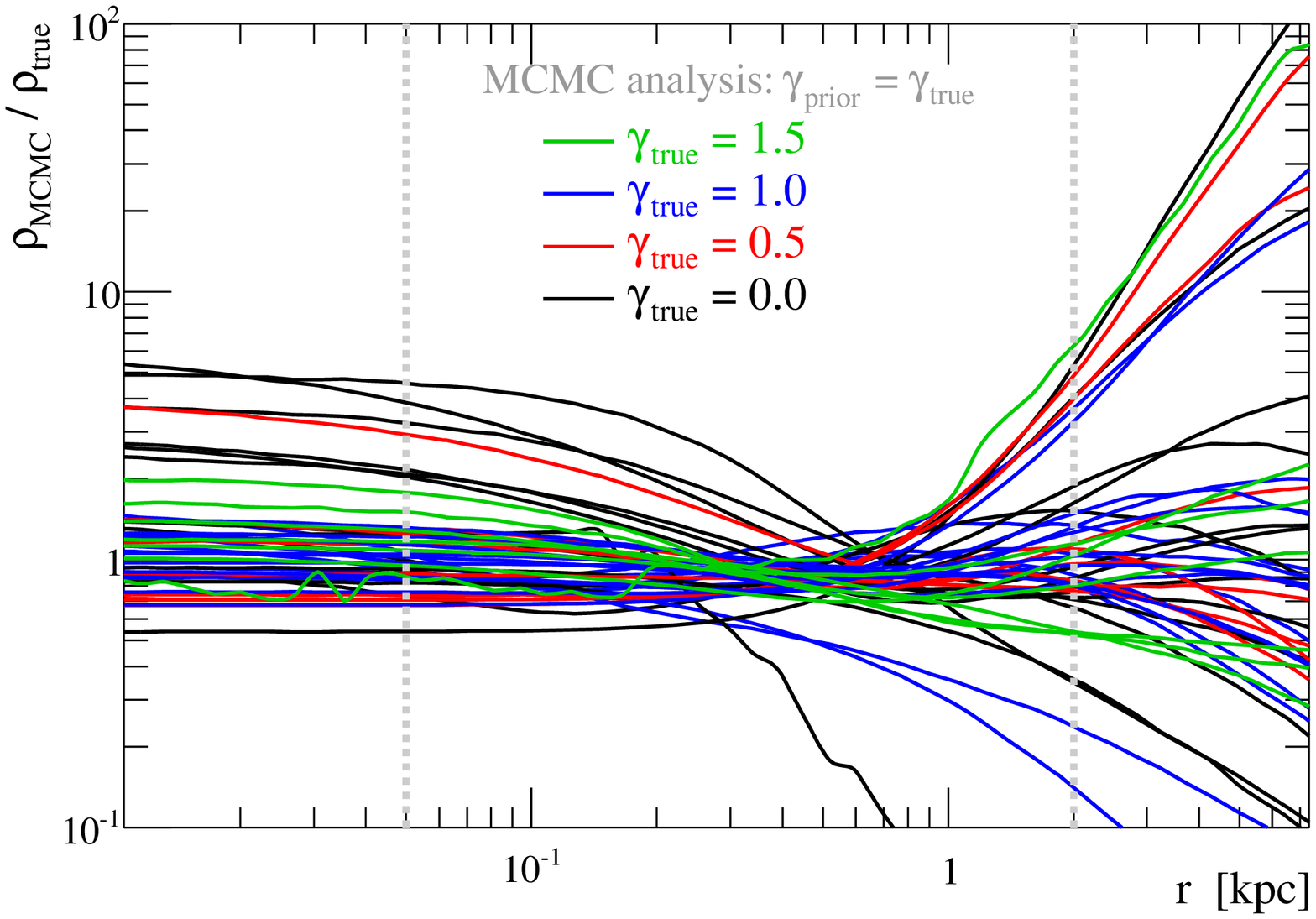}
\includegraphics[width=\linewidth]{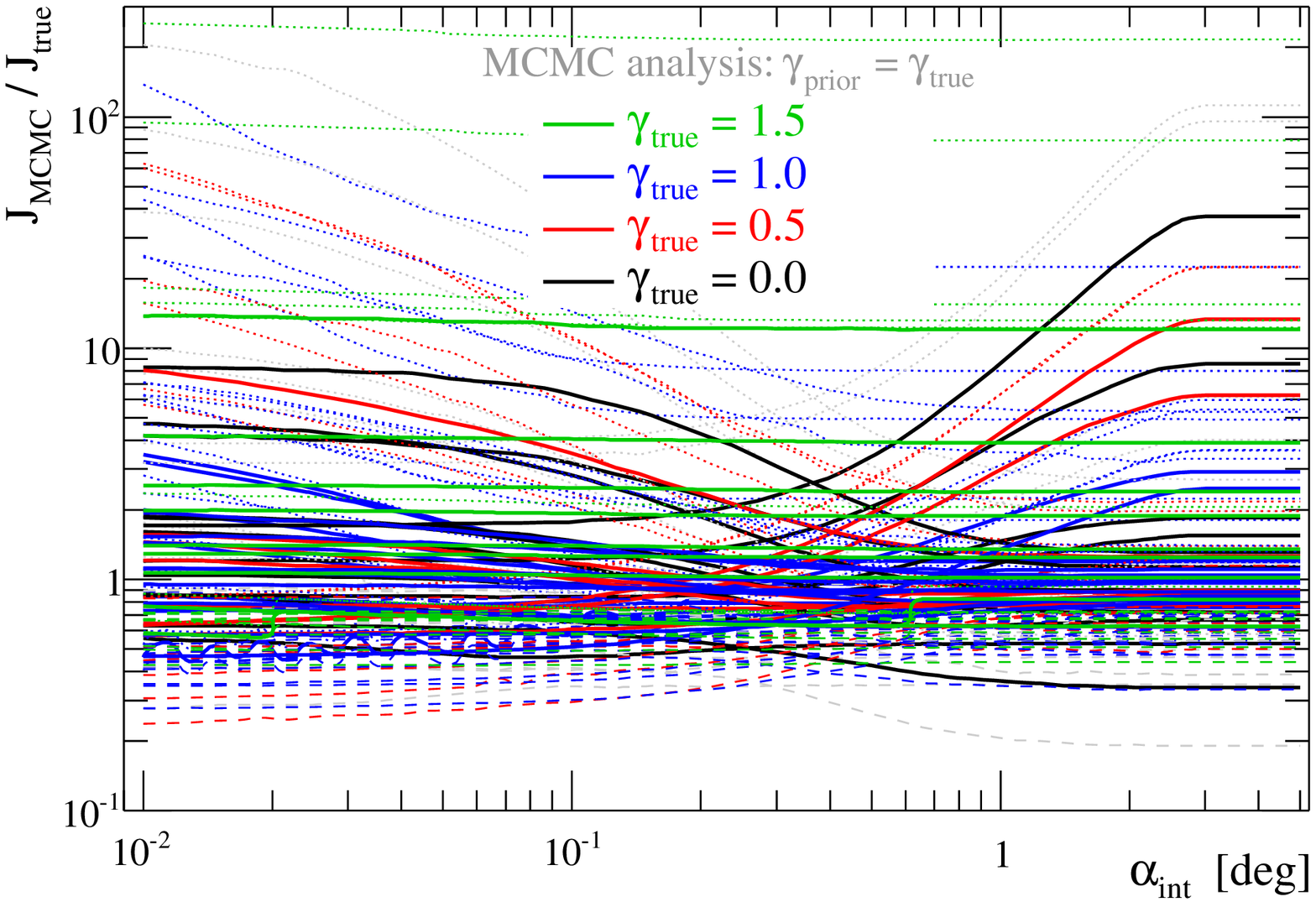}
\caption{Fixed $\gamma_{\rm prior}$ MCMC analysis. {\bf Top:} $\rho(r)$.
{\bf Bottom:} $J(\alpha_{\rm int})$.}
\label{fig:fake_J_par5}
\end{figure}
In Fig.~\ref{fig:fake_J_par5} below, we use a prior $\gamma_{\rm
  prior}=0$ for models having $\gamma_{\rm true}=0$, a prior
$\gamma_{\rm prior}=0.5$ for models having $\gamma_{\rm true}=0.5$,
etc.

A comparison of Figs.~\ref{fig:fake_par6} (using $0\leq\gamma_{\rm
  prior}\leq1$ or $0\leq\gamma_{\rm prior}\leq2$) and
\ref{fig:fake_J_par5} (fixed $\gamma_{\rm prior}$) shows that the latter
prior only slightly improves the precision of
the $J$-factor reconstruction for $\gamma_{\rm true}=0$, $\gamma_{\rm
true}=0.5$, and $\gamma_{\rm true}=1$. However, if $\gamma_{\rm
true}=1.5$ (green curves), although the corresponding $J$-factor is
now better reconstructed than when using the prior $0\leq\gamma_{\rm
prior}\leq2$ (Figs~\ref{fig:fake_par6}, top panel), it is surprisingly
less reliable than the strongly biased $0\leq\gamma_{\rm prior}\leq1$
prior.

The main conclusion is that the knowledge of $\gamma_{\rm true}$
does not help providing tighter constraints on $J$: the uncertainty remains a
factor of a few, except when the inner profile is really cuspy
($\gamma_{\rm true}=1.5$), in which case it becomes strongly biased/unreliable.

\section{Other reconstruction 'biases' on the $J$-factor}
\label{app:biases}

In this Appendix, the MCMC analysis is performed
 based on the prior $0\leq\gamma_{\rm prior}\leq1$, for which the analysis
 is found to be the most robust (see previous Appendix).

\subsection{Impact of the binning of the stars}
 \label{app:binning}

\begin{figure*}
\includegraphics[width=0.495\linewidth]{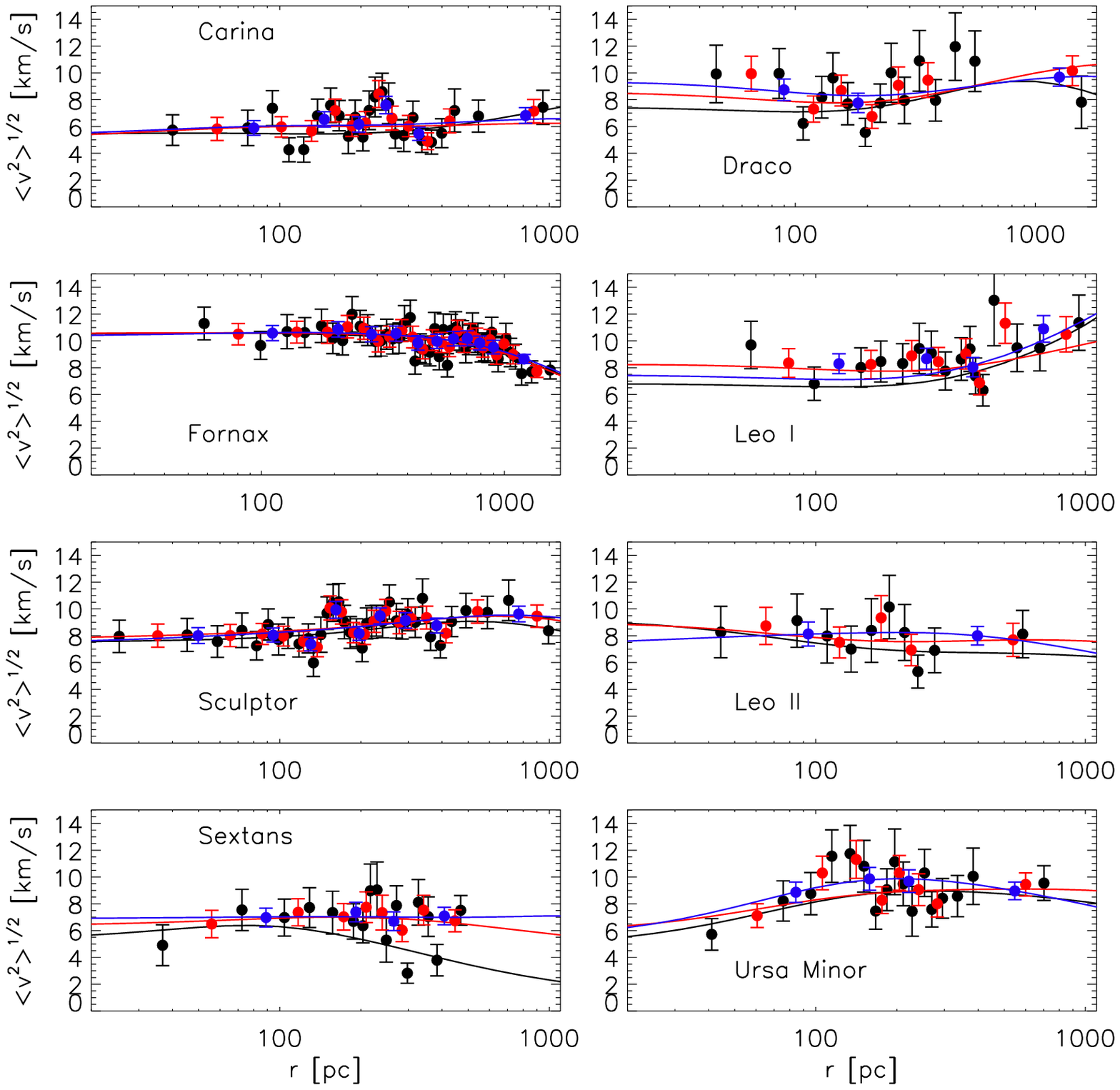}
\includegraphics[width=0.48\linewidth]{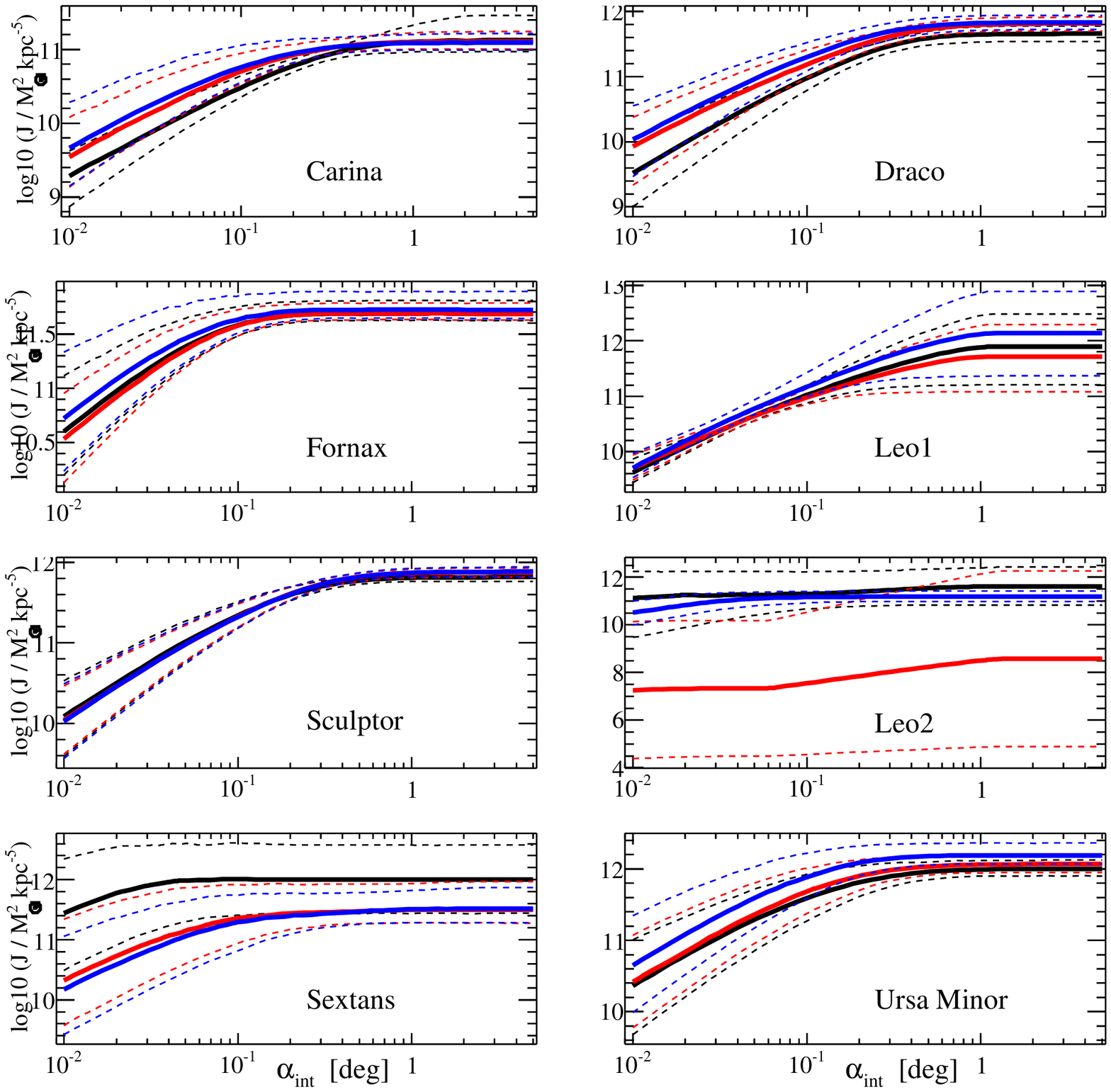}
\caption{For the 8 classical dSphs, impact of using several binnings
  of the stars: black is the binning used throughout this paper
  (i.e. $\sqrt{N}$ bins), red has $\sqrt{N}/2$ bins, and blue has
  $\sqrt{N}/4$ bins. {\bf Left:} velocity dispersion as a function of
  $\log(r)$ (symbols are data, lines are the MCMC median values based
  on the data). {\bf Right:} corresponding median values and 68\% CI
  for $J(\alpha_{\rm int})$.}
\label{fig:momentprofiles_binning}
\end{figure*}

 Figure~\ref{fig:momentprofiles_binning} shows the impact of using
 different binnings in the MCMC analysis. The left panel shows the
 reconstructed (median) value of the velocity dispersion as a function
 of the logarithm of $r$ (to emphasise the differences at small
 radii), for a binning used in this paper (black; where each of
 $\sqrt{N}$ bins has $\sqrt{N}$ member stars, where $N$ is the total
 number of members), a binning with two times (red) and four times
 (blue) fewer bins. For Fornax and Sculptor, the profiles are
 insensitive to the binning chosen, so that the reconstruction of the
 $J$ values median and 68\% CLs (right panel) is robust. For other
 dSphs, either the adjusted velocity dispersion profile is affected at
 small radii, or at large radii.  In the latter case, the $J$
 calculation should not be affected, as the outer
 part does not contribute much to the annihilation signal. In the
 former case, a deviation even at small radii can affect the
 associated $J$ by a factor of a few. The exact impact depends on the
 integration angle, the distance to the dSph (which corresponds to a
 given radius), and the 'cuspiness' of the reconstructed profile (the
 $J$ value of a core profile will be less sensitive to differences in
 the inner parts than would be a cuspy profile).  For instance, Draco
 and Leo1 both have a 2 km/s uncertainty below 100 pc, but Draco is
 three times closer than Leo1: their $J$ for a given $\alpha_{\rm
   int}$ have different behaviours (right panel).  The strongest
 impact is for Leo1 that have the fewest data.  The flatness of the
 $J$ curve seems to indicate a cuspy profile (all the signal in the
 very inner parts), which we know are the least well
 reconstructed ones (see Appendix~\ref{app:biases1}). 
 Leo1 is thus the most sensitive dSph to the binning, for which
 a balance between a sufficient
 coverage over $r$ and small error bars cannot be achieved. The
 ultra-faints dSphs are expected to have even fewer stars, so that
 their $J$ calculation is expected to be even more uncertain.
 
 Overall, the choice of the binning can produce an additional bias of
 a few on the $J$ reconstruction. This is an extra uncertainty factor
 that makes Fornax and Sculptor the more robust targets with respect
 to their annihilation signal. Surveys in the inner parts and outer
 parts of Carina, Draco, Sextans, Leo I, Leo II and Ursa Minor are
 desired to get rid of this binning bias.

%

\subsection{Impact of the choice of the light profile}
 \label{app:light_profile}

\begin{figure*}
\includegraphics[width=0.7\linewidth]{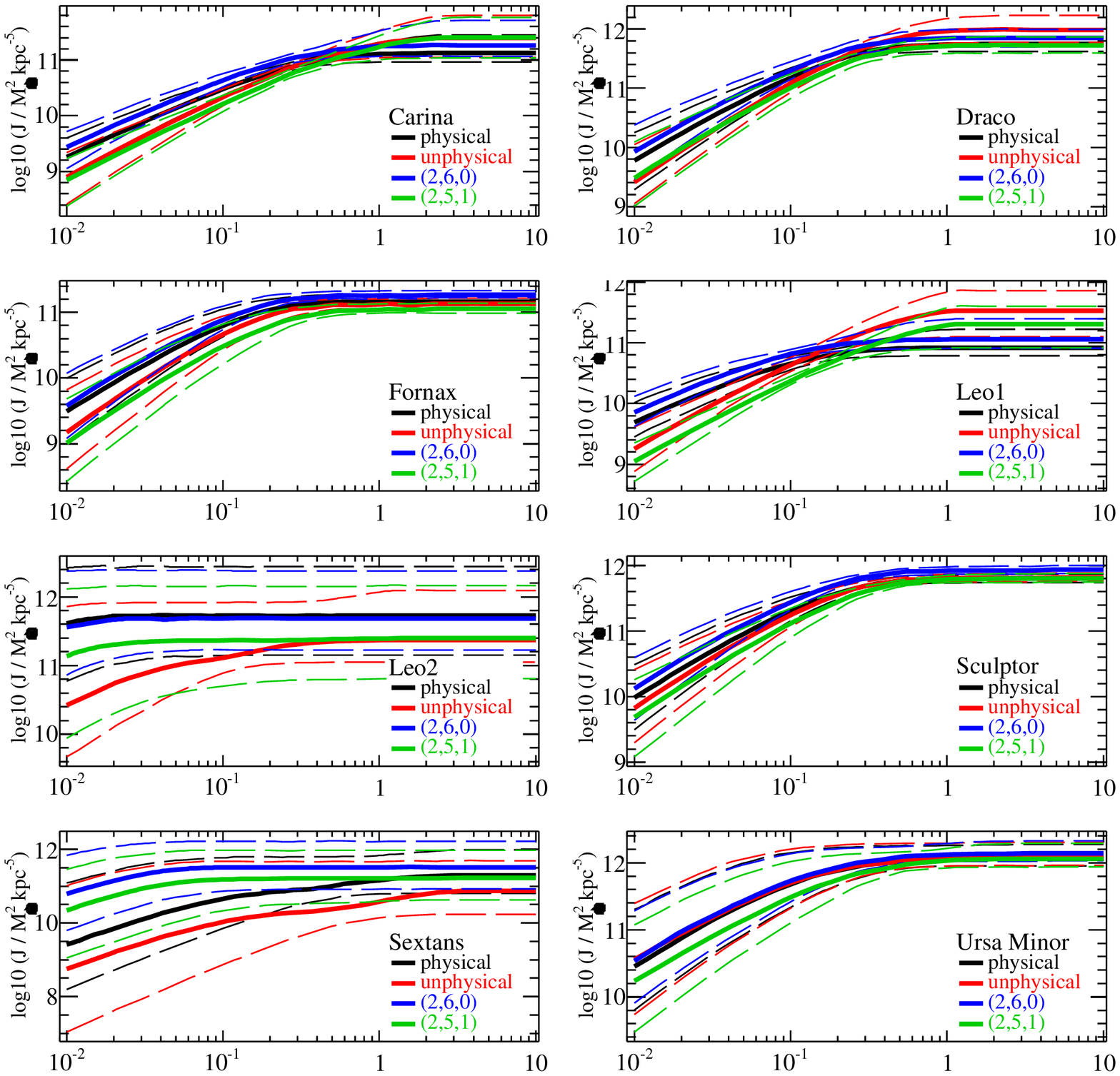}
\caption{For the 8 classical dSphs, impact of using different light-profiles (see text).
Solid lines are the MCMC median values and dashed lines the 68\% CI
for $J(\alpha_{\rm int})$.}
\label{fig:biases}
\end{figure*}

Figure~\ref{fig:biases} shows the various median values
and 68\% CIs of J when changing the assumptions made on the light profile. The black lines
labeled 'physical' correspond to the Plummer model used for the main analysis
(see Eq.~\ref{eq:plummer}); the red lines labeled 'unphysical' are also Plummer,
but the physical constraint given by Eq.~(\ref{eq:anevans}) is relaxed; the blue lines and green lines
correspond respectively to a light profile modeled with an $(\alpha,\,\beta,\,\gamma)$
profile in order to get a steeper outer slope $(2,6,0)$ or a steeper inner slope
$(2,5,1)$ with respect to the Plummer profile.
Regardless of the light profile used, we recover similar critical angles for which
$J$ is the most constrained. The impact on the $J$ value is strongest for the
least-well measured profiles (Leo 2 and Sextans), but is contained within the
95\% CI and marginally within the 68\% CL.

\label{lastpage}
\bibliography{jbar_project}

\begin{thebibliography}{}

\bibitem[\protect\citeauthoryear{{Abdo} et~al.,}{{Abdo}
  et~al.}{2010}]{2010ApJ...712..147A}
{Abdo} A.~A.,  et~al., 2010, \apj, 712, 147

\bibitem[\protect\citeauthoryear{{Abramowski} et~al.,}{{Abramowski}
  et~al.}{2011a}]{2011APh....34..608H}
{Abramowski} A.,  et~al., 2011a, Astroparticle Physics, 34, 608

\bibitem[\protect\citeauthoryear{{Abramowski} et~al.,}{{Abramowski}
  et~al.}{2011b}]{2011PhRvL.106p1301A}
{Abramowski} A.,  et~al., 2011b, Physical Review Letters, 106, 161301

\bibitem[\protect\citeauthoryear{{Acciari} et~al.,}{{Acciari}
  et~al.}{2010}]{2010ApJ...720.1174A}
{Acciari} V.~A.,  et~al., 2010, \apj, 720, 1174

\bibitem[\protect\citeauthoryear{{AGIS Collaboration}}{{AGIS
  Collaboration}}{2010}]{agis:website}
{AGIS Collaboration}, 2010, AGIS

\bibitem[\protect\citeauthoryear{{Aharonian} et~al.,}{{Aharonian}
  et~al.}{2004}]{2004A&A...425L..13A}
{Aharonian} F.,  et~al., 2004, \aap, 425, L13

\bibitem[\protect\citeauthoryear{{Albert} et~al.,}{{Albert}
  et~al.}{2008}]{Albert2008}
{Albert} J.,  et~al., 2008, \apj, 679, 428

\bibitem[\protect\citeauthoryear{{Amorisco} \& {Evans}}{{Amorisco} \&
  {Evans}}{2011}]{amorisco10}
{Amorisco} N.~C.,  {Evans} N.~W.,  2011, \mnras, 411, 2118

\bibitem[\protect\citeauthoryear{{An} \& {Evans}}{{An} \&
  {Evans}}{2006}]{2006ApJ...642..752A}
{An} J.~H.,  {Evans} N.~W.,  2006, \apj, 642, 752

\bibitem[\protect\citeauthoryear{{Battaglia}, {Helmi}, {Tolstoy}, {Irwin},
  {Hill} \& {Jablonka}}{{Battaglia} et~al.}{2008}]{2008ApJ...681L..13B}
{Battaglia} G.,  {Helmi} A.,  {Tolstoy} E.,  {Irwin} M.,  {Hill} V.,
  {Jablonka} P.,  2008, \apjl, 681, L13

\bibitem[\protect\citeauthoryear{{Berezinsky}, {Gurevich} \&
  {Zybin}}{{Berezinsky} et~al.}{1992}]{1992PhLB..294..221B}
{Berezinsky} V.~S.,  {Gurevich} A.~V.,    {Zybin} K.~P.,  1992, Physics Letters
  B, 294, 221

\bibitem[\protect\citeauthoryear{{Berge}, {Funk} \& {Hinton}}{{Berge}
  et~al.}{2007}]{berge_bg}
{Berge} D.,  {Funk} S.,    {Hinton} J.,  2007, \aap, 466, 1219

\bibitem[\protect\citeauthoryear{{Bergstr{\"o}m} \& {Hooper}}{{Bergstr{\"o}m}
  \& {Hooper}}{2006}]{2006PhRvD..73f3510B}
{Bergstr{\"o}m} L.,  {Hooper} D.,  2006, \prd, 73, 063510

\bibitem[\protect\citeauthoryear{{Bergstr{\"o}m} \& {Ullio}}{{Bergstr{\"o}m} \&
  {Ullio}}{1997}]{1997NuPhB.504...27B}
{Bergstr{\"o}m} L.,  {Ullio} P.,  1997, Nuclear Physics B, 504, 27

\bibitem[\protect\citeauthoryear{{Bergstr{\"o}m}, {Ullio} \&
  {Buckley}}{{Bergstr{\"o}m} et~al.}{1998}]{1998APh.....9..137B}
{Bergstr{\"o}m} L.,  {Ullio} P.,    {Buckley} J.~H.,  1998, Astroparticle
  Physics, 9, 137

\bibitem[\protect\citeauthoryear{{Bernl{\"o}hr}, {Carmona} \&
  {Schweizer}}{{Bernl{\"o}hr} et~al.}{2008}]{2008ICRC....3.1469B}
{Bernl{\"o}hr} K.,  {Carmona} E.,    {Schweizer} T.,  2008, in International
  Cosmic Ray Conference Vol.~3 of International Cosmic Ray Conference, {MC
  Simulation and Layout Studies for a future Cherenkov Telescope Array}.
pp 1469--1472

\bibitem[\protect\citeauthoryear{{Bertone}, {Hooper} \& {Silk}}{{Bertone}
  et~al.}{2005}]{2005PhR...405..279B}
{Bertone} G.,  {Hooper} D.,    {Silk} J.,  2005, \physrep, 405, 279

\bibitem[\protect\citeauthoryear{{Binney} \& {Tremaine}}{{Binney} \&
  {Tremaine}}{2008}]{bt08}
{Binney} J.,  {Tremaine} S.,  2008, {Galactic Dynamics: Second Edition}.
Princeton University Press

\bibitem[\protect\citeauthoryear{{Boyarsky}, {Neronov}, {Ruchayskiy},
  {Shaposhnikov} \& {Tkachev}}{{Boyarsky} et~al.}{2006}]{boyarsky06}
{Boyarsky} A.,  {Neronov} A.,  {Ruchayskiy} O.,  {Shaposhnikov} M.,
  {Tkachev} I.,  2006, Physical Review Letters, 97, 261302

\bibitem[\protect\citeauthoryear{{Bringmann}, {Bergstr{\"o}m} \&
  {Edsj{\"o}}}{{Bringmann} et~al.}{2008}]{2008JHEP...01..049B}
{Bringmann} T.,  {Bergstr{\"o}m} L.,    {Edsj{\"o}} J.,  2008, Journal of High
  Energy Physics, 1, 49

\bibitem[\protect\citeauthoryear{{Bringmann}, {Doro} \& {Fornasa}}{{Bringmann}
  et~al.}{2009}]{2009JCAP...01..016B}
{Bringmann} T.,  {Doro} M.,    {Fornasa} M.,  2009, Journal of Cosmology and
  Astro-Particle Physics, 1, 16

\bibitem[\protect\citeauthoryear{{Bullock}, {Kolatt}, {Sigad}, {Somerville},
  {Kravtsov}, {Klypin}, {Primack} \& {Dekel}}{{Bullock}
  et~al.}{2001}]{2001MNRAS.321..559B}
{Bullock} J.~S.,  {Kolatt} T.~S.,  {Sigad} Y.,  {Somerville} R.~S.,  {Kravtsov}
  A.~V.,  {Klypin} A.~A.,  {Primack} J.~R.,    {Dekel} A.,  2001, \mnras, 321,
  559

\bibitem[\protect\citeauthoryear{{Cannoni}, {G{\'o}mez}, {S{\'a}nchez-Conde},
  {Prada} \& {Panella}}{{Cannoni} et~al.}{2010}]{2010PhRvD..81j7303C}
{Cannoni} M.,  {G{\'o}mez} M.~E.,  {S{\'a}nchez-Conde} M.~A.,  {Prada} F.,
  {Panella} O.,  2010, \prd, 81, 107303

\bibitem[\protect\citeauthoryear{{Cole}, {Dehnen} \& {Wilkinson}}{{Cole}
  et~al.}{2011}]{2011arXiv1105.4050C}
{Cole} D.,  {Dehnen} W.,    {Wilkinson} M.,  2011, ArXiv e-prints

\bibitem[\protect\citeauthoryear{{CTA Consortium}}{{CTA
  Consortium}}{2010}]{2010arXiv1008.3703C}
{CTA Consortium} 2010, ArXiv e-prints

\bibitem[\protect\citeauthoryear{{Cuddeford}}{{Cuddeford}}{1991}]{1991MNRAS.25%
3..414C}
{Cuddeford} P.,  1991, \mnras, 253, 414

\bibitem[\protect\citeauthoryear{{de Blok}}{{de Blok}}{2010}]{deblok10}
{de Blok} W.~J.~G.,  2010, Advances in Astronomy, 2010

\bibitem[\protect\citeauthoryear{{Dehnen}}{{Dehnen}}{1993}]{1993MNRAS.265..250%
D}
{Dehnen} W.,  1993, \mnras, 265, 250

\bibitem[\protect\citeauthoryear{{Diemand}, {Kuhlen}, {Madau}, {Zemp}, {Moore},
  {Potter} \& {Stadel}}{{Diemand} et~al.}{2008}]{2008Natur.454..735D}
{Diemand} J.,  {Kuhlen} M.,  {Madau} P.,  {Zemp} M.,  {Moore} B.,  {Potter} D.,
     {Stadel} J.,  2008, \nat, 454, 735

\bibitem[\protect\citeauthoryear{{Diemand}, {Moore} \& {Stadel}}{{Diemand}
  et~al.}{2005}]{2005Natur.433..389D}
{Diemand} J.,  {Moore} B.,    {Stadel} J.,  2005, \nat, 433, 389

\bibitem[\protect\citeauthoryear{{Essig}, {Sehgal} \& {Strigari}}{{Essig}
  et~al.}{2009}]{2009PhRvD..80b3506E}
{Essig} R.,  {Sehgal} N.,    {Strigari} L.~E.,  2009, \prd, 80, 023506

\bibitem[\protect\citeauthoryear{Evans, Ferrer \& Sarkar}{Evans
  et~al.}{2004}]{Evans:2003sc}
Evans N.~W.,  Ferrer F.,    Sarkar S.,  2004, Phys. Rev., D69, 123501

\bibitem[\protect\citeauthoryear{{Fermi-LAT Collaboration}}{{Fermi-LAT
  Collaboration}}{2010}]{fermi:website}
{Fermi-LAT Collaboration}, 2010, LAT Performance

\bibitem[\protect\citeauthoryear{{Fornengo}, {Pieri} \& {Scopel}}{{Fornengo}
  et~al.}{2004}]{2004PhRvD..70j3529F}
{Fornengo} N.,  {Pieri} L.,    {Scopel} S.,  2004, \prd, 70, 103529

\bibitem[\protect\citeauthoryear{{Funk}, {Reimer}, {Torres} \& {Hinton}}{{Funk}
  et~al.}{2008}]{2008ApJ...679.1299F}
{Funk} S.,  {Reimer} O.,  {Torres} D.~F.,    {Hinton} J.~A.,  2008, \apj, 679,
  1299

\bibitem[\protect\citeauthoryear{{Goerdt}, {Moore}, {Read} \&
  {Stadel}}{{Goerdt} et~al.}{2010}]{2010ApJ...725.1707G}
{Goerdt} T.,  {Moore} B.,  {Read} J.~I.,    {Stadel} J.,  2010, \apj, 725, 1707

\bibitem[\protect\citeauthoryear{{Goerdt}, {Moore}, {Read}, {Stadel} \&
  {Zemp}}{{Goerdt} et~al.}{2006}]{2006MNRAS.368.1073G}
{Goerdt} T.,  {Moore} B.,  {Read} J.~I.,  {Stadel} J.,    {Zemp} M.,  2006,
  \mnras, 368, 1073

\bibitem[\protect\citeauthoryear{{Governato}, {Brook}, {Mayer}, {Brooks},
  {Rhee}, {Wadsley}, {Jonsson}, {Willman}, {Stinson}, {Quinn} \&
  {Madau}}{{Governato} et~al.}{2010}]{2010Natur.463..203G}
{Governato} F.,  {Brook} C.,  {Mayer} L.,  {Brooks} A.,  {Rhee} G.,  {Wadsley}
  J.,  {Jonsson} P.,  {Willman} B.,  {Stinson} G.,  {Quinn} T.,    {Madau} P.,
  2010, \nat, 463, 203

\bibitem[\protect\citeauthoryear{{Gunn}, {Lee}, {Lerche}, {Schramm} \&
  {Steigman}}{{Gunn} et~al.}{1978}]{1978ApJ...223.1015G}
{Gunn} J.~E.,  {Lee} B.~W.,  {Lerche} I.,  {Schramm} D.~N.,    {Steigman} G.,
  1978, \apj, 223, 1015

\bibitem[\protect\citeauthoryear{{Hargreaves}, {Gilmore} \&
  {Annan}}{{Hargreaves} et~al.}{1996}]{hargreaves96b}
{Hargreaves} J.~C.,  {Gilmore} G.,    {Annan} J.~D.,  1996, \mnras, 279, 108

\bibitem[\protect\citeauthoryear{{Hastings}}{{Hastings}}{1970}]{hastings70}
{Hastings} W.~K.,  1970, \textit{Biometrika}, 57, 97

\bibitem[\protect\citeauthoryear{{Hernquist}}{{Hernquist}}{1990}]{hernquist90}
{Hernquist} L.,  1990, \apj, 356, 359

\bibitem[\protect\citeauthoryear{{Hisano}, {Matsumoto} \& {Nojiri}}{{Hisano}
  et~al.}{2004}]{2004PhRvL..92c1303H}
{Hisano} J.,  {Matsumoto} S.,    {Nojiri} M.~M.,  2004, Physical Review
  Letters, 92, 031303

\bibitem[\protect\citeauthoryear{{Hisano}, {Matsumoto}, {Nojiri} \&
  {Saito}}{{Hisano} et~al.}{2005}]{2005PhRvD..71f3528H}
{Hisano} J.,  {Matsumoto} S.,  {Nojiri} M.~M.,    {Saito} O.,  2005, \prd, 71,
  063528

\bibitem[\protect\citeauthoryear{{Hogan} \& {Dalcanton}}{{Hogan} \&
  {Dalcanton}}{2000}]{2000PhRvD..62f3511H}
{Hogan} C.~J.,  {Dalcanton} J.~J.,  2000, \prd, 62, 063511

\bibitem[\protect\citeauthoryear{{Horns}}{{Horns}}{2005}]{hesspsf}
{Horns} D.,  2005, Physics Letters B, 607, 225

\bibitem[\protect\citeauthoryear{{Irwin} \& {Hatzidimitriou}}{{Irwin} \&
  {Hatzidimitriou}}{1995}]{ih95}
{Irwin} M.,  {Hatzidimitriou} D.,  1995, \mnras, 277, 1354

\bibitem[\protect\citeauthoryear{{Ishiyama}, {Makino} \&
  {Ebisuzaki}}{{Ishiyama} et~al.}{2010}]{2010ApJ...723L.195I}
{Ishiyama} T.,  {Makino} J.,    {Ebisuzaki} T.,  2010, \apjl, 723, L195

\bibitem[\protect\citeauthoryear{{Jungman}, {Kamionkowski} \&
  {Griest}}{{Jungman} et~al.}{1996}]{1996PhR...267..195J}
{Jungman} G.,  {Kamionkowski} M.,    {Griest} K.,  1996, \physrep, 267, 195

\bibitem[\protect\citeauthoryear{{King}}{{King}}{1962}]{king62}
{King} I.,  1962, \aj, 67, 471

\bibitem[\protect\citeauthoryear{{Kleyna}, {Wilkinson}, {Gilmore} \&
  {Evans}}{{Kleyna} et~al.}{2003}]{2003ApJ...588L..21K}
{Kleyna} J.~T.,  {Wilkinson} M.~I.,  {Gilmore} G.,    {Evans} N.~W.,  2003,
  \apjl, 588, L21

\bibitem[\protect\citeauthoryear{{Koch}, {Wilkinson}, {Kleyna}, {Gilmore},
  {Grebel}, {Mackey}, {Evans} \& {Wyse}}{{Koch} et~al.}{2007}]{koch07b}
{Koch} A.,  {Wilkinson} M.~I.,  {Kleyna} J.~T.,  {Gilmore} G.~F.,  {Grebel}
  E.~K.,  {Mackey} A.~D.,  {Evans} N.~W.,    {Wyse} R.~F.~G.,  2007, \apj, 657,
  241

\bibitem[\protect\citeauthoryear{{Kuhlen}}{{Kuhlen}}{2010}]{2010AdAst2010E..45%
K}
{Kuhlen} M.,  2010, Advances in Astronomy, 2010

\bibitem[\protect\citeauthoryear{{Kuhlen}, {Diemand} \& {Madau}}{{Kuhlen}
  et~al.}{2008}]{2008ApJ...686..262K}
{Kuhlen} M.,  {Diemand} J.,    {Madau} P.,  2008, \apj, 686, 262

\bibitem[\protect\citeauthoryear{{Lake}}{{Lake}}{1990}]{1990Natur.346...39L}
{Lake} G.,  1990, \nat, 346, 39

\bibitem[\protect\citeauthoryear{{Lavalle}, {Yuan}, {Maurin} \& {Bi}}{{Lavalle}
  et~al.}{2008}]{2008A&A...479..427L}
{Lavalle} J.,  {Yuan} Q.,  {Maurin} D.,    {Bi} X.,  2008, \aap, 479, 427

\bibitem[\protect\citeauthoryear{{Lewis} \& {Bridle}}{{Lewis} \&
  {Bridle}}{2002}]{lewis02}
{Lewis} A.,  {Bridle} S.,  2002, \prd, 66, 103511

\bibitem[\protect\citeauthoryear{{Mamon} \& {{\L}okas}}{{Mamon} \&
  {{\L}okas}}{2005}]{mamon05}
{Mamon} G.~A.,  {{\L}okas} E.~L.,  2005, \mnras, 363, 705

\bibitem[\protect\citeauthoryear{{Martinez}, {Bullock}, {Kaplinghat},
  {Strigari} \& {Trotta}}{{Martinez} et~al.}{2009}]{2009JCAP...06..014M}
{Martinez} G.~D.,  {Bullock} J.~S.,  {Kaplinghat} M.,  {Strigari} L.~E.,
  {Trotta} R.,  2009, Journal of Cosmology and Astro-Particle Physics, 6, 14

\bibitem[\protect\citeauthoryear{{Mashchenko}, {Wadsley} \&
  {Couchman}}{{Mashchenko} et~al.}{2008}]{2008Sci...319..174M}
{Mashchenko} S.,  {Wadsley} J.,    {Couchman} H.~M.~P.,  2008, Science, 319,
  174

\bibitem[\protect\citeauthoryear{{Mateo}, {Olszewski} \& {Walker}}{{Mateo}
  et~al.}{2008}]{mateo08}
{Mateo} M.,  {Olszewski} E.~W.,    {Walker} M.~G.,  2008, \apj, 675, 201

\bibitem[\protect\citeauthoryear{{Mateo}}{{Mateo}}{1998}]{mateo98}
{Mateo} M.~L.,  1998, \araa, 36, 435

\bibitem[\protect\citeauthoryear{{McConnachie} \& {C{\^o}t{\'e}}}{{McConnachie}
  \& {C{\^o}t{\'e}}}{2010}]{2010ApJ...722L.209M}
{McConnachie} A.~W.,  {C{\^o}t{\'e}} P.,  2010, \apjl, 722, L209

\bibitem[\protect\citeauthoryear{{Merritt}, {Graham}, {Moore}, {Diemand} \&
  {Terzi{\'c}}}{{Merritt} et~al.}{2006}]{2006AJ....132.2685M}
{Merritt} D.,  {Graham} A.~W.,  {Moore} B.,  {Diemand} J.,    {Terzi{\'c}} B.,
  2006, \aj, 132, 2685

\bibitem[\protect\citeauthoryear{{Metropolis}, {Rosenbluth}, {Teller} \&
  {Teller}}{{Metropolis} et~al.}{1953}]{metropolis53}
{Metropolis} A.~W.,  {Rosenbluth} M.~N.,  {Teller} A.~H.,    {Teller} E.,
  1953, \textit{Journal of Chemical Physics}, 21, 1087

\bibitem[\protect\citeauthoryear{{Moore}, {Gelato}, {Jenkins}, {Pearce} \&
  {Quilis}}{{Moore} et~al.}{2000}]{Moore:2000fp}
{Moore} B.,  {Gelato} S.,  {Jenkins} A.,  {Pearce} F.~R.,    {Quilis} V.,
  2000, \apj, 535, L21

\bibitem[\protect\citeauthoryear{{Navarro}, {Eke} \& {Frenk}}{{Navarro}
  et~al.}{1996}]{1996MNRAS.283L..72N}
{Navarro} J.~F.,  {Eke} V.~R.,    {Frenk} C.~S.,  1996, \mnras, 283, L72

\bibitem[\protect\citeauthoryear{{Navarro, Frenk \& White}}{{Navarro, Frenk \&
  White}}{1996}]{navarro96}
{Navarro, Frenk \& White} 1996, \apj, 462, 563

\bibitem[\protect\citeauthoryear{{Navarro, Frenk \& White}}{{Navarro, Frenk \&
  White}}{1997}]{navarro97}
{Navarro, Frenk \& White} 1997, \apj, 490, 493

\bibitem[\protect\citeauthoryear{{Olszewski}, {Pryor} \&
  {Armandroff}}{{Olszewski} et~al.}{1996}]{edo96}
{Olszewski} E.~W.,  {Pryor} C.,    {Armandroff} T.~E.,  1996, \aj, 111, 750

\bibitem[\protect\citeauthoryear{{Palomares-Ruiz} \&
  {Siegal-Gaskins}}{{Palomares-Ruiz} \&
  {Siegal-Gaskins}}{2010}]{2010JCAP...07..023P}
{Palomares-Ruiz} S.,  {Siegal-Gaskins} J.~M.,  2010, \jcap, 7, 23

\bibitem[\protect\citeauthoryear{{Pieri}, {Bertone} \& {Branchini}}{{Pieri}
  et~al.}{2008}]{2008MNRAS.384.1627P}
{Pieri} L.,  {Bertone} G.,    {Branchini} E.,  2008, \mnras, 384, 1627

\bibitem[\protect\citeauthoryear{Pieri et~al.,}{Pieri
  et~al.}{2009}]{2009A&A...496..351P}
Pieri L.,  et~al., 2009, \aap, 496, 351

\bibitem[\protect\citeauthoryear{{Pieri}, {Lattanzi} \& {Silk}}{{Pieri}
  et~al.}{2009}]{2009MNRAS.399.2033P}
{Pieri} L.,  {Lattanzi} M.,    {Silk} J.,  2009, \mnras, 399, 2033

\bibitem[\protect\citeauthoryear{{Plummer}}{{Plummer}}{1911}]{plummer11}
{Plummer} H.~C.,  1911, \mnras, 71, 460

\bibitem[\protect\citeauthoryear{{Putze}, {Derome}, {Maurin}, {Perotto} \&
  {Taillet}}{{Putze} et~al.}{2009}]{2009A&A...497..991P}
{Putze} A.,  {Derome} L.,  {Maurin} D.,  {Perotto} L.,    {Taillet} R.,  2009,
  \aap, 497, 991

\bibitem[\protect\citeauthoryear{{Read} \& {Gilmore}}{{Read} \&
  {Gilmore}}{2005}]{2005MNRAS.356..107R}
{Read} J.~I.,  {Gilmore} G.,  2005, \mnras, 356, 107

\bibitem[\protect\citeauthoryear{{S{\'a}nchez-Conde}, {Prada}, {{\L}okas},
  {G{\'o}mez}, {Wojtak} \& {Moles}}{{S{\'a}nchez-Conde}
  et~al.}{2007}]{2007PhRvD..76l3509S}
{S{\'a}nchez-Conde} M.~A.,  {Prada} F.,  {{\L}okas} E.~L.,  {G{\'o}mez} M.~E.,
  {Wojtak} R.,    {Moles} M.,  2007, \prd, 76, 123509

\bibitem[\protect\citeauthoryear{{Sersic}}{{Sersic}}{1968}]{sersic68}
{Sersic} J.~L.,  1968, {Atlas de galaxias australes}.
Cordoba, Argentina: Observatorio Astronomico, 1968

\bibitem[\protect\citeauthoryear{{Silk} \& {Bloemen}}{{Silk} \&
  {Bloemen}}{1987}]{1987ApJ...313L..47S}
{Silk} J.,  {Bloemen} H.,  1987, \apjl, 313, L47

\bibitem[\protect\citeauthoryear{{Springel}, {Wang}, {Vogelsberger}, {Ludlow},
  {Jenkins}, {Helmi}, {Navarro}, {Frenk} \& {White}}{{Springel}
  et~al.}{2008}]{2008MNRAS.391.1685S}
{Springel} V.,  {Wang} J.,  {Vogelsberger} M.,  {Ludlow} A.,  {Jenkins} A.,
  {Helmi} A.,  {Navarro} J.~F.,  {Frenk} C.~S.,    {White} S.~D.~M.,  2008,
  \mnras, 391, 1685

\bibitem[\protect\citeauthoryear{{Stecker}}{{Stecker}}{1978}]{1978ApJ...223.10%
32S}
{Stecker} F.~W.,  1978, \apj, 223, 1032

\bibitem[\protect\citeauthoryear{{Strigari}, {Bullock}, {Kaplinghat},
  {Diemand}, {Kuhlen} \& {Madau}}{{Strigari}
  et~al.}{2007}]{2007ApJ...669..676S}
{Strigari} L.~E.,  {Bullock} J.~S.,  {Kaplinghat} M.,  {Diemand} J.,  {Kuhlen}
  M.,    {Madau} P.,  2007, \apj, 669, 676

\bibitem[\protect\citeauthoryear{{Strigari}, {Bullock}, {Kaplinghat}, {Simon},
  {Geha}, {Willman} \& {Walker}}{{Strigari} et~al.}{2008}]{2008Natur.454.1096S}
{Strigari} L.~E.,  {Bullock} J.~S.,  {Kaplinghat} M.,  {Simon} J.~D.,  {Geha}
  M.,  {Willman} B.,    {Walker} M.~G.,  2008, \nat, 454, 1096

\bibitem[\protect\citeauthoryear{{Strigari}, {Frenk} \& {White}}{{Strigari}
  et~al.}{2010}]{2010MNRAS.408.2364S}
{Strigari} L.~E.,  {Frenk} C.~S.,    {White} S.~D.~M.,  2010, \mnras, 408, 2364

\bibitem[\protect\citeauthoryear{{Strigari}, {Koushiappas}, {Bullock} \&
  {Kaplinghat}}{{Strigari} et~al.}{2007}]{2007PhRvD..75h3526S}
{Strigari} L.~E.,  {Koushiappas} S.~M.,  {Bullock} J.~S.,    {Kaplinghat} M.,
  2007, \prd, 75, 083526

\bibitem[\protect\citeauthoryear{{Ullio} \& {Bergstr{\"o}m}}{{Ullio} \&
  {Bergstr{\"o}m}}{1998}]{1998PhRvD..57.1962U}
{Ullio} P.,  {Bergstr{\"o}m} L.,  1998, \prd, 57, 1962

\bibitem[\protect\citeauthoryear{{Valkenburg}, {Krauss} \&
  {Hamann}}{{Valkenburg} et~al.}{2008}]{2008PhRvD..78f3521V}
{Valkenburg} W.,  {Krauss} L.~M.,    {Hamann} J.,  2008, \prd, 78, 063521

\bibitem[\protect\citeauthoryear{{Walker}, {Combet}, {Hinton}, {Maurin} \&
  {Wilkinson}}{{Walker} et~al.}{2011}]{letter_dsph}
{Walker} M.~G.,  {Combet} C.,  {Hinton} J.,  {Maurin} D.,    {Wilkinson} M.~I.,
   2011, ApJL, accepted

\bibitem[\protect\citeauthoryear{{Walker}, {Mateo} \& {Olszewski}}{{Walker}
  et~al.}{2009}]{2009AJ....137.3100W}
{Walker} M.~G.,  {Mateo} M.,    {Olszewski} E.~W.,  2009, \aj, 137, 3100

\bibitem[\protect\citeauthoryear{{Walker}, {Mateo}, {Olszewski}, {Gnedin},
  {Wang}, {Sen} \& {Woodroofe}}{{Walker} et~al.}{2007}]{walker07b}
{Walker} M.~G.,  {Mateo} M.,  {Olszewski} E.~W.,  {Gnedin} O.~Y.,  {Wang} X.,
  {Sen} B.,    {Woodroofe} M.,  2007, \apjl, 667, L53

\bibitem[\protect\citeauthoryear{{Walker}, {Mateo}, {Olszewski},
  {Pe{\~n}arrubia}, {Wyn Evans} \& {Gilmore}}{{Walker}
  et~al.}{2009}]{2009ApJ...704.1274W}
{Walker} M.~G.,  {Mateo} M.,  {Olszewski} E.~W.,  {Pe{\~n}arrubia} J.,  {Wyn
  Evans} N.,    {Gilmore} G.,  2009, \apj, 704, 1274

\bibitem[\protect\citeauthoryear{{Walker}, {Mateo}, {Olszewski}, {Sen} \&
  {Woodroofe}}{{Walker} et~al.}{2009}]{walker09b}
{Walker} M.~G.,  {Mateo} M.,  {Olszewski} E.~W.,  {Sen} B.,    {Woodroofe} M.,
  2009, \aj, 137, 3109

\bibitem[\protect\citeauthoryear{{Wolf}, {Martinez}, {Bullock}, {Kaplinghat},
  {Geha}, {Mu{\~n}oz}, {Simon} \& {Avedo}}{{Wolf}
  et~al.}{2010}]{2010MNRAS.406.1220W}
{Wolf} J.,  {Martinez} G.~D.,  {Bullock} J.~S.,  {Kaplinghat} M.,  {Geha} M.,
  {Mu{\~n}oz} R.~R.,  {Simon} J.~D.,    {Avedo} F.~F.,  2010, \mnras, 406, 1220

\bibitem[\protect\citeauthoryear{{Zhao}}{{Zhao}}{1996}]{zhao96}
{Zhao} H.,  1996, \mnras, 278, 488

\end{thebibliography}
\end{document}